\newif\ifjournal
\newif\ifdoublecol

\journalfalse
\doublecolfalse

\ifdoublecol
\documentclass[10pt,twocolumn,twoside]{IEEEtran}
\else	\documentclass[11pt,draftclsnofoot,onecolumn,twoside,romanappendices]{IEEEtran}
\fi

\usepackage[intlimits,cmex10]{amsmath} 
\usepackage{amsxtra}     
\usepackage{amsthm}
\usepackage{amssymb}
\usepackage{amsfonts}
\usepackage{dsfont}
\usepackage{graphicx}    
\usepackage{multirow}
\usepackage{epsfig}
\usepackage[caption=false]{subfig}   
\usepackage{verbatim}
\usepackage{dbl12}
\usepackage{cite}
\usepackage{bm}
\usepackage[usenames]{color}
\usepackage{algorithm}
\usepackage{algpseudocode}

\usepackage{colortbl}
\usepackage{rotating}
\usepackage{fixltx2e}

\ifjournal
	\ifdoublecol
		\def\heuristicFigHeight{0.15\textwidth}
		\def\policyFigHeight{.18\textwidth}
		\newcommand{\policyLineBreak}{}
		\def\figwidth{0.17\textwidth}
		
		\def\bigfigwidth{.48\textwidth}
		\def\theoryfigheight{3.5cm}
        \def\legendwidth{0.7\textwidth}
        \def\onecolumnlegendwidth{0.48\textwidth}
	\else
		\def\heuristicFigHeight{0.25\textwidth}
		\def\policyFigHeight{.25\textwidth}
		\newcommand{\policyLineBreak}{\\}
		\def\figwidth{0.3\textwidth}
		
		\def\bigfigwidth{.88\textwidth}
		\def\theoryfigheight{4.75cm}
        \def\legendwidth{\textwidth}
        \def\onecolumnlegendwidth{0.7\textwidth}
	\fi
\else
	\def\heuristicFigHeight{0.375\textwidth}
	\def\policyFigHeight{.374\textwidth}
	\newcommand{\policyLineBreak}{\newline}
	\def\figwidth{0.3\textwidth}
	
	\def\bigfigwidth{.88\textwidth}
	\def\theoryfigheight{4.75cm}
\fi

\newcommand{\generalci}[2]{#2}

\newcommand{\set}[1]{\left\{#1\right\}}
\newcommand{\ignore}[1]{}

\newcommand{\vkap}{{\bm \kappa}}
\newcommand{\vkapt}{{\tilde{\bm \kappa}}}
\newcommand{\vy}{{\bm y}}
\newcommand{\vY}{{\bm Y}}
\newcommand{\vS}{{\bm S}}

\newcommand{\vx}{{\bm x}}

\newcommand{\vO}{{\bm O}}

\newcommand{\calX}{{\cal X}}

\newcommand{\vlam}{{\bm \lambda}}
\newcommand{\vlamu}[1]{\bm \lambda}
\newcommand{\vpi}{{\bm \phi}}
\newcommand{\vwi}{{\bm \omega}}

\newcommand{\var}{{\rm var}}

\newcommand{\vyi}{{\bf y}_i}
\newcommand{\vxi}{{\bf \xi}}
\newcommand{\sn}{s^{(n)}}
\newcommand{\snn}[1]{s^{(#1)}}
\newcommand{\xn}{x^{(n)}}

\newcommand{\indic}[1]{\mathbf{1}_{\{#1\}}}

\definecolor{firstcol}{RGB}{251,236,93}
\definecolor{firstsubcol}{RGB}{251,236,200}
\definecolor{firstrow}{RGB}{0,0,139}
\definecolor{linecolor}{named}{White}
\definecolor{textRowHeader}{named}{White}
\definecolor{textColumnHeader}{named}{Black}
\newcolumntype{A}{>{\color{textRowHeader}\columncolor{firstrow}}c}	
\newcolumntype{B}{>{\color{textColumnHeader}\columncolor{firstcol}}l}	
\newcolumntype{C}{>{\color{textColumnHeader}\columncolor{firstsubcol}}l}	

\newenvironment{changemargin}[2]{%
  \begin{list}{}{%
    \setlength{\topsep}{0pt}%
    \setlength{\leftmargin}{#1}%
    \setlength{\rightmargin}{#2}%
    \setlength{\listparindent}{\parindent}%
    \setlength{\itemindent}{\parindent}%
    \setlength{\parsep}{\parskip}%
  }%
  \item[]}{\end{list}}

\newcommand{\E}{\mathbb{E}}
\newcommand{\uni}{u}
\newcommand{\omni}{o}
\newcommand{\semi}{s}
\newcommand{\cbar}{\overline{c}}
\newcommand{\ccrit}{c_{\mathrm{crit}}}
\newcommand{\steady}{\mathrm{ss}}
\newcommand{\Lambdab}{\bar{\Lambda}}
\newcommand{\ptilde}{\tilde{p}_{0}}

\theoremstyle{plain}
\newtheorem{theorem}{Theorem}
\newtheorem{prop}{Proposition}
\newtheorem{corollary}[theorem]{Corollary}
\newtheorem{lemma}{Lemma}

\theoremstyle{definition}
\newtheorem{assumption}{Assumption}

\theoremstyle{remark}
\newtheorem*{remark}{Remark}

\begin{document}
\ifjournal
\else	
\title{Resource-Constrained Adaptive Search and Tracking for Sparse Dynamic Targets}
\fi

\author{Gregory~Newstadt,~\IEEEmembership{Member,~IEEE,}~Dennis~Wei,~\IEEEmembership{Member,~IEEE,}~and~Alfred~O.~Hero~III,~\IEEEmembership{Fellow,~IEEE}
\thanks{Gregory Newstadt and Alfred Hero are with the Dept. of Electrical Engineering and Computer Science, University of Michigan, Ann
Arbor, E-mail: (\{newstage\},\{hero\}@umich.edu).}
\thanks{Dennis Wei is with the IBM T.~J.~Watson Research Center, Yorktown Heights, NY 10598, USA, E-mail: dwei@us.ibm.com.}
\thanks{The research in this paper was partially supported by Air Force Office of Scientific Research award FA9550-06-1-0324, by Air Force Research Laboratory award FA8650-07-D-1221-TO1, and by Army Research Office MURI grant
number W911NF-11-1-0391.}
\thanks{This work was presented in part at IEEE CAMSAP 2013 and Asilomar Conference on Signals, Systems and Computers 2011.}
\\ \today}

\maketitle

\begin{abstract}
This paper considers the problem of resource-constrained and noise-limited localization and estimation of 
dynamic targets that are sparsely distributed over a large area.  
We generalize an existing framework [Bashan et al, 2008] for adaptive allocation of sensing resources to the dynamic case, accounting for 
time-varying target behavior such as transitions to neighboring cells and varying amplitudes over a potentially long time horizon.  
The proposed adaptive sensing policy is driven by minimization of 
a surrogate function for 
mean squared error within locations containing targets.  We provide theoretical upper bounds on the performance of adaptive sensing policies by analyzing solutions with oracle knowledge of target locations, gaining insight into the effect of target motion and amplitude variation as well as sparsity.  Exact minimization of the multi-stage objective function is infeasible, but myopic optimization yields a closed-form solution.  We propose a simple non-myopic extension, the Dynamic Adaptive Resource Allocation Policy (D-ARAP), 
that allocates a fraction 
of resources for exploring all locations rather than solely exploiting the current belief state.  Our numerical studies indicate that D-ARAP has the following advantages: (a) it is more robust than the myopic policy to noise, missing data, and model mismatch; (b) it performs comparably to well-known approximate dynamic programming solutions but at significantly lower computational complexity; and (c) it improves greatly upon non-adaptive uniform resource allocation in terms of estimation error and probability of detection.  

\end{abstract}

\section{Introduction}
\label{darap-chap-sec:intro}
Systems for wide area surveillance such as the Gotcha synthetic aperture radar and the Angel Fire electro-optical sensor currently offer 
near real-time imaging and surveillance of city-sized scenes.  Generally, these systems perform continuous collection of data, followed by forensic analysis to detect and track targets within the scene. The amount of raw data collected by these systems is often very large, often collected uniformly over a large area, even though the interesting features (e.g. moving targets, etc.) exist only in a few locations. 
Due to such data collection inefficiencies, there is a high likelihood that much sensing and/or computational resources may be wasted by searching areas where targets are not located.  Abidi et al \cite{abidi2008survey} provides a comprehensive survey of recent work in wide-area surveillance, including topics in coverage analysis, optimal sensor positioning and sensor fusion.  They stress the need for intelligent data collection and system analysis  in order to deal with such data collection inefficiencies. 

An alternative is adaptive sampling, for which past observations are used 
to inform the collection 
of future observations, with the goal of focusing effort onto the ``interesting'' regions of the search space. Adaptive sampling can be an important 
tool for efficiently managing the collection inefficiency problem faced 
by wide-area surveillance systems.  Previous work has shown that, when constrained to use equal resources, adaptive sampling can significantly improve target localization performance \cite{Castro-04-coarse-to-fine,Castro-05-faster-rate-reg-via-act-learn, Bashan-08-opt-2-stage-search, bashan2011marapTSP, Hitchings_Castanon_AdaptiveSensing2010, haupt2011distilled, haupt2012sequentially, Wei-Hero-multistage-adaptive-estimation-of-sparse-signals,malloy2012sequential,MalloyAdaptiveCompressed13} in comparison to a uniform policy that uses equal sensing effort across the scene.  Benefits of adaptive sensing include: gains in estimation precision \cite{Bashan-08-opt-2-stage-search, bashan2011marapTSP}; provable detection of the targets often at faster convergence rates \cite{haupt2011distilled,malloy2012sequential}; and improved robustness in detection performance as measured by the minimum detectable amplitudes \cite{haupt2012sequentially}.  

Bashan et al \cite{Bashan-08-opt-2-stage-search} provided an optimal two-stage policy called ARAP for adaptively localizing targets and estimating their amplitudes in noise under effort budget constraints.  A  multiscale approach was subsequently introduced in order to further reduce the total number of 
measurements \cite{bashan2011marapTSP}.  Hitchings and Castanon \cite{Hitchings_Castanon_AdaptiveSensing2010} provided an online modification to \cite{Bashan-08-opt-2-stage-search} through Lagrangian constraint relaxation.  Additionally, \cite{Wei-Hero-multistage-adaptive-estimation-of-sparse-signals} extended the two-stage policy to an arbitrary number of stages using approximate dynamic programming, 
and generalized the framework to allow for a variety of measurement/estimation loss models.  Using a similar model called Distilled Sensing, \cite{haupt2011distilled,haupt2012sequentially} specified a methodology for locating targets in noise at much lower signal-to-noise ratios (SNR) than non-adaptive methods. Malloy et al \cite{malloy2012sequential} extends Distilled Sensing to non-Gaussian models, while \cite{MalloyAdaptiveCompressed13,haupt2012sequentially} consider compressive measurements.

In the above methods, it is assumed that the targets of interest remain stationary across sensing/observation epochs.  In wide-area surveillance, however, many targets 
exhibit complex dynamic behavior such as movement, entering/leaving the scene, and  obscuration.   Krishnamurthy \cite{Krishnamurthy-01-hmm-mab-beam-sched} considers the problem of selecting the direction to point an agile sensor in order to track $P$ moving targets among a finite number of cells.  When the state is fully observable, the problem can be posed as a Markov decision process (MDP).  Krishnamurthy formulates the problem as a hidden Markov model (HMM) tracking problem in the more challenging case where the state is observed with noise. 
He 
discusses an optimal policy which depends on the individual target's belief state - the conditional density of the state given the observation history.  Moreover, a suboptimal approach to approximating the optimal selection criterion  is provided to combat the prohibitive computational complexity of the optimal solution.

Chong et al. \cite{chong2008monte} show that many adaptive sensing problems can be formulated as partially observable Markov decision processes (POMDPs).  This general framework is concerned with selecting actions to maximize the sum of an immediate reward and an expected future reward, given a system with Markovian evolution but only partially observable state (e.g.\ due to noise).  Unfortunately, 
the expected future reward is often difficult to compute, while the tractability of the optimal solution is further impaired when the state space and action space are large. Chong et al. \cite{chong2008monte} propose approximate methods that include parametric approximations, reinforcement learning techniques, and rollout policies.  The last of these 
assumes that a base policy is available that may not be optimal but is simple to compute.   Rollout policies then ensure policy improvement, i.e., they are guaranteed to do at least as well as the base policy.  However, rollout policies still remain impractical in cases where the action space is large. 

\ignore{Informally, a POMDP consists of
\begin{itemize}
\item a set of possible states (the state space).
\item a set of possible actions (the action space)
\item a set of possible observations (the observation space)
\item  $r(b_t,a)$, a reward (or cost) function received given an action $a$ taken at a state $b_t$.
\item a state transition law (i.e., distribution)
\item an observation law specifying the distribution of observations given an action taken at a state.
\end{itemize}
Given the current belief state, Bellman's principle states that the optimal objective function value starting from belief state $b_0$ is given by
\begin{equation}
V_T^*(b_0) = \max\limits_a \left( r(b_0,a) + \mathbb{E}[V_{T-1}^*(b_1) | b_0,a]\right).
\end{equation}
Note that $V_T^*(b_0)$ is defined recursively where each subsequent stage chooses the optimal action $a$.  The $Q$-value of an action $a$ at time $t$ with belief state $b_t$ is defined as
\begin{equation}
Q_{T-t}(b_t,a) = r(b_t,a) + \mathbb{E}[V_{T-t-1}^*(b_{t+1}) | b_t,a].
\end{equation}
Bellman's principle can be rewritten as stating the optimal action as being the action that maximizes the $Q$-value:
\begin{equation}
\pi_t^*(b_t) = \arg\max\limits_a Q_{T-t}(b_t,a).
\end{equation}
The $Q$-values are composed of two parts: (a) the immediate reward of taking action $a$, and (b) the expected cumulative reward for the remaining stages assuming optimal actions are taken.  Generally, the first part can be computed in a straightforward manner, but the latter part requires significant and possibly intractable computation.  , particularly when the belief state space and action spaces are large.  Chong et al. \cite{chong2008monte} describes approximate methods in detail that include parametric approximations, reinforcement learning techniques, and rollout policies.   In particular, rollout policies assume that a base policy is available that may not be optimal (as in Bellman's principle), but is simple to compute.  In this case, the $Q$-values can be approximated as 
\begin{equation}
Q^{base}(b,a) = r(b,a) + \mathbb{E}[V^{base}(b')|b,a].
\end{equation}
Computation of the approximate $Q$-values still involves expectations over remaining stages, but removes the degrees of freedom in finding the ``optimal'' actions. } 

In this paper, we extend the Bayesian formulation of adaptive sampling  \cite{Bashan-08-opt-2-stage-search} to targets that are dynamic as well as sparsely located in the scene.  Our model encompasses target motion, appearance and disappearance, and target amplitude variation.  This formulation can simultaneously account for multiple targets as well as allocation of continuous-valued sensing resources, such as energy, in contrast to the discrete resource allocation formulation in \cite{Krishnamurthy-01-hmm-mab-beam-sched}.  Our formulation is based on a simple approximation
to the target posterior distribution and a multistage extension of the cost function in \cite{Bashan-08-opt-2-stage-search}.  
In the context of this framework we then introduce the Dynamic Adaptive Resource Allocation Policy (D-ARAP), a non-myopic policy for adaptive sampling that achieves a favorable trade-off between performance and planning complexity.  Our analysis suggests that as compared to approximate POMDP solutions, in particular rollout policies, D-ARAP performs well but at a fraction of the computational cost.  Compared to myopic policies, D-ARAP has increased robustness to noise, missing data, and model mismatch.  Lastly, compared to the non-adaptive uniform policy, D-ARAP continues to yield large improvements in estimation and detection performance similar to the static case \cite{Bashan-08-opt-2-stage-search,Wei-Hero-multistage-adaptive-estimation-of-sparse-signals}.

For static targets, it is known  \cite{Bashan-08-opt-2-stage-search,bashan2011marapTSP,Wei-Hero-guarantees-adaptive-estimation-sparse-signals} that estimation gains due to adaptive sampling increase in the sparse target regime, where targets occupy few locations in the search area.  The present work shows that the benefits of adaptive sensing framework \cite{Bashan-08-opt-2-stage-search} can be extended to the dynamic setting.
We derive upper bounds on adaptive performance gains 
through analysis of omniscient and semi-omniscient policies with complete or partial knowledge of target locations over time, respectively.  The bounds confirm the benefit of the sparse target regime for dynamic targets.  They also characterize the effect of target motion and amplitude variation on the potential gain.  
\ignore{
More specifically, we first consider an ``omniscient'' policy, which has complete knowledge of 
target locations at all times, to explore the limiting effect of target sparsity. 
The omniscient policy however is non-causal when 
targets may enter, leave, or transition between cells, implying that the omniscient upper bound is not attainable even in the limit of high SNR or number of planning stages.  Accordingly, we also consider a ``semi-omniscient'' policy that has only causal knowledge of target locations. 
While analysis of the semi-omniscient policy involves some additional assumptions compared to the omniscient policy, the resulting performance bounds are tighter.}
We show simulations that indicate that D-ARAP can approach the semi-omniscient performance as the SNR and number of stages increase. 
Furthermore, comparison of the omniscient and semi-omniscient policies allows us to quantify the effect of partial and causal knowledge of target motion. 

\ignore{The current paper improves upon our previous work \cite{NewstadtWeiHeroCAMSAP13,NewstadtBashanHeroAsilomar2011} by providing an offline rollout policy in Section \ref{darap-sec:adaptive-policy} whose training complexity scales linearly with the number of stages, with no noticeable performance degradation compared to the nested policy in \cite{NewstadtWeiHeroCAMSAP13} that had quadratic complexity.  In addition, Section \ref{darap-sec:theory} on theoretical performance bounds is new, simulation results have been significantly expanded, and the presentation of our framework has been refined. }


The rest of this paper is organized as follows.  
We formalize the problem in Section \ref{darap-sec:problem-formulation} and present adaptive sensing policies in Section \ref{darap-sec:adaptive-policy}.  We 
derive performance bounds for adaptive sensing of dynamic targets in Section \ref{darap-sec:theory}. Numerical performance analysis is given in Section \ref{darap-sec:performance-analysis}.  In Section \ref{darap-sec:conclusion}, we conclude and discuss future directions.


\section{Problem formulation}
\label{darap-sec:problem-formulation}
We consider a space $\calX = \set{1,2,\dots,Q}$ containing $Q$ cells 
and a time-varying region of interest (ROI) $\Psi(t) \subset \calX$, $t=1,\ldots,T$. Let $i$ be a location in $\calX$ and define $I_i(t)$ to be the indicator that $i$ is in the ROI at time $t$, i.e., $I_i(t)=1$ if $i \in \Psi(t)$ and $I_i(t)=0$ otherwise. We use a probabilistic target model in which $I_i(1) = 1$ with prior probability $p_i(1)$, independently of the other indicators. For $I_{i}(t) = 0$, the corresponding signal amplitude $\theta_{i}(t)$ is taken to be zero, while for $I_{i}(t) = 1$, the amplitude $\theta_i(t)$ is modeled as a Gaussian random variable.  The initial amplitudes $\theta_{i}(1)$, $i \in \Psi(1)$ are drawn independently with means $\mu_i(1)$ and variances $\sigma_i^2(1)$. As in previous work \cite{Bashan-08-opt-2-stage-search, Wei-Hero-multistage-adaptive-estimation-of-sparse-signals}, a non-informative uniform prior on target locations/amplitudes is assumed with $p_i(1) = p_0$, $\mu_i(1) = \mu_0$ and $\sigma_i^2(1)=\sigma_0^2$ for all $i$, although non-uniform priors could also be accommodated.  

We generalize previous work \cite{Bashan-08-opt-2-stage-search,Wei-Hero-multistage-adaptive-estimation-of-sparse-signals} by introducing a dynamic target state model with state transitions and a birth-death model for target appearance/disappearance. 
To describe the model, we index the targets by target number instead of by cell: 
Let $\sn(t)\in{\Psi}(t),\ n=1,\dots,|{\Psi}(t)|$ be the position of the $n$-th target at time $t$ and $\vartheta^{(n)}(t)=\theta_{\sn(t)}(t)$ be its associated amplitude.  Let $\alpha$ be the probability that 
each target is removed from the scene at each time.
The target transition model and amplitude update for remaining targets is 
\begin{equation}\label{eqn:targetTrans}
\begin{split}
\Pr(\sn(t+1)=i&|\sn(t)=j) 
\\&=
\begin{cases}
(1-\alpha)\pi_0,&i=j\\
\dfrac{(1-\alpha)(1-\pi_0)}{|G(j)|},&i\in G(j)\\
\end{cases},
\end{split}
\end{equation}
\begin{equation}
\label{darap-eq:variance-update}
\vartheta^{(n)}(t+1)=\vartheta^{(n)}(t) + Z^{(n)}(t)
\end{equation}
where $\pi_0$ is the probability that a target remains in the same location,  $G(j)$ is the set of cells that are neighbors of cell $j$, and $Z^{(n)}(t)$ is a zero mean white Gaussian noise with variance $\Delta^2$:
\begin{equation}
Z^{(n)}(t)\sim \mathcal{N}(0,\Delta^2).
\end{equation}
$\Delta^2>0$ captures the variance of random perturbations to the target amplitudes.  The model (\ref{darap-eq:variance-update}) can be used to approximate the effect of model mismatch, target fluctuations, or scattering of the radar signal.  In each of these cases, measurements at each stage are discounted by the increase in uncertainty due to the error sources.

Let $B(t)$ be the event that a single new target enters the scene at time $t$ with probability $\beta$.  Then conditioned on $B(t)$, 
\begin{equation}
\begin{split}
\snn{|{\Psi}(t)|+1}(t+1)|B(t)&\sim\mathrm{Uniform}\{1,2,\dots,Q\},\\
\vartheta^{(|{\Psi}(t)|+1)}(t+1)|B(t)&\sim{\mathcal{N}}(\mu_0,\sigma_0^2).
\end{split}
\end{equation}
We restrict our attention to the case where at most one target occupies a cell at any instant.  
In the sparse situations considered here (i.e. $p_0 \ll 1$), this occurs with high probability.

Observations are made in $T$ stages with effort levels $\lambda_i(t)$
that vary with location $i$ and time $t$. In general, effort might be computing power, complexity, cost, or energy that is allocated to probing a
particular cell. It is assumed that the quality of an observation increases with effort. Given $\lambda_i(t)$, the corresponding observation $y_i(t)$ takes the form
\begin{equation}\label{darap-eq:obs}
y_i(t) = \sqrt{\lambda_i(t)}I_i(t)\theta_i(t) + n_i(t), \quad t = 1,\dots,T,
\end{equation}
where $n_{i}(t)$ represents i.i.d.~zero-mean Gaussian noise with variance $\sigma^{2}$.  The total effort in each stage is constrained as $\sum_{i=1}^{Q} \lambda_{i}(t) \leq \Lambda(t)$.

The goal is to estimate the target state $\set{{\Psi}(t), \set{\theta_{i}(t)}_{i\in\Psi(t)}}_{t=1,\dots,T}$ over 
$T$ stages.  
The posterior distribution of the target state conditioned on measurements is therefore of central interest.  When the targets are static, the posterior distribution factors by cell and can be exactly represented by the posterior mean/variance of the target amplitude and the posterior probability of target existence in 
each cell. In the dynamic case, there is no simple factorization that allows for efficient exact estimation of the posterior distribution, partly due to the fact that the posterior distribution of the amplitudes becomes a Gaussian mixture (due to nonzero transition probabilities to neighboring cells) rather than a simple univariate Gaussian.  
As an alternative, the posterior distribution may be approximated using several standard approaches 
including particle filters, extended Kalman filters, and Unscented Kalman filters, with varying tradeoffs between accuracy and computational burden.

Here we propose a simple 
approximation to the posterior that is accurate under two conditions: (a) at most one target occupies the vicinity of a cell at any one time; and (b) the Gaussian mixture is well represented by the most likely Gaussian mixture component; i.e., the component belonging to the most probable trajectory of the target given the measurements.  Further details on this approximation are available in Chapter 3 of \cite{newstadtThesis2013}.  The first condition is valid when 
there is very low probability that targets will cross tracks.  In practice, one could relax this condition by using methods such as the Joint Multitarget Probability Density Filter (JMPD) \cite{Kreucher-05-multi-tar-track-jpdf}, which independently tracks targets when they are far apart, while jointly tracking targets that are close to each other.   However, we do not address this generalization in this paper. The second condition is equivalent to the existence of a dominant mode in the Gaussian mixture characterizing the posterior density.

\ignore{Although this assumption may not be valid in general, it works well empirically for adaptive policies that are designed to concentrate SNR in locations surrounding targets.  }

Under conditions (a) and (b), the posterior distribution can be approximated with the following:
\begin{align}
p_i(t) &= \Pr(I_i(t)=1|{\bm Y}(t-1)),\label{eqn:p_i}\\
{\mu}_i(t) &= \mathbb{E}[\theta_i(t)|I_i(t)=1,{\bm Y}(t-1)],\\
{\sigma}_i^2(t) &= \mathrm{var}[\theta_i(t)|I_i(t)=1,{\bm Y}(t-1)].\label{eqn:sigma_i}
\end{align}
where ${\bm Y}(T) = \set{y_i(t)}_{i\in\calX,t\in\set{1,\dots,T}}$ is the sequence of observations. For brevity, we denote the collection of posterior probabilities, means, and variances as
\begin{equation}
{\bm x}(t) = \set{p_i(t),{\mu}_i(t),{\sigma}_i^2(t)}_{i=1}^Q
\end{equation}

\ignore{For brevity, we simplify the notation so that $p_i(t)=p_i(t|t-1)$, $\mu_i(t)=\mu_i(t|t-1)$, and $\sigma_i^2(t)=\sigma_i^2(t|t-1)$.}

This representation may be combined with a particle filter, e.g. one using the JMPD\cite{Kreucher-05-multi-tar-track-jpdf}, for target state estimation, while using the above model for resource planning.

\section{Search policy for dynamic targets under resource constraints}
\label{darap-sec:adaptive-policy}
\ifjournal
\else	
	In this section, we provide methods for determining for a sequence of effort allocations $\vlamu{T}=\set{\vlam(t)}_{t=1}^T$ where $\vlam(t)=\set{\lambda_1(t),\dots,\lambda_Q(t)}$. $\vlam(t)$ is a mapping from the previous observations $\mathbf Y(t-1)$ to $[0,\Lambda(t)]^Q$ and is called the allocation policy.

\subsection{Optimization objective}
The following is a multistage extension of the cost function in \cite{Bashan-08-opt-2-stage-search,Wei-Hero-multistage-adaptive-estimation-of-sparse-signals}: 
\begin{equation}
\label{darap-eq:DARAP-cost}
J_T(\vlamu{T}) = \mathbb{E}\left[\sum\limits_{t=1}^T \gamma(t)\sum\limits_{i=1}^Q \dfrac{p_i(t)}{\sigma^2/\sigma_i^2(t)+\lambda_i(t)} \right],
\end{equation}
where $\set{\gamma(t)}_{t=1}^T$ is a set of known 
weights on different planning stages.  
The cost function (\ref{darap-eq:DARAP-cost}) corresponds exactly to the MSE for estimating target amplitudes $\set{\theta_{i}(t)}_{i \in \Psi(t)}$ in two cases: (a) when targets are stationary (but amplitudes may vary); 
and (b) when target locations may change but are known exactly (i.e.\ $p_i(t)=I_i(t)$). 
We define the per-stage cost:
\begin{equation}
\label{darap-eq:myopic-cost}
M_t(\vlam;\vx(t))=\sum\limits_{i=1}^Q \dfrac{p_i(t)}{\sigma^2/\sigma_i^2(t)+\lambda_i(t)}.
\end{equation}
Recalling from \eqref{eqn:p_i} that $p_{i}(t) = \E[I_{i}(t)=1 \mid \mathbf{Y}(t-1)]$,  the expected per-stage cost can also be expressed as 
\begin{equation}\label{eqn:myopicCost2}
\E\left[ M_{t}(\boldsymbol{\lambda};\vx(t)) \right] 
= \E\left\{ \sum_{i\in\Psi(t)} \frac{1}{\sigma^{2}/\sigma_{i}^{2}(t) + \lambda_{i}(t)} \right\},
\end{equation}
where the expectation is taken over both $\Psi(t)$ and $\mathbf{Y}(t-1)$. 

\generalci{
\renewcommand{\arraystretch}{1.5}
\begin{table*}
\caption{Precision parameters $c_i(T)$ as function of the target state model for the objective function given by equation (\ref{darap-eq:DARAP-cost}) when $I_i(t)=I_i$, $\gamma(T)=1$ and $\gamma(t)=0$ for $t<T$}
\label{darap-table:c_i-parameters}
\centering
\begin{tabular}{*{4}{|c}|}
\hline
\multirow{2}{*} {Model} & {Variance of CME,}  &\multirow{2}{*}{$c_i(T)$} & Recursive form \\
{} & {$\sigma_i^2(T|T)$} & {} & {for $c_i(T)$}\\
\hline
\hline
$\theta_i(t)=\mu_i$ & \multirow{2}{*}{$\sigma^2\left( \frac{\sigma^2}{\sigma_\theta^2}+\sum\limits_{t=1}^T \lambda_i(t)\right)^{-1}$} & \multirow{2}{*}{$\frac{\sigma^2}{\sigma_\theta^2}+\sum\limits_{t=1}^{T-1}\lambda_i(t)$} & \multirow{2}{*}{$c_i(T-1)+\lambda_i(t)$} \\
$\mu_i\sim\mathcal{N}(\mu_\theta,\sigma_\theta^2)$ & {} & {} & \\
\hline
\ignore{$\theta_i(t)\sim\mathcal{N}(\mu_i,\Delta^2)$ & \multirow{2}{*}{$\frac{\sigma^2}{\Delta^2}\left( \frac{\sigma^2}{\Delta^2}+\sum\limits_{t=1}^T \lambda_i(t)\right)^{-1}$} & \multirow{2}{*}{$\frac{\sigma^2}{\Delta^2}+\sum\limits_{t=1}^{T-1}\lambda_i(t)$} & \multirow{2}{*}{$c_i(T-1)+\lambda_i(t)$} \\
$\mu_i\sim\mathcal{N}(\mu_\theta,\sigma_\theta^2)$ & {} & {} & \\
\hline}
\multirow{2}{*}{$\theta_i(t)\sim\mathcal{N}(\mu_\theta,\sigma_\theta^2)$} & \multirow{2}{*}{$\sigma^2\left( \frac{\sigma^2}{\sigma_\theta^2}+ \lambda_i(t)\right)^{-1}$} & \multirow{2}{*}{$\frac{\sigma^2}{\sigma_\theta^2}$} & \multirow{2}{*}{$c_i(T-1)$} \\
 & {} & {} & \\
\hline
$\theta_i(t) = \theta_i(t-1) + \delta_i(t)$ & $\sigma^2\left( \frac{\sigma^2}{\sigma_i^2(T|T-1)}+\lambda_i(t)\right)^{-1}$  & \multirow{3}{*}{$\frac{\sigma^2}{\sigma_i^2(T|T)}$} & \multirow{3}{*}{$\left(\frac{\Delta^2}{\sigma^2}+\frac{1}{c_i(T-1)}\right)^{-1}$} \\
$\delta_i(t)\sim\mathcal{N}(0,\Delta^2)$ & {where $\sigma_i^2(t+1|t)$} & {} & \\
{$\theta_i(0)\sim\mathcal{N}(\mu_\theta,\sigma_\theta^2)$} & {$=\sigma_i^2(t|t)+\Delta^2$} & {} & {}\\
\hline
\end{tabular}
\end{table*}
\renewcommand{\arraystretch}{1.0}
}{}
\subsection{Optimal dynamic programming solution}
The optimal effort allocation problem can be stated as 
\begin{equation}
\{\hat{\lambda}_i(t)\}_{i,t} = \arg\min_{{\vlam}} J_T(\vlamu{T}),
\end{equation}
where $\{\hat{\lambda}_i(t)\}_{i}$ is a function of $\vY(t-1)$ and
\begin{equation}
\sum_{i=1}^Q \lambda_i(t) \leq \Lambda(t),\quad t=1,2,\dots,T.
\end{equation} 
Dynamic programming (DP) can be used to exactly obtain an optimal policy that minimizes equation (\ref{darap-eq:DARAP-cost}).  In the case when $\gamma(T)=1$ and $\gamma(t)=0,t<T$, this policy is given by a sequence of recursive minimizations that proceed as follows\footnote{As a minor technical point, if we consider general values for the weights $\gamma(t)$, then equation (\ref{darap-eq:dp-formulation2}) requires an additional term for the current cost at stage $t$.}
\begin{equation}
\label{darap-eq:dp-formulation1}
K_T(\vx(T)) = \min\limits_{\vlam(T)} M_T(\vlam;\vx(T)), \ \sum\limits_{i=1}^Q \lambda_i(T) = \Lambda(T)
\end{equation}
and define recursively for $t=T-1,T-2,\dots,1$
\begin{equation}
\label{darap-eq:dp-formulation2}
\begin{split}
K_t(\vx(t)) &= \min\limits_{\vlam(t)} \mathbb{E}\left[K_{t+1}(\vx(t+1)) \Big| \vx(t), \vlam(t)\right],\\
&s.t.\quad \sum\limits_{i=1}^Q \lambda_i(t) = \Lambda(t).
\end{split}
\end{equation}
Wei and Hero \cite{Wei-Hero-multistage-adaptive-estimation-of-sparse-signals} show that this solution is only tractable for $T\leq 2$.  This is an artifact of the difficulty in computing the expectation in \eqref{darap-eq:dp-formulation2}, which is generally approximated with Monte Carlo samples, as well as the fact that $\vlam(t)$ lies in a multi-dimensional space for $t>1$.  For $T>2$, we therefore have to consider approximations to the optimal policy.  In the next sections, we provide a myopic solution that optimizes $M_t(\vlam;\vx(t))$ for $t=1,2,\dots,T$ without recursion (i.e., assuming that $t$ is the last stage) and an alternative policy that improves upon the myopic solution with low additional computational cost.  

\subsection{Myopic policy}
The myopic optimization problem at time $t$ is given by
\begin{equation}
\label{darap-eq:greedy-optimization-definition}
\min\limits_{\vlam(t)} M_t(\vlam(t);\vx(t)) \qquad \mathrm{s.t} \qquad \sum\limits_{i=1}^Q \lambda_i(t) = \Lambda(t)
\end{equation}
where $\vlam(t)$ depends on previous observations ${\bf Y}(t-1)$ through $\vx(t)$. The optimal solution, similar to the one given in \cite{Wei-Hero-multistage-adaptive-estimation-of-sparse-signals}, begins by defining $\chi$ to be an index permutation that sorts the quantities $\sqrt{p_i(t)}\sigma_i^2(t)$ in non-increasing rank order:
\begin{equation}\label{darap-eq:pi}
\sqrt{p_{\chi(1)}(t)}\sigma_{\chi(1)}^2(t)\geq \cdots \geq \sqrt{p_{\chi(Q)}(t)}\sigma_{\chi(Q)}^2(t).
\end{equation}
Let $c_i(t)=\sigma^2/\sigma_i^2(t)$. Then define $g(k)$ to be the monotonically non-decreasing function of $k=0,1,\dots,Q$ with $g(0)=0$, $g(Q)=\infty$, and
\begin{equation}\label{darap-eq:g(k)}
g(k) = \frac{c_{\chi(k+1)}(t)}{\sqrt{p_{\chi(k+1)}(t)}} \sum\limits_{i=1}^k\sqrt{p_{\chi(i)}(t)}- \sum\limits_{i=1}^kc_{\chi(i)}(t)
\end{equation}
for $k=1,\dots,Q-1$.  Then the solution to (\ref{darap-eq:greedy-optimization-definition}) is
\begin{equation}
\label{darap-eq:myopic-solution}
\lambda_{\chi(i)}^{m}(t)=\left(\Lambda(t)+\sum\limits_{j=1}^{k^*} c_{\chi(j)}(t)\right)\frac{\sqrt{p_{\chi(i)}(t)}}{\sum_{j=1}^{k^*}\sqrt{p_{\chi(j)}(t)}}-c_{\chi(i)}(t),
\end{equation}
for $i=1,\dots,k^*$ and $\lambda_{\chi(i)}^{m}(t)=0$ for $i=k^*+1,\dots,Q$.  The number of nonzero components $k^*$ is determined by the interval $(g(k-1),g(k)]$ to which the budget parameter $\Lambda(t)$ belongs.  Since $g(k)$ is monotonic, the mapping from $\Lambda(t)$ to $k^*$ is well-defined.

\subsection{Non-myopic extension}
\ignore{Myopic policies have the drawback that they are overly aggressive in the allocation of resources.  This may lead to missed or lost targets as well as a lack of robustness to model mismatch.  Chong \cite{chong2008monte} shows that there are significant gains to be had by using non-myopic policies, i.e. policies which trade off short-term performance gains for long term benefits.  There may also be advantages in many cases such as target motion, where there is potential benefit for sensing the target before it becomes unresolvable, or environment variability, when some locations may become unobservable at a particular epoch.}

We propose a simple improvement to the myopic policy that 
combines exploitation of the current belief state 
and exploration of the scene at large.  The proposed non-myopic allocation policy is called the Dynamic Adaptive Resource Allocation Policy, or D-ARAP,  and is defined by
\begin{equation}
\label{darap-eq:nonmyopic-allocation}
\lambda_i^{d}(t;\kappa(t)) = [\kappa(t)]\lambda^{u}(t)+[1-\kappa(t)]{\lambda_i^{m}}(t),
\end{equation}
where $\kappa(t)\in[0,1]$ is the exploration coefficient, $\lambda^{u}(t)=\Lambda(t)/Q$ is the uniform allocation policy, and ${\lambda_i^{m}}(t)$ is given by (\ref{darap-eq:myopic-solution}).  Note that the first term in \eqref{darap-eq:nonmyopic-allocation} allocates a percentage of the resources uniformly to the scene, while the second term weights allocations according to the myopic solution \eqref{darap-eq:myopic-solution}.  

We define the full set of exploration coefficients for a $T$-stage policy as $\vkap(T) =\{\kappa_T(t)\}_{t=1}^{T}$, where the subscript $T$ indicates the number of stages when needed for clarity.  In the rest of the paper, we often use vector notation to represent the allocations to all $T$ stages, defined as
\begin{equation}
\vlamu{T}^{d}(\vkap(T))=\set{\lambda_i^{d}(t;\kappa(t))}_{i=1,\dots,Q,t=1,\dots,T}.
\end{equation}

Without prior knowledge on the location of targets, the first stage should be purely exploratory, i.e., 
$\kappa_T(1)=1$. In addition, since the last stage should be purely exploitative or myopic, we set $\kappa_T(T) = 0$.  
To determine $\vkap(T)$, we consider both offline policies, which are determined prior to collecting observations, and online policies, which are determined adaptively as measurements are collected.  Note that $\lambda_i^{m}(t)$ is a function of previous measurements.  Thus, the offline policies can still be data-dependent as long as $\kappa_T(t)<1$.  

\subsection{Rollout policy}
We first describe an offline policy called the ``offline rollout policy'' that is recursive in the sense that a $T$-stage policy is created by building upon a previously defined $(T-1)$-stage policy.  This method also requires a "base" policy, $\vpi(T_0)$ which pre-defines the last $T_0$ stages of a $T$-stage policy.  Rollout policies for general dynamic programming problems are discussed in great detail in \cite{bertsekas1999rollout}.   In this context, the simplest rollout policy is just $\vpi(1)=\set{0}$, which indicates that the last stage should be purely exploitative.  The pseudocode for the offline rollout policy is given in Fig. \ref{darap-alg:kappa-selection-offline-rollout} and yields policies for $\set{\vkap(\tau)}_{\tau=T_0+1}^T$ from $(T_0+1)$ to $T$ inclusive.  Define 
$\vwi_{\tau-1}(t) = \set{\kappa_{\tau-1}(t')}_{t'=1}^{t}$ to be the first $t$ values of the previous policy $\vkap(\tau-1)$.  Then in each iteration, a $\tau$-stage policy is constructed as $\vkap(\tau)=\set{\vwi_{\tau-1}(\tau-1-T_0),\kappa(\tau-T_0),\vpi(T_0)}$ where  
$\kappa(\tau-T_0)$ is a single parameter which we search over.  The values of $\kappa(\tau-T_0)$ are chosen to minimize the full non-myopic cost in (\ref{darap-eq:DARAP-cost}).  

The expectation in (\ref{darap-eq:DARAP-cost}) is approximated with Monte Carlo samples from the belief state $\vx(t)$ for $t=1,2,\dots,\tau$.  This process can be done efficiently by noting that the first $\tau-T_0-1$ stages remain the same for $\vkap(\tau-1)$ and $\vkap(\tau)$.  Therefore, we only need to draw samples for the last $T_0 +1$ stages at each iteration, as well as perform a line search over the single parameter $\kappa(\tau-T_0)$.  Thus, the offline rollout policy requires $\mathcal{O}(TT_0)$ Monte Carlo simulations to determine policies for $\set{\vkap(\tau)}_{\tau=T_0+1}^T$.  This improves upon the approach \cite{NewstadtWeiHeroCAMSAP13} where a nested optimization procedure (aka, the ``nested policy'') required $\mathcal{O}(T^2)$ calculations.  In our experiments (not shown), the offline rollout policy performed just as well as the nested policy, though with reduced computational complexity.  We do not further discuss the nested policy.  

\floatname{algorithm}{Procedure}
\renewcommand{\algorithmicrequire}{\textbf{Input:}}
\renewcommand{\algorithmicensure}{\textbf{Output:}}
\algrenewcommand{\algorithmiccomment}[1]{\hfill //\emph{\small{#1}}}

\begin{figure}[t]
\fbox{
\begin{minipage}{0.95\columnwidth}
\begin{changemargin}{-.15cm}{0cm}
{\bf procedure} $\set{\vkap(\tau)}_{\tau=T_0+1}^{T}=$ OfflineRolloutPolicy$(\vpi(T_0))$
\end{changemargin}
\begin{changemargin}{-.2cm}{0cm}
\begin{algorithmic}
\State Set $\vwi(1)=1$, $\vkap(T_0+1)=\set{\vwi(1),\vpi(T_0)}$.
\For{$\tau = 
T_0+2,\dots,T$}
\For{each $\kappa(\tau-T_0)\in(0,1]$\ignore{\footnote{Defined on a linear grid.}}}
\State Set 
$\vkapt(\tau) = \set{\vwi_{\tau-1}(\tau-1-T_0), \kappa(\tau-T_0), \vpi(T_0)}$.
\State Calculate\ignore{\footnote{With Monte Carlo simulation}} $C(\kappa(\tau-T_0))=J_\tau(\vlam^{d}(\vkapt(\tau)))$.
\EndFor
\State Choose
$\hat{\kappa}(\tau-T_0) = \arg\min\limits_{\kappa(\tau-T_0)} C(\kappa(\tau-T_0))$.
\State Set $\vwi_\tau(\tau-T_0)=\set{\vwi_{\tau-1}(\tau-1-T_0),\hat{\kappa}(\tau-T_0)}$.
\State Set $\vkap(\tau) = \set{\vwi_{\tau}(\tau-T_0), \vpi(T_0)}$.
\EndFor
\State Return 
$\set{\vkap(\tau)}_{\tau=T_0+1}^{T}$.
\end{algorithmic}
\end{changemargin}
\begin{changemargin}{-.15cm}{0cm}
{\bf end procedure}
\end{changemargin}
\end{minipage}
}
\caption{Offline rollout policy pseudocode for determining exploration parameters $\vkap$}          
\label{darap-alg:kappa-selection-offline-rollout}                           
\vspace{-0mm}
\end{figure}

\subsection{Myopic+ policy}
To further reduce the computational burden, we consider another policy which we call the ``myopic+ policy'' which requires only $\mathcal{O}(T)$ expectations to be calculated (once again through Monte Carlo approximation.) Similar to the offline rollout policy, this policy is built in a sequential fashion.  Whereas the $\tau$-stage offline rollout policy iteratively optimizes over $\kappa_\tau(\tau-T_0)$ followed by a $T_0$-stage base policy, the myopic+ policy chooses $\kappa_\tau(\tau)$ directly without any subsequent rollout.  In particular, we define $\vkapt(\tau)=\set{\vkapt(\tau-1),\kappa_\tau(\tau)}$.  Note that given $\vkapt(\tau-1)$, the current state $\vx(\tau)$ is random only through the noisy measurements $\vY(\tau-1)$.  Additionally, given $\vkapt(\tau-1)$, minimization of $J_{\tau}(\vlam^d(\vkapt(\tau)))$ is equivalent to minimization of the following over the single exploration coefficient $\kappa_\tau(\tau)$:
\begin{equation}
\label{darap-eq:myopic-cost-kappa}
\mathbb{E}_{\vY(\tau-1)}\set{M_\tau(\vlam^d(\set{\vkapt(\tau-1),\kappa_\tau(\tau)});\vx(\tau))}
\end{equation} 
Note that the quantity within the expectation is always minimized by $\kappa_\tau(\tau)=0$, since, by definition, this value optimizes the myopic cost.  To promote exploration, i.e. $\kappa_\tau(\tau)>0$, we adopt a $(1+\rho)$-optimality criterion:
\begin{align}
\label{eq:heuristic-rule}
\hat{\kappa}(\tau) &= \\
\nonumber\max\limits_{\kappa}&\set{\kappa: B_\tau^D(\tilde{\vkap}(\tau-1),\kappa) \leq (1+\rho)B_\tau^D(\tilde{\vkap}(\tau-1),0)},
\end{align}
where $\rho>0$ is a tolerance and
\begin{equation}
\label{darap-eq:heuristic-cost}
B_\tau^D(\tilde{\vkap}(\tau-1),\kappa) = \mathbb{E}_{\vY(\tau-1)}\set{M_\tau\left(\vlam^{d}(\set{\tilde{\vkap}(\tau-1),\kappa});\vx(\tau)\right)},
\end{equation}
where $\vx(\tau)$ is a function of $\vkapt(\tau-1)$ through the measurements $\vY(\tau-1)$. Since $\kappa_\tau(\tau)=0$ optimizes the last-stage cost by definition, 
\eqref{eq:heuristic-rule} results in a policy that is within 
$(1+\rho)$ of the expected minimum myopic cost at each stage.  Observe that \eqref{darap-eq:myopic-cost-kappa}-\eqref{darap-eq:heuristic-cost} are used to build a $T$-stage policy in an iterative fashion and have computational complexity $\mathcal{O}(T)$.  The iterative process is repeated for $\tau=2,3,\dots,T-1$, but the final stage is given by the analytical solution $\kappa_T(T)=0$. Pseudocode for the myopic+ policy is given in  Fig. \ref{darap-alg:kappa-selection-greedy}.

To understand the optimality-criterion in equation (\ref{eq:heuristic-rule}), it is illustrative to look at Fig. \ref{darap-fig:alpha-tradeoff} which plots $B_2^D(1,\kappa)$ as a function of $\kappa$ for low, medium, and high values of $\Lambda(2)$ in (a), (b), and (c), respectively.  It is seen that in all cases, the myopic cost is optimized when $\kappa(2)=0$. \ignore{ However, lower SNR values can tolerate a larger value of $\kappa(2)$ and only have a small deviation in cost.}  To encourage exploration, (\ref{eq:heuristic-rule}) increases $\kappa(2)$.  The amount of increase becomes larger as the SNR decreases. The red dotted line shows a deviation of 10\% from the minimum cost, while the yellow circle marks the point where $\kappa$ attains the maximum deviation.

The offline rollout and myopic+ 
policy parameters are shown in Fig. \ref{darap-fig:alpha-function} for various values of SNR\footnote{SNR is defined 
in terms of the budget per stage $\Lambda(t)$ and the 
noise variance $\sigma^2$ as 
SNR$(\Lambda(t))=10\log_{10}(\Lambda(t)/(Q\sigma^2))$.}, 
$T=20$, and model parameters given by Table \ref{darap-table:simulation-parameters}. 
It should be noted that the offline rollout policies require numerical optimization over the $\kappa(t)$ parameters, which tend to be noisy unless a large number of Monte Carlo realizations are used.  In contrast,  experiments in Section \ref{darap-sec:performance-analysis} indicate that the myopic+ policy parameters tend to be less sensitive to noise and mismodeling errors.
\ignore{For $t>2$, the myopic+ policy parameters are nearly monotonically decreasing in both SNR and stage $t$.  This motivates the ``functional'' policy (Fig. \ref{darap-fig:alpha-function}(e)) approximation, defined as
\begin{equation}
\label{darap-eq:functional-approximation-to-kappa}
\kappa_{functional}(t) = f\left(\vkap(T);SNR^{(cum)}(t)\right),
\end{equation}
where $f(\vkap(T);SNR^{(cum)}(t))$ is a polynomial function that is fit to the observed data $\vkap(T)$ and $SNR^{(cum)}(t)$ is the cumulative SNR\footnote{
Cumulative SNR is defined as SNR$(\sum_{\tau=1}^t \Lambda(\tau))$.}. Note that this functional approximation reduces the computational burden for finding $\vkapt(T)$ to just $O(1)$ simulations (i.e., it is does not grow as a function of $T$).  For large $T$, this computational savings may be significant.}  	
\ignore{
This modification reduces the optimization problem to choosing the exploration/exploitation tuning knobs at each stage, $\vkap^{(T)}=\set{\kappa(t)}_{t=1}^T$.  Without prior knowledge on the location of targets, it is clear that $\kappa(1)=1$.  However, finding optimal choices for the remaining parameters, $\vkapt^{(T)}=\set{\kappa(t)}_{t=2}^T$  is still a difficult problem that requires approximate methods when the state space is large (i.e., when $Q$ is large).  }

\floatname{algorithm}{Procedure}
\renewcommand{\algorithmicrequire}{\textbf{Input:}}
\renewcommand{\algorithmicensure}{\textbf{Output:}}
\algrenewcommand{\algorithmiccomment}[1]{\hfill //\emph{\small{#1}}}
\begin{figure}
\fbox{
\begin{minipage}{0.95\columnwidth}
\begin{changemargin}{-.15cm}{0cm}
{\bf procedure} $\set{\vkap(\tau)}_{\tau=1}^T=$ Myopic+Policy($\rho$)
\end{changemargin}
\begin{changemargin}{-.2cm}{0cm}
\begin{algorithmic}
\State Set $\kappa(1)=1, \vkapt(1) = \set{\kappa(1)}$.
\For{$\tau = 2,3,\dots,T-1$}
\For{each $\kappa(\tau)\in(0,1]$\ignore{\footnote{Defined on a linear grid.}}}
\State Calculate\ignore{\footnote{With Monte Carlo simulation}} $B_\tau^D(\tilde{\vkap}(\tau-1),\kappa(\tau))$ according to \eqref{darap-eq:heuristic-cost}.
\EndFor
\State Choose
$\hat{\kappa}(\tau)$ according to \eqref{eq:heuristic-rule}.
\State Set $\vkapt(\tau) = \set{\vkapt(\tau-1),\hat{\kappa}(\tau)}$.
\EndFor
\State Return 
$\set{\vkap(\tau) = \set{\vkapt(\tau-1), 0}}_{\tau=2}^{T}$, $\vkap(1)=1$.
\end{algorithmic}
\end{changemargin}
\begin{changemargin}{-.15cm}{0cm}
{\bf end procedure}
\end{changemargin}
\end{minipage}
}
\caption{Myopic+ policy pseudocode for determining exploration parameters $\vkap$}          
\label{darap-alg:kappa-selection-greedy}                          
\end{figure}

\def\subfigwidth{\heuristicFigWidth\textwidth}
\begin{figure}
\centering
\subfloat[SNR = 0 dB]{
\label{darap-fig:K_vs_kappa-SNR0}
\includegraphics[height=\heuristicFigHeight]{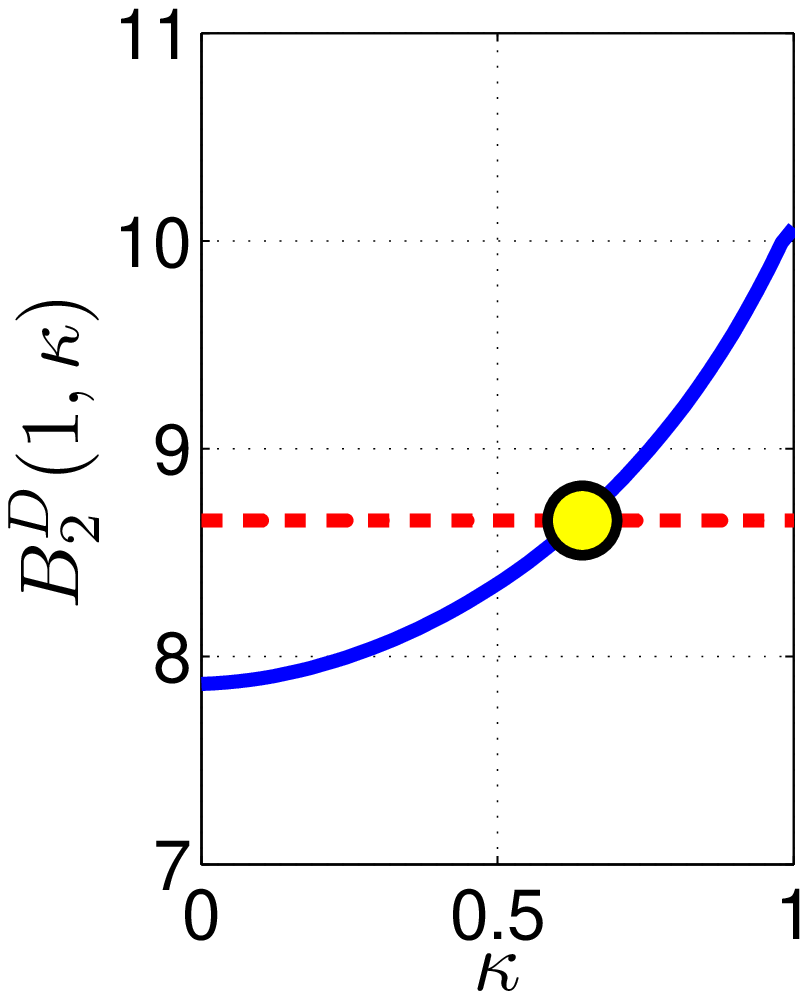}
}
\subfloat[SNR = 5 dB]{
\label{darap-fig:K_vs_kappa-SNR5}
\includegraphics[height=\heuristicFigHeight]{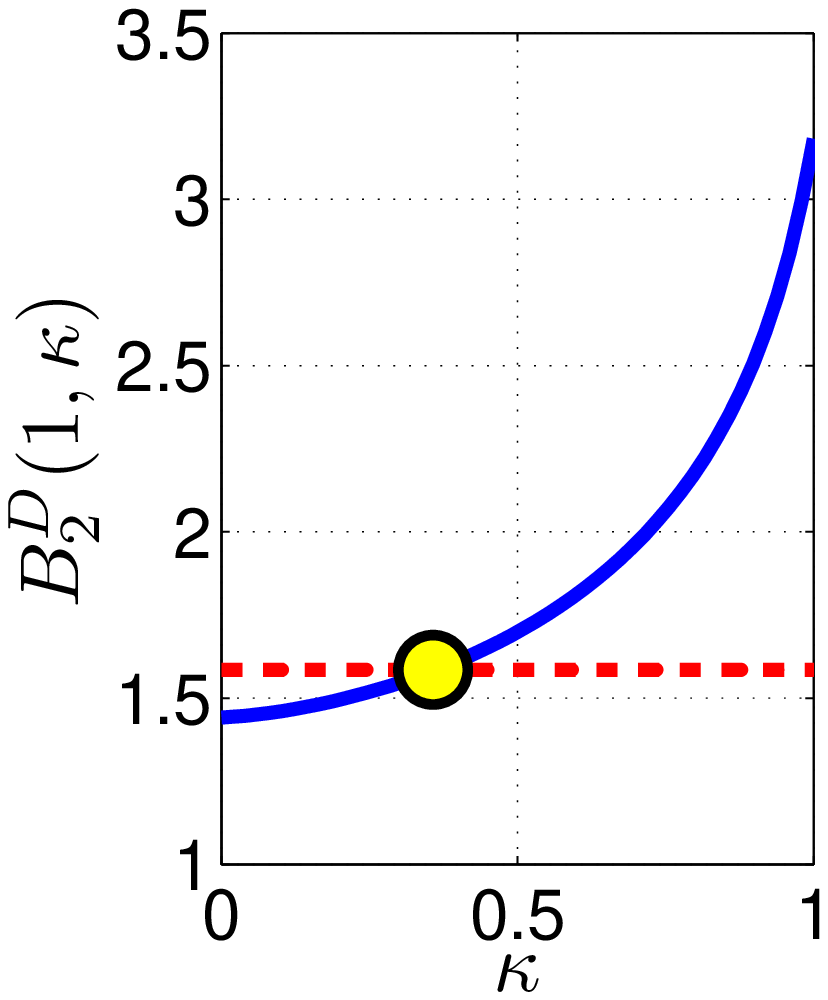}
}
\subfloat[SNR = 10 dB]{
\label{darap-fig:K_vs_kappa-SNR10}
\includegraphics[height=\heuristicFigHeight]{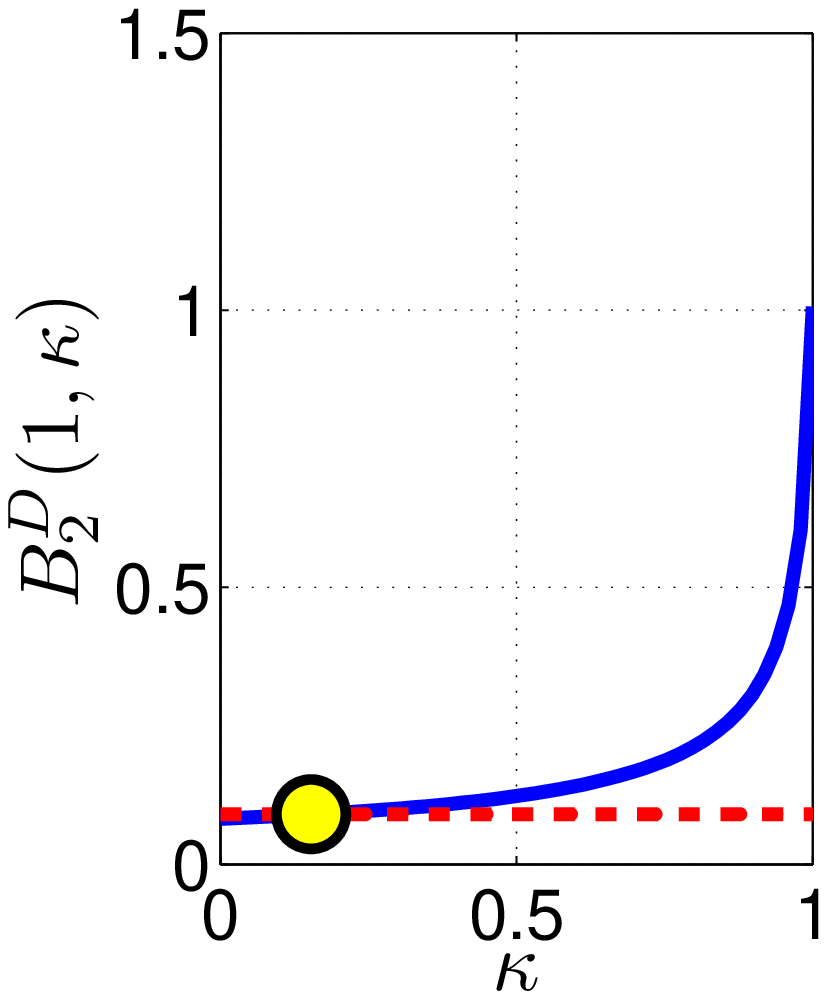}
}
\caption{We plot the myopic cost $B_2^D(1,\kappa)$ as given by equation (\ref{darap-eq:heuristic-cost}) as a function of the exploitive vs. explorative parameter $\kappa$.  $B_{2}^D(1,\kappa)$ is shown for low, medium, and high values of $\Lambda(2)$ in (a), (b), and (c), respectively.  In all cases, the myopic cost is optimized when $\kappa=0$.  However, lower SNR values can tolerate a larger value of $\kappa$ and only have a small deviation in cost.  The red dotted line shows a deviation of 10\% from the minimum cost, while the yellow circle marks the point where $\kappa$ attains this value.}
\label{darap-fig:alpha-tradeoff}
\end{figure}

\renewcommand{\arraystretch}{1.5}
\begin{table}
\caption{Parameters used for simulation analysis}
\label{darap-table:simulation-parameters}
\centering
\begin{tabular}{|l|c|c|}
\hline
Parameter & Variable Name &  Value\\
\hline
\hline
Number of locations & $Q$ & $1,000$ \\
Prior sparsity & $p_i(1)=p_0$ & $0.01$\\
\ignore{Expected number of targets & $E[\tilde{N}(t)]$ & 10\\}
Target amplitude mean & $\mu_i(1)=\mu_0$ & 1 \\
Target amplitude std. deviation (prior) & $\sigma_i(1)=\sigma_0$ & $1/6$\\
Target amplitude std. deviation (update) & $\Delta$ & $1/20$\\
Noise variance & $\sigma^2$ & 1\\
Stationary probability & $\pi_0$ & 1/3\\
Death probability & $\alpha$ & 0\\
Birth probability & $\beta$ & 0\\
Number of neighbors & $|G|$ &2\\
Stage weights & $\set{\gamma(t)}_{t=1}^T$ & $\set{0,\dots,0,1}$\\
\hline
\end{tabular}
\end{table}
\renewcommand{\arraystretch}{1.0}

\begin{figure*}
\centering
\subfloat[Offline Rollout ($T_0=1$)]{
\label{darap-fig:offline-rollout-policy-T01}
\includegraphics[height=\policyFigHeight]{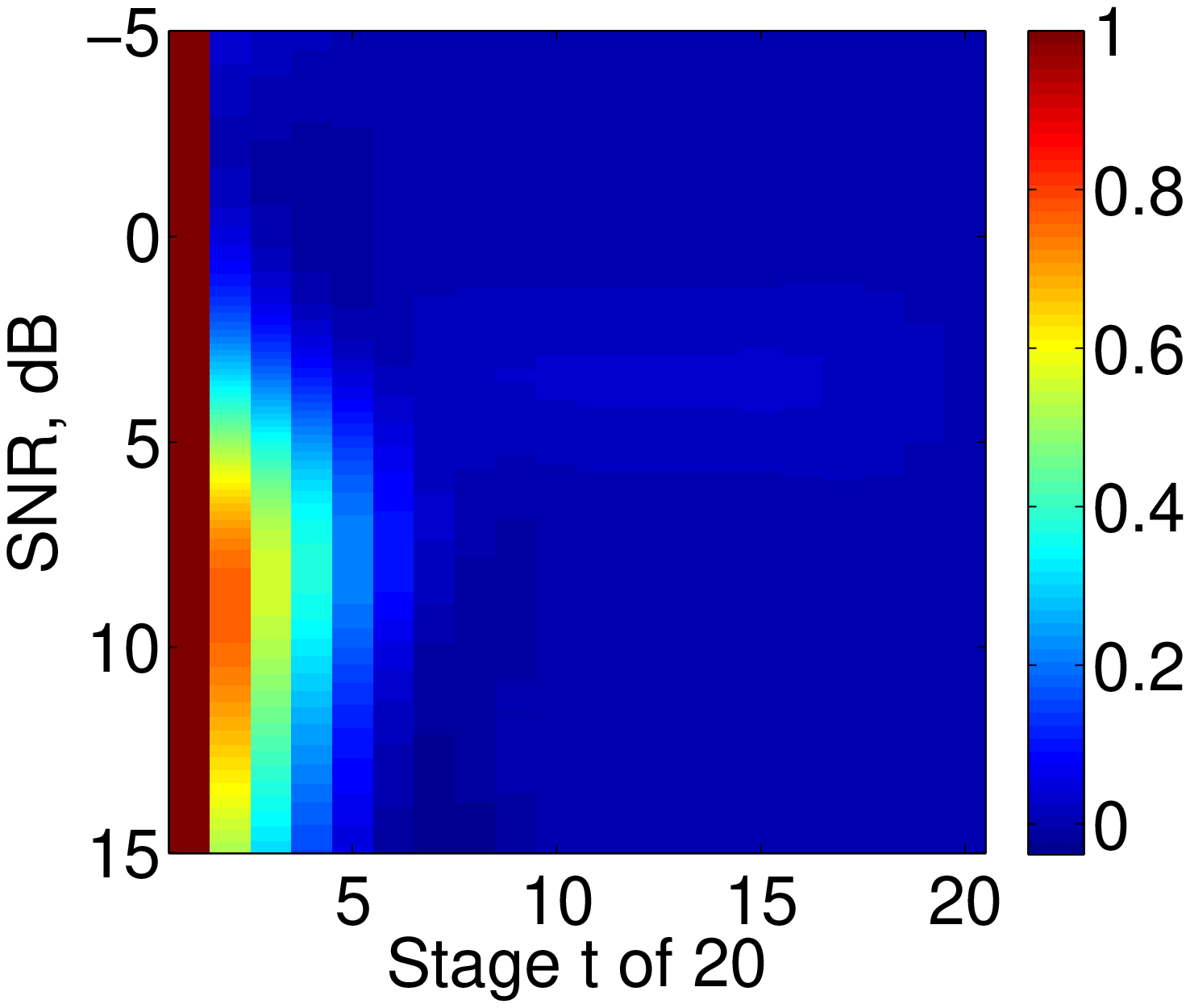}
}\subfloat[Offline Rollout ($T_0=2$)]{
\label{darap-fig:offline-rollout-policy-T02}
\includegraphics[height=\policyFigHeight]{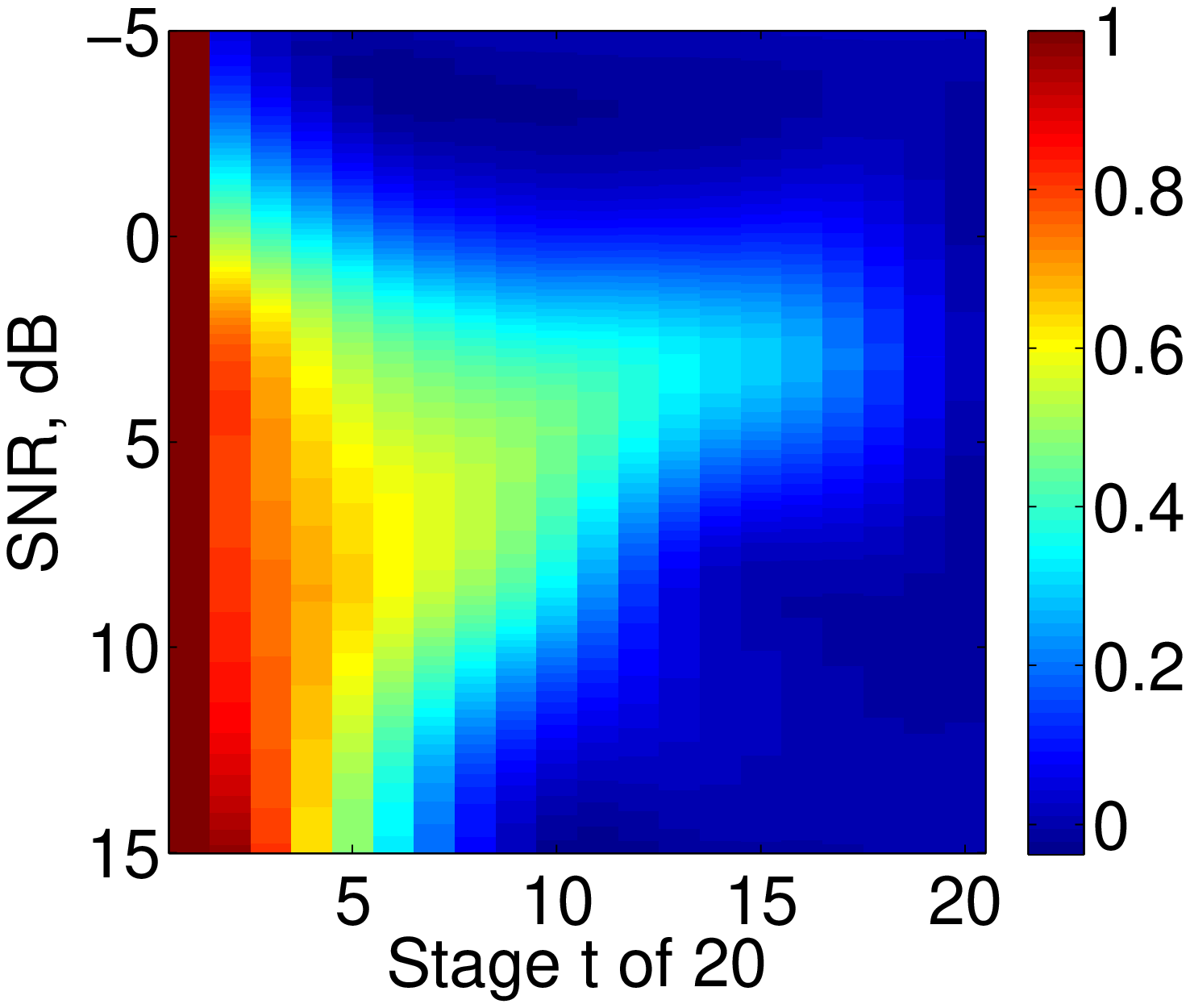}
}
\ifdoublecol
\else
\\
\fi
\subfloat[Offline Rollout ($T_0=5$)]{
\label{darap-fig:offline-rollout-policy-T05}
\includegraphics[height=\policyFigHeight]{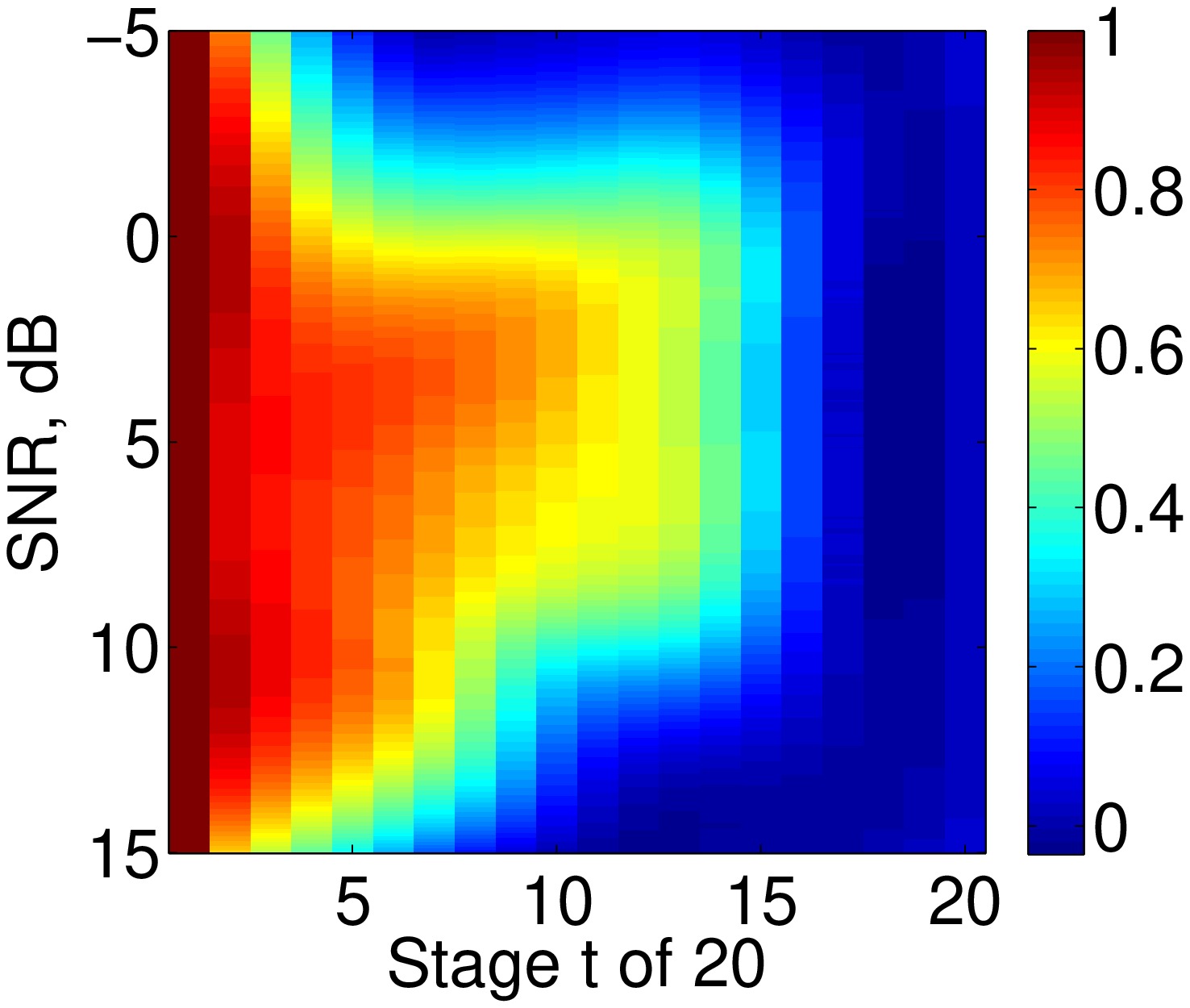}
}
\subfloat[Myopic+ Policy]{
\label{darap-fig:heuristic-policy}
\includegraphics[height=\policyFigHeight]{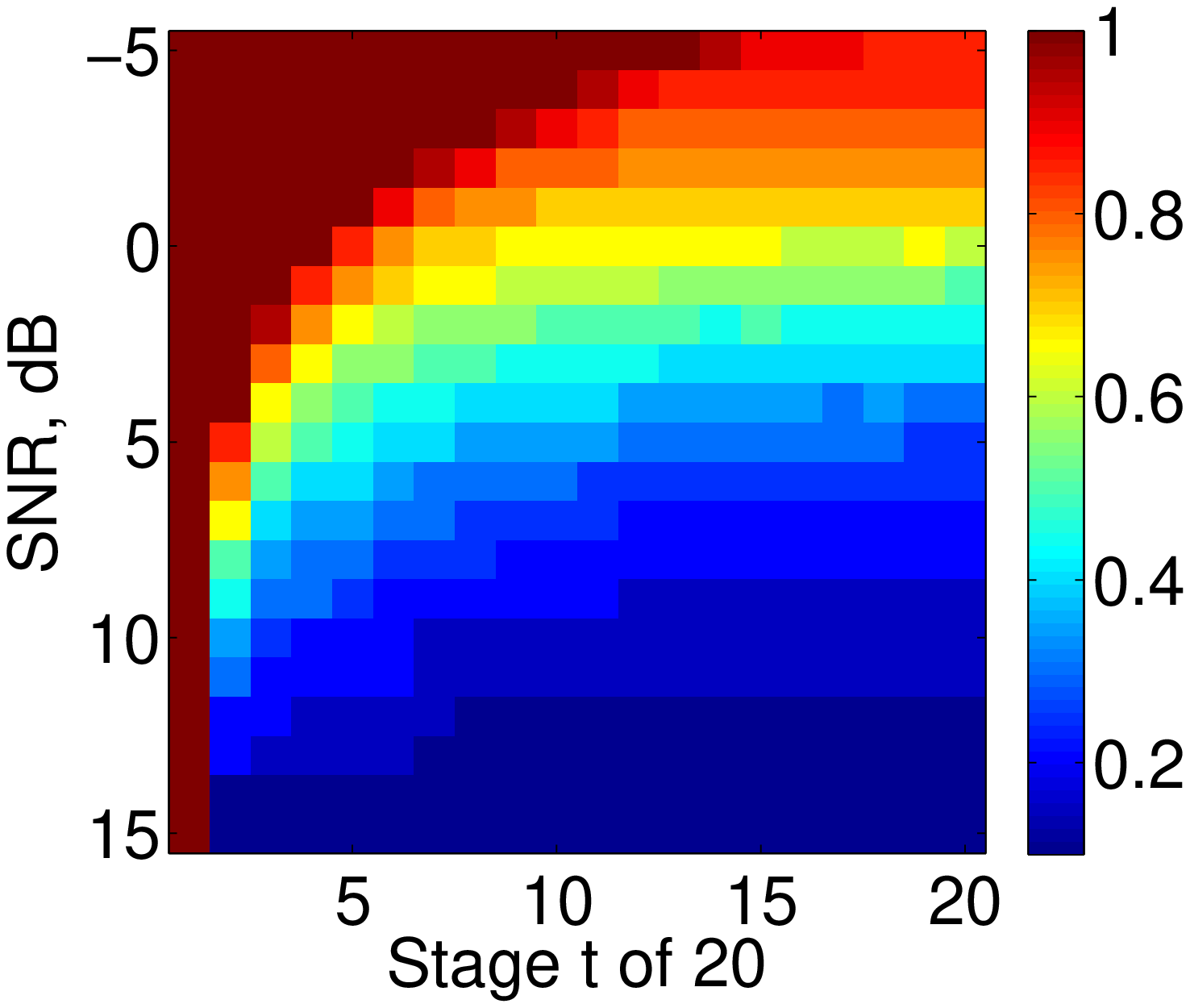}
}
\caption{The heat maps show the selection of the exploration coefficient $\vkap(T)$ according to Algorithms \ref{darap-alg:kappa-selection-offline-rollout} (offline rollout) and \ref{darap-alg:kappa-selection-greedy} (myopic+) for policies of length $T=20$.  The rollout policies use a myopic base policy (i.e. $\kappa(t)=0$) of lengths $T_0=1,2,5$ in (a), (b), and (c), respectively.  \ignore{The heat maps indicate that larger $T_0$ leads to more exploratory strategies (i.e., $\kappa(t)$ closer to 1).}  The myopic+ policy is given in (d), which performs more exploration in general for lower SNR than the rollout policies.  In all four policies, $\kappa(t)$ is nearly monotonically decreasing in $t$.  The offline rollout policies all exhibit a phase transition from a low SNR regime (where $\kappa(t)\approx 0$) to a higher SNR regime where significant exploration occurs.  The heatmaps indicate that higher values of $T_0$ lead to more exploratory strategies at low SNR and $t$.  The myopic+ policy is monotonically decreasing in both SNR and $t$. \ignore{A functional approximation to the heuristic policy is shown in (e), and plotted over both SNR and stage in (f).}}
\label{darap-fig:alpha-function}
\end{figure*}

\ignore{\begin{figure*}
\centering
\subfloat[Nested Policy]{
\centering
\includegraphics[width=.8\columnwidth]{Images/nested-flow}
}
\qquad
\subfloat[Heuristic Policy]{
\centering
\includegraphics[width=.8\columnwidth]{Images/heuristic-flow}
}
\caption{{\color{red}Flow diagram comparison between nested and heuristic methods.  Elements within ellipses denoted parameters that are selected at that stage, while elements within rectangles represent previously computed policies.}}
\label{darap-fig:flow}
\end{figure*}}

\ignore{
\subsection{Online policies}
\label{darap-subsec:pomdp-policies}

As discussed in the introduction, the full POMDP solution to this adaptive sensing problem is generally intractable due to the size of belief state and action spaces.  As an alternative, we consider an approximate POMDP solution, namely the rollout policy, in order to compare D-ARAP to online solutions.  Note that in online solutions, the optimal action at each stage must be chosen separately for each realization of the model.  Thus, the online rollout policy likely will incur significant computational costs in comparison to the offline policies presented in this paper. 

}

\fi

\section{Performance bounds}
\label{darap-sec:theory}
\ifjournal
\else	

In this section, we develop bounds on the performance gain that can be achieved with D-ARAP, and more generally any adaptive policy, compared to non-adaptive uniform allocation policies.  The gain is measured using the cost function \eqref{darap-eq:DARAP-cost}.  \ignore{Our aim is to understand the potential benefits of adaptive sensing for targets that are dynamic as well as sparse, thus extending \cite{Bashan-08-opt-2-stage-search, bashan2011marapTSP,Wei-Hero-multistage-adaptive-estimation-of-sparse-signals}. }
The bounds result from analyzing two oracle policies that have exact 
knowledge of target locations. 
The first of these, the omniscient policy, has access to the 
target locations $\Psi(t)$ for all $t$ and is discussed in Sections \ref{subsec:omni0} and \ref{subsec:omni+}.  \ignore{Analysis of the omniscient policy results in Propositions \ref{prop:omni0} and \ref{prop:omni+}, which show that the potential gain due to adaptation increases with sparsity, as is the case for static targets \cite{Bashan-08-opt-2-stage-search, bashan2011marapTSP,Wei-Hero-multistage-adaptive-estimation-of-sparse-signals}.  However, when 
targets may enter, leave, or transition between cells, the omniscient policy is non-causal and hence not attainable, even in the asymptotic limit of high SNR or number of stages $T$.  Accordingly, we also consider a} 
The second, the semi-omniscient policy, 
has access to only the previous locations $\Psi(t-1)$ at stage $t$ and is considered in Sections \ref{subsec:semi0} and \ref{subsec:semi+}.  

We distinguish two qualitatively different cases corresponding to either constant or increasing target amplitude variance, characterized by 
the increment $\Delta^{2} = 0$ or $\Delta^{2} > 0$ respectively.  For oracle policies, the definitions of the state variables \eqref{eqn:p_i}--\eqref{eqn:sigma_i} are modified by augmenting the observation history ${\bm Y}(t-1)$ with the exact target positions $\Psi(t-1)$, i.e., ${\bm Y}(t-1) \to \{ {\bm Y}(t-1), \Psi(t-1) \}$.  In this case, it can be shown that the posterior variances 
evolve according to  
\begin{equation}\label{eqn:sigma+}
\sigma_{i}^{2}(t+1) = \frac{\sigma^{2}}{\sigma^{2} + \lambda_{s^{(n)}(t)}(t) \sigma_{s^{(n)}(t)}^{2}(t)} \sigma_{s^{(n)}(t)}^{2}(t) + \Delta^{2}, \quad i \in H\left(s^{(n)}(t)\right),
\end{equation}
%
where $H(j) = \{j\} \cup G(j)$.  Hence in the case of static target amplitudes ($\Delta^{2} = 0$, Sections \ref{subsec:omni0} and \ref{subsec:semi0}), the posterior variances 
decay to zero as $t$ increases, while for $\Delta^{2} > 0$ (Sections \ref{subsec:omni+} and \ref{subsec:semi+}), the posterior variances reach a nonzero steady state. 
For simplicity, we make the following assumption 
for derivation of the performance bounds:
\begin{assumption}\label{ass:numTargets}
The number of targets $\lvert\Psi(t)\rvert$ is constant, i.e., $\alpha = \beta = 0$.
\end{assumption}

\subsection{Omniscient policy, $\Delta^{2} = 0$}
\label{subsec:omni0}

In Sections \ref{subsec:omni0} and \ref{subsec:semi0} we make the additional assumption that the target amplitudes are constant: 
\begin{assumption}\label{ass:Delta}
The variance increment $\Delta^{2}$ is zero. 
\end{assumption}
\noindent In this case, 
\eqref{eqn:sigma+} reduces to a simple recursion for the posterior precisions $c_i(t) = \sigma^{2} / \sigma_{i}^{2}(t)$:
\begin{equation}\label{eqn:c0}
c_{i}(t+1) = c_{s^{(n)}(t)}(t) + \lambda_{s^{(n)}(t)}(t), \quad i \in H\left(s^{(n)}(t)\right), 
\end{equation}
where $c_{i}(1) = \sigma^{2} / \sigma_{0}^{2}$ for all $i$. 


The omniscient policy has perfect knowledge of the target locations $\Psi(t)$ at all times. 
Conditioned on $\Psi(t)$, it follows that the target probabilities are atomic, $p_{i}(t) = I_{i}(t)$, and the omniscient policy allocates effort solely and uniformly to targets: 
\begin{equation}\label{darap-eq:omn-policy-defn}
\lambda_i^{\omni}(t) =
\begin{cases}
\Lambda(t)/|\Psi(1)|, &i\in \Psi(t)\\
0, &i \notin \Psi(t),
\end{cases}
\end{equation}
noting that $\lvert\Psi(t)\rvert = \lvert\Psi(1)\rvert$ under Assumption \ref{ass:numTargets}.  Given \eqref{eqn:c0} and \eqref{darap-eq:omn-policy-defn}, the posterior precisions also remain uniform over targets:
\begin{equation}\label{eqn:cOmni0}
c_{i}(t) = \frac{\sigma^{2}}{\sigma_{0}^{2}} + 
\frac{\Lambdab(t-1)}{\lvert\Psi(1)\rvert}, \quad i \in \Psi(t) \;\; \forall \; t,
\end{equation}
where $\Lambdab(t) = \sum_{\tau=1}^{t} \Lambda(\tau)$. 
To verify \eqref{darap-eq:omn-policy-defn}, we begin with $t = 1$, in which case $c_{i}(1)$ is uniform over $i \in \Psi(1)$.  Specializing the optimal allocation given by \eqref{darap-eq:pi}--\eqref{darap-eq:myopic-solution} to the case $p_{i}(1) = I_{i}(1)$, it is seen that the sequence $g(k)$ \eqref{darap-eq:g(k)} is equal to $0$ for $k < \lvert\Psi(1)\rvert$ and $\infty$ for $k \geq \lvert\Psi(1)\rvert$.  Hence the number of nonzero allocations $k^{\ast} = \lvert\Psi(1)\rvert$ and \eqref{darap-eq:omn-policy-defn} follows from \eqref{darap-eq:myopic-solution}.  For $t > 1$, \eqref{darap-eq:omn-policy-defn} continues to hold by induction since $c_{i}(t)$ remains uniform over $i \in \Psi(t)$. 

We define the gain of a policy with respect to the uniform allocation policy as 
\begin{equation}\label{darap-eq:gain-cost}
\Gamma_{T}(\boldsymbol{\lambda}) = \frac{J_{T}(\boldsymbol{\lambda}^{\uni})}{J_{T}(\boldsymbol{\lambda})}.
\end{equation}
Using \eqref{darap-eq:omn-policy-defn} and \eqref{eqn:cOmni0}, the gain of the omniscient policy is characterized in Proposition \ref{prop:omni0}. 
The following assumption is used to obtain a more interpretable expression.
\begin{assumption}\label{ass:gamma}
The stage weights $\gamma(t)$ decay to zero as $t$ decreases from $T$.  
\end{assumption}
\noindent This assumption ensures that as $T \to \infty$, the cost \eqref{darap-eq:DARAP-cost} becomes dominated by terms at large $t$.  The assumption is satisfied by common ``forgetting'' schemes 
that emphasize performance in later stages.  

\begin{prop}\label{prop:omni0}
Let $r_{0}(t) = \sigma^{2} Q / (\sigma_{0}^{2} \Lambdab(t))$.  Under Assumptions \ref{ass:numTargets} and \ref{ass:Delta}, the gain of the omniscient policy relative to uniform allocation is bounded from above as 
\begin{multline*}
\Gamma_{T}(\boldsymbol{\lambda}^{\omni}) \leq 
\left. \left( \sum_{t=1}^{T} \frac{\gamma(t)}{\Lambdab(t)} \frac{1}{1 + r_{0}(t)} \right) \right/\\
\left( \sum_{t=1}^{T} \frac{\gamma(t)}{\Lambdab(t)} \left[ \frac{p_{0}}{1 + p_{0} r_{0}(t)} + \frac{1-p_{0}}{Q} \frac{1}{(1 + p_{0} r_{0}(t))^{3}} - \frac{(1-p_{0}) (1-2p_{0})}{Q^{2}} 
\frac{r_{0}(t)}{(1 + p_{0} r_{0}(t))^{4}} \right] \right). 
\end{multline*}
In the high-SNR limit ($\sigma^{2} \to 0$) or if Assumption \ref{ass:gamma} holds and the number of stages $T \to \infty$, then $r_{0}(t) \to 0$ and the above expression simplifies to 
\[
\Gamma_{T}(\boldsymbol{\lambda}^{\omni}) = \frac{1}{p_{0} + (1-p_{0})/Q} \left( 1 - O(r_{0}) \right).
\]
\end{prop}
\begin{IEEEproof}
See Appendix \ref{app:omni0}.
\end{IEEEproof}

Proposition \ref{prop:omni0} shows that the omniscient gain is proportional to the sparsity of the scene, similar to \cite{Bashan-08-opt-2-stage-search, bashan2011marapTSP,Wei-Hero-multistage-adaptive-estimation-of-sparse-signals}.  In other words, the potential gain due to adaptation is higher when there are fewer targets.  

\ignore{
\begin{corollary}
Under Assumptions \ref{ass:numTargets}--\ref{ass:gamma}, at high SNR or numbers of stages $T$ the gain of the omniscient policy is given by 
\[
\Gamma_{T}(\boldsymbol{\lambda}^{\omni}) = \frac{1}{p_{0} + (1-p_{0})/Q} \left( 1 - O\left( \frac{\sigma^{2} Q}{\sigma_{0}^{2} \Lambdab(T)} \right) \right).
\]
\end{corollary}
\begin{IEEEproof}
As an alternative to Jensen's inequality, we may exploit convexity to bound the right-hand side of \eqref{eqn:myopicOmni0} directly by its first-order expansion:
\[
M_{t}(\boldsymbol{\lambda}^{\omni}) \geq \frac{\lvert\Psi(1)\rvert^{2}}{\Lambdab(t)} \left( 1 - \frac{\sigma^{2} \lvert\Psi(1)\rvert}{\sigma_{0}^{2} \Lambdab(t)} \right),
\]
with equality holding asymptotically as $\sigma^{2} \lvert\Psi(1)\rvert / (\sigma_{0}^{2} \Lambdab(t)) \to 0$. Taking expectations and summing over $t$ yields 
\begin{equation}\label{eqn:JOmni2}
J_{T}(\boldsymbol{\lambda}^{\omni}) \geq p_{0} Q (p_{0} Q + 1 - p_{0}) \left[ \sum_{t=1}^{T} \frac{\gamma(t)}{\Lambdab(t)} - \frac{\ptilde \sigma^{2} Q}{\sigma_{0}^{2}} \sum_{t=1}^{T} \frac{\gamma(t)}{\Lambdab(t)^{2}} \right], 
\end{equation}
where $\E\{ \lvert\Psi(1)\rvert^{2} \} = p_{0} Q (p_{0} Q + 1 - p_{0})$ and $\ptilde = (1/Q) \E\{ \lvert\Psi(1)\rvert^{3} \} / \E\{ \lvert\Psi(1)\rvert^{2} \} = p_{0} + O(1/Q)$ by straightforward algebra.  A similar expansion of \eqref{eqn:myopicUni0} leads to  
\begin{equation}\label{eqn:JUni2}
J_{T}(\boldsymbol{\lambda}^{\uni}) = p_{0} Q^{2} \left[ \sum_{t=1}^{T} \frac{\gamma(t)}{\Lambdab(t)} - \frac{\sigma^{2} Q}{\sigma_{0}^{2}} \sum_{t=1}^{T} \frac{\gamma(t)}{\Lambdab(t)^{2}} + O\left( \left( \frac{\sigma^{2} Q}{\sigma_{0}^{2}} \right)^{2} \right) \sum_{t=1}^{T} \frac{\gamma(t)}{\Lambdab(t)^{3}} \right]. 
\end{equation}
As the SNR increases, i.e., $\sigma^{2}/\sigma_{0}^{2} \to 0$, it can be seen that the ratio of \eqref{eqn:JUni2} to \eqref{eqn:JOmni2} has the desired form in the corollary statement.  Furthermore, under Assumption \ref{ass:gamma}, as $T \to \infty$ the summations over $t$ become dominated by terms at large $t$, where $\Lambdab(t)$ also becomes large.  Hence the asymptotic form also holds as $T \to \infty$. 
\end{IEEEproof}
}

\subsection{Semi-omniscient policy, $\Delta^{2} = 0$}
\label{subsec:semi0}

We now turn to the semi-omniscient policy, which in stage $t$ has knowledge only of the previous target locations $\Psi(t-1)$. 
In the semi-omniscient case, the target probabilities $p_{i}(t) = \Pr(I_{i}(t)=1 \mid \Psi(t-1))$ are no longer binary but are given by the target dynamics \eqref{eqn:targetTrans} as 
\begin{equation}\label{eqn:pSemi}
p_{i}(t) = \begin{cases}
\pi_{0}, & i \in \Psi(t-1),\\
\frac{1-\pi_{0}}{\lvert G \rvert}, & i \in G\left(\Psi(t-1)\right),\\
0 & \text{otherwise},
\end{cases}
\end{equation}
where $G(\Psi(t-1)) = \bigcup_{i\in\Psi(t-1)} G(i)$ is the set of neighbors of all targets.  
We assume that the probability of target transitions is bounded.
\begin{assumption}\label{ass:pi0}
The probability of a target remaining in the same location is no smaller than the probability of it transitioning to any one neighboring cell,
\[
\pi_{0} \geq \frac{1-\pi_{0}}{\lvert G \rvert}.
\]
\end{assumption}

Unlike in the omniscient case, under the semi-omniscient policy the posterior precisions $c_{i}(t)$ 
become random and non-uniform for $t > 1$ over the set of locations $H(\Psi(t-1)) = \Psi(t-1) \cup G(\Psi(t-1))$ where $p_{i}(t) > 0$.  The non-uniformity arises because $H(\Psi(t-1))$ contains both target and non-target locations, 
and even among targets, the precisions 
differ randomly depending on the number of times a target has stayed in the same cell or moved to a different one.  This 
makes it difficult to determine the allocations analytically via \eqref{darap-eq:pi}--\eqref{darap-eq:myopic-solution}.  
As an alternative, we focus on 
developing an upper bound $\cbar(t)$ on the expected precisions $\E[ c_{i}(t) \mid \lvert\Psi(1)\rvert ]$, $i \in H(\Psi(t-1))$, 
conditioned on the number of targets $\lvert\Psi(1)\rvert$.  
For $t = 1$, $\cbar(t)$ is defined as $\cbar(1) = c_{i}(1) = \sigma^{2} / \sigma_{0}^{2}$, satisfying the upper bound property.  For $t > 1$, $\cbar(t)$ is defined by the recursion 
\begin{equation}\label{eqn:cbar}
\cbar(t+1) =
\begin{cases}
\frac{ \pi_{0}^{3/2} + \frac{1}{\sqrt{\lvert G \rvert}} (1-\pi_{0})^{3/2} }{ \sqrt{\pi_{0}} + \sqrt{\lvert G\rvert (1-\pi_{0})} } \left( \left(1 + \lvert G \rvert\right) \cbar(t) + \frac{\Lambda}{\lvert\Psi(1)\rvert} \right), & \cbar(t) < \ccrit,\\
\cbar(t) + \frac{\pi_{0} \Lambda}{\lvert\Psi(1)\rvert}, & \cbar(t) \geq \ccrit,
\end{cases}
\end{equation}
where the threshold $\ccrit$ is defined as 
\begin{equation}\label{eqn:ccrit}
\ccrit = \frac{\Lambda}{\lvert\Psi(1)\rvert e(\pi_0,G)},
\end{equation}
and $e(\pi_0,G)=\sqrt{\frac{\lvert G\rvert \pi_{0}}{1 - \pi_{0}}} - 1 \geq 0$. 

We also use the following assumptions to determine the number of nonzero allocations under the semi-omniscient policy:
\begin{assumption} 
The posterior precisions are uniform in the vicinity of targets, 
\label{ass:c}
\[
c_{i}(t) = \cbar(t), \quad i \in H\left(\Psi(t-1)\right), \;\; t > 1.
\]
\end{assumption}
\begin{assumption}\label{ass:Lambda}
The per-stage effort budget $\Lambda(t)$ is constant, $\Lambda(t) = \Lambda$. 
\end{assumption}
\noindent Assumption \ref{ass:c} replaces $c_{i}(t)$ with an upper bound on its expected value and is therefore an optimistic approximation consistent with deriving the upper bound $\cbar(t)$. 
As $t$ increases, the short-term deviations of $c_{i}(t)$ from its mean decrease relative to the long-term increase of the mean and the approximation corresponds to an upper bound on $c_{i}(t)$ itself with high probability.  
We note that Assumption \ref{ass:c} is used primarily to determine the number of nonzero allocations and only indirectly to determine the amount allocated.

Given Assumptions \ref{ass:pi0}--\ref{ass:Lambda}, the following lemma proves that the recursion in \eqref{eqn:cbar} yields a valid upper bound on $\E [ c_{i}(t) \mid \lvert\Psi(1)\rvert ]$. 

\begin{lemma}\label{lem:cbar}
Under Assumptions \ref{ass:numTargets}, \ref{ass:Delta}, and \ref{ass:pi0}--\ref{ass:Lambda}, the expected posterior precisions for the semi-omniscient policy satisfy 
\[
\E \left[ c_{i}(t) \mid \lvert\Psi(1)\rvert \right] \leq \cbar(t), \quad i \in H(\Psi(t-1)), \quad t \geq 1,
\]
where $\cbar(t)$ is defined by the recursion \eqref{eqn:cbar}. 
\end{lemma}
\begin{IEEEproof}
See Appendix \ref{app:cbar}.
\end{IEEEproof}

\ignore{  
Given \eqref{eqn:pSemi} and Assumption \ref{ass:c}, the permutation $\pi$ in \eqref{darap-eq:pi} ranks all of the indices $i \in \Psi(t-1)$ equally, followed by $i \in G(\Psi(t-1))$, again all equally.  It is then straightforward to see that the sequence $g(k)$ \eqref{darap-eq:g(k)} that determines the number of nonzero allocations is as follows: 
\begin{equation}\label{eqn:gSemi}
g(k) = \begin{cases}
0, & k = 1, \ldots, \lvert\Psi(1)\rvert - 1,\\
\cbar(t) \lvert\Psi(1)\rvert \left( \sqrt{\frac{\lvert G\rvert \pi_{0}}{1 - \pi_{0}}} - 1 \right), & k = \lvert\Psi(1)\rvert, \ldots, \lvert H(\Psi(1)) \rvert - 1,\\
\infty & k \geq \lvert H(\Psi(1)) \rvert. 
\end{cases}
\end{equation}
%
Comparing $g(k)$ to the effort budget $\Lambda$ shows that there are two regimes to consider depending on whether $\cbar(t)$ is lower or higher than the critical value 
\begin{equation}\label{eqn:ccrit}
\ccrit = \frac{\Lambda}{\lvert\Psi(1)\rvert} \left( \sqrt{\frac{\lvert G\rvert \pi_{0}}{1 - \pi_{0}}} - 1 \right)^{-1}. 
\end{equation}
The definition and behavior of $\cbar(t)$ in the first regime $\cbar(t) < \ccrit$ are summarized below.

\begin{lemma}\label{lem:cbarLow0}
Under Assumptions \ref{ass:numTargets}, \ref{ass:Delta}, \ref{ass:c}, and \ref{ass:Lambda}, in the low-precision regime $\cbar(t) < \ccrit$ the expected target precision conditioned on $\lvert\Psi(1)\rvert$ under the semi-omniscient policy may be bounded as 
\[
\E \left[ c_{s^{(n)}(t)}(t) \mid \lvert\Psi(1)\rvert \right] \leq \cbar(t), \quad t = 1, 2, \ldots,
\]
where $\cbar(1) = c_{i}(1) = \sigma^{2} / \sigma_{0}^{2}$ and $\cbar(t)$ satisfies the following recursion for $t > 1$:
\[
\cbar(t+1) = \frac{ \pi_{0}^{3/2} + \frac{1}{\sqrt{\lvert G \rvert}} (1-\pi_{0})^{3/2} }{ \sqrt{\pi_{0}} + \sqrt{\lvert G\rvert (1-\pi_{0})} } \left( \left(1 + \lvert G \rvert\right) \cbar(t) + \frac{\Lambda}{\lvert\Psi(1)\rvert} \right).
\]
\end{lemma}
\begin{IEEEproof}
By definition of $\cbar(1)$, the bound holds for $t = 1$.  We proceed by induction.  Under Assumption \ref{ass:c} with $\cbar(t) < \ccrit$, \eqref{eqn:gSemi} implies that all locations in $H(\Psi(t-1))$ receive nonzero allocations.  Combining \eqref{eqn:c0} and \eqref{darap-eq:myopic-solution}, the target precisions $c_{s^{(n)}(t+1)}(t+1)$ are given by 
\begin{align}
c_{s^{(n)}(t+1)}(t+1) 
&= \left( \Lambda + \sum_{j\in H(\Psi(t-1))} c_{j}(t) \right) \frac{\sqrt{p_{s^{(n)}(t)}(t)}}{ \sum_{j\in H(\Psi(t-1))} \sqrt{p_{j}(t)} }\nonumber\\
&= \left( \Lambda + \sum_{j\in H(\Psi(t-1))} c_{j}(t) \right) \frac{\sqrt{p_{s^{(n)}(t)}(t)}}{\lvert\Psi(1)\rvert \left( \sqrt{\pi_{0}} + \sqrt{\lvert G\rvert (1-\pi_{0})} \right)},\label{eqn:cLow}
\end{align}
using \eqref{eqn:pSemi} in the second equality.  
%
%
Conditioned on $\lvert\Psi(1)\rvert$, there are two random quantities in \eqref{eqn:cLow}: the probability $p_{s^{(n)}(t)}(t)$, which depends on whether the target moves or remains stationary between stages $t-1$ and $t$, and the precisions $c_{j}(t)$, which depend on target motion up to stage $t-1$.  These quantities are independent according to the target model.  Thus taking the conditional expectation of both sides of \eqref{eqn:cLow} yields 
\begin{subequations}
\begin{align}
\E \left[ c_{s^{(n)}(t+1)}(t+1) \mid \lvert\Psi(1)\rvert \right] &= \left( \Lambda + \sum_{j\in H(\Psi(t-1))} \E \left[ c_{j}(t) \mid \lvert\Psi(1)\rvert \right] \right) \frac{\E \left[ \sqrt{p_{s^{(n)}(t)}(t)} \mid \lvert\Psi(1)\rvert \right]}{\lvert\Psi(1)\rvert \left( \sqrt{\pi_{0}} + \sqrt{\lvert G\rvert (1-\pi_{0})} \right)}\nonumber\\
&= \left( \Lambda + \sum_{j\in H(\Psi(t-1))} \E \left[ c_{j}(t) \mid \lvert\Psi(1)\rvert \right] \right) \frac{\pi_{0}^{3/2} + \frac{1}{\sqrt{\lvert G \rvert}} (1-\pi_{0})^{3/2}}{\lvert\Psi(1)\rvert \left( \sqrt{\pi_{0}} + \sqrt{\lvert G\rvert (1-\pi_{0})} \right) }\label{eqn:cbarLow1}\\
&\leq \left( \frac{\Lambda}{\lvert\Psi(1)\rvert} + \left(1 + \lvert G \rvert\right) \cbar(t) \right) \frac{\pi_{0}^{3/2} + \frac{1}{\sqrt{\lvert G \rvert}} (1-\pi_{0})^{3/2}}{\sqrt{\pi_{0}} + \sqrt{\lvert G\rvert (1-\pi_{0})} }\label{eqn:cbarLow2}\\
&\equiv \cbar(t+1).\label{eqn:cbarLow3}
\end{align}
\end{subequations}
The second line \eqref{eqn:cbarLow1} follows from \eqref{eqn:pSemi} and because $s^{(n)}(t) = s^{(n)}(t-1)$ with probability $\pi_{0}$ and $s^{(n)}(t) \in G(s^{(n)}(t-1))$ with probability $1 - \pi_{0}$.  In the third line \eqref{eqn:cbarLow2}, we have used the fact that the expected precision for non-targets is lower than for targets, the inductive assumption $\cbar(t) \geq \E \left[ c_{s^{(n)}(t)}(t) \mid \lvert\Psi(1)\rvert \right]$, and the equality $\lvert H(\Psi(t-1)) \rvert = (1 + \lvert G \rvert) \lvert\Psi(1)\rvert$. 
Lastly in \eqref{eqn:cbarLow3}, the recursive definition of $\cbar(t+1)$ ensures that it remains an upper bound on $\E \left[ c_{s^{(n)}(t+1)}(t+1) \mid \lvert\Psi(1)\rvert \right]$, thus completing the induction.  
\end{IEEEproof}
}
\begin{remark}
It can be shown that for $\cbar(t) < \ccrit$, the coefficient multiplying $\cbar(t)$ in \eqref{eqn:cbar} is 
greater than or equal to $1$, with equality if and only if Assumption \ref{ass:pi0} holds with equality. 
Hence $\cbar(t)$ increases geometrically with $t$ if the inequality in Assumption \ref{ass:pi0} is strict. 
A closed-form expression can be derived for $\cbar(t)$ in the regime $\cbar(t) < \ccrit$, for example by viewing \eqref{eqn:cbar} 
as specifying a first-order recursive system driven by a step input, but we do not pursue this here.
\end{remark}

Using Lemma \ref{lem:cbar} and taking the limit $t \to \infty$, we arrive at a simple characterization of the semi-omniscient policy. 
\begin{prop}\label{prop:semi0}
Under Assumptions \ref{ass:numTargets}, \ref{ass:Delta}, and \ref{ass:pi0}--\ref{ass:Lambda}, in the limit $t \to \infty$ the expected per-stage cost of the semi-omniscient policy is bounded as 
\[
\E\left[ M_{t}(\boldsymbol{\lambda}^{\semi}) \right] \geq \frac{p_{0} Q (p_{0} Q + 1-p_{0})}{\pi_{0} \Lambda t} + O\left(\frac{1}{t^{2}}\right).
\]
\end{prop}
\begin{IEEEproof}
See Appendix \ref{app:semi0}. 
\end{IEEEproof}
\ignore{
\begin{IEEEproof}
First we define a recursion for $\cbar(t)$ in the regime $\cbar(t) \geq \ccrit$ and show by induction that $\cbar(t)$ continues to be an upper bound on the conditional expected precision $\E \left[ c_{s^{(n)}(t)}(t) \mid \lvert\Psi(1)\rvert \right]$.  Given Assumption \ref{ass:c} and $\cbar(t) \geq \ccrit$, \eqref{eqn:gSemi} indicates that only the previous target locations $i \in \Psi(t-1)$ are allocated nonzero effort.  In the case $s^{(n)}(t) = s^{(n)}(t-1)$, which occurs with probability $\pi_{0}$, we proceed as in the proof of Lemma \ref{lem:cbarLow0} to obtain 
\[
c_{s^{(n)}(t+1)}(t+1) = \frac{1}{\lvert\Psi(1)\rvert} \left( \Lambda + \sum_{j\in \Psi(t-1)} c_{j}(t) \right).
\]
In the other case $s^{(n)}(t) \in G(s^{(n)}(t-1))$ with probability $1-\pi_{0}$, $\lambda_{s^{(n)}(t)}(t) = 0$ and $c_{s^{(n)}(t+1)}(t+1) = c_{s^{(n)}(t)}(t)$.  Therefore the expected target precision conditioned on $\lvert\Psi(1)\rvert$ is given by 
\begin{align}
\E \left[ c_{s^{(n)}(t+1)}(t+1) \mid \lvert\Psi(1)\rvert \right] &= \frac{\pi_{0}}{\lvert\Psi(1)\rvert} \left( \Lambda + \sum_{j\in \Psi(t-1)} \E \left[ c_{j}(t) \mid \lvert\Psi(1)\rvert \right] \right) + (1-\pi_{0}) \E \left[ c_{s^{(n)}(t)}(t) \mid \lvert\Psi(1)\rvert \right]\nonumber\\
&\leq \cbar(t) + \frac{\pi_{0} \Lambda}{\lvert\Psi(1)\rvert}\nonumber\\
&\equiv \cbar(t+1),\label{eqn:cbarHigh0}
\end{align}
where in the second line we use the inductive assumption on $\cbar(t)$ (initialized from $t=1$ or Lemma \ref{lem:cbarLow0}) 
and the fact that the expected precision for non-targets is lower than for targets, and in the third line we complete the induction.  The recursion for $\cbar(t)$ in \eqref{eqn:cbarHigh0} implies that 
\begin{equation}\label{eqn:cbarInf}
\cbar(t) = \frac{\pi_{0} \Lambda}{\lvert\Psi(1)\rvert} t + O(1) 
\end{equation}
%
as $t \to \infty$. 

We now rewrite the expected per-stage cost by combining \eqref{eqn:myopicCost2} and \eqref{eqn:c0} and iterating expectations to yield 
\[
\E\left[ M_{t}(\boldsymbol{\lambda}) \right] = \E\left\{ \sum_{n=1}^{\lvert\Psi(1)\rvert} \E\left[ \frac{1}{c_{s^{(n)}(t+1)}(t+1)} \mid \lvert\Psi(1)\rvert \right] \right\}.
\]
Using the convexity of the function $1/x$ and Jensen's inequality, this may be bounded from below as 
\[
\E\left[ M_{t}(\boldsymbol{\lambda}) \right] \geq \E\left\{ \sum_{n=1}^{\lvert\Psi(1)\rvert} \frac{1}{\E \left[ c_{s^{(n)}(t+1)}(t+1) \mid \lvert\Psi(1)\rvert \right]} \right\}.
\]
Substituting \eqref{eqn:cbarHigh0}, \eqref{eqn:cbarInf} and recalling that $\lvert\Psi(1)\rvert$ is binomially distributed with parameters $Q$ and $p_{0}$, 
\begin{align*}
\E\left[ M_{t}(\boldsymbol{\lambda}^{\semi}) \right] 
&\geq \E\left\{ \frac{\lvert\Psi(1)\rvert}{\cbar(t+1)} \right\}\\
&= \E\left\{ \frac{\lvert\Psi(1)\rvert^{2}}{\pi_{0} \Lambda t} + O\left(\frac{1}{t^{2}}\right) \right\}\\
&= \frac{p_{0} Q (p_{0} Q + 1-p_{0})}{\pi_{0} \Lambda t} + O\left(\frac{1}{t^{2}}\right).
\end{align*}
\end{IEEEproof}
}

Proposition \ref{prop:semi0} can be used to determine the gain of the semi-omniscient policy relative to uniform allocation in the limit $T \to \infty$, again invoking Assumption \ref{ass:gamma} so that the total cost $J_{T}(\boldsymbol{\lambda})$ is dominated by terms at large $t$.  In the special case $\gamma(T) = 1$, $\gamma(t) = 0$ for $t < T$, the gain reduces to the ratio of the expected final-stage costs.  From Proposition \ref{prop:semi0} and using \eqref{eqn:myopicUni0} for the per-stage cost of the uniform policy with $\Lambdab(t) = \Lambda t$, the gain is bounded as 
\begin{align}
\Gamma_{T}(\boldsymbol{\lambda}^{\semi}) &\leq 
\frac{\pi_{0}}{p_{0} + (1-p_{0})/Q} + O\left(\frac{1}{T}\right) \approx \frac{\pi_{0}}{p_{0}}.\label{eqn:gainSemi0}
\end{align}
Compared to Proposition \ref{prop:omni0} in the limit $T \to \infty$, the analogous result for the omniscient policy, \eqref{eqn:gainSemi0} shows that the performance of the semi-omniscient policy is discounted by the probability $\pi_{0}$ that target locations are constant from stage to stage. 

\subsection{Omniscient policy, $\Delta^{2} > 0$}
\label{subsec:omni+}


In the remainder of this section, we relax Assumption \ref{ass:Delta} on the variance increment $\Delta^{2}$.  For $\Delta^{2} > 0$, the evolution equation for posterior variances reverts to \eqref{eqn:sigma+}, from which 
it is difficult to obtain a closed-form expression for $\sigma_{s^{(n)}(t)}^{2}(t)$, in contrast to the case $\Delta^{2} = 0$. 
We focus instead on the steady-state behavior in the limit of large $t$. 
Using Assumption \ref{ass:gamma}, in the limit $T \to \infty$ the cost $J_{T}(\boldsymbol{\lambda})$ becomes well-approximated by a sum of terms at large $t$, each of which is proportional to the steady-state expected per-stage cost $\lim_{t\to\infty} \E\left[ M_{t}(\boldsymbol{\lambda}) \right]$.  This simplification allows us to obtain the following bound on the gain of the omniscient policy.  
%
%
%
%
%
\begin{prop}\label{prop:omni+}
Let $r_{+} = \sigma^{2} Q / (\Delta^{2} \Lambda)$.  Under Assumptions \ref{ass:numTargets}, \ref{ass:gamma}, and \ref{ass:Lambda}, in the steady-state limit $T \to \infty$ the gain of the omniscient policy relative to uniform allocation is bounded from above as 
\begin{align*}
\lim_{T\to\infty} \Gamma_{T}(\boldsymbol{\lambda}^{\omni}) \leq 
&\left. \left( \frac{\sqrt{1 + 4r_{+}} - 1}{2r_{+}} \right) \right/\\
&\qquad \left( 
\frac{\sqrt{1 + 4p_{0} r_{+}} - 1}{2r_{+}} + \frac{1-p_{0}}{Q} \frac{1 + 3p_{0} r_{+}}{(1 + 4p_{0} r_{+})^{3/2}} 
- \frac{(1-p_{0}) (1-2p_{0})}{Q^{2}} \frac{r_{+} (1 + 2p_{0} r_{+})}{(1 + 4p_{0} r_{+})^{5/2}} \right)\\
= &\frac{1}{p_{0} + (1-p_{0})/Q} \left( 1 - O(r_{+}) \right). 
\end{align*}
\ignore{
At high SNR (
For small $\sigma^{2} Q / (\Delta^{2} \Lambda)$, the omniscient gain is given by $\lim_{T\to\infty} \Gamma_{T}(\boldsymbol{\lambda}^{\omni}) =$
\begin{equation*}
\begin{split}
\frac{1}{p_{0} + (1-p_{0})/Q} 
\left[ 1 - (1 - \ptilde) \frac{\sigma^{2} Q}{\Delta^{2} \Lambda} + O\left( \frac{\sigma^{2} Q}{\Delta^{2} \Lambda} \right)^{2} \right],
\end{split}
\end{equation*}
where $\ptilde = (1/Q) \E\{ \lvert\Psi(1)\rvert^{3} \} / \E\{ \lvert\Psi(1)\rvert^{2} \} = p_{0} + O(1/Q)$.
}
\end{prop}
\begin{IEEEproof}
See Appendix \ref{app:omni+}.
\end{IEEEproof}
\ignore{
\begin{IEEEproof}
Given Assumption \ref{ass:gamma}, the gain reduces to a ratio of steady-state expected per-stage costs.  To compute the per-stage cost, we first note that the denominator $c_{i}(t) + \lambda_{i}(t)$ in \eqref{eqn:myopicCost2} corresponds to the posterior variance after measurement but before the increment $\Delta^{2}$, while $\sigma_{\steady}^{2}$ in \eqref{eqn:sigmaSS} is the steady-state variance after the increment.  Hence the steady-state per-stage cost conditioned on $\Psi(t)$ is 
\begin{equation}\label{eqn:gSS}
\lim_{t\to\infty} \sum_{i\in\Psi(t)} \frac{1}{c_{i}(t) + \lambda_{i}(t)} 
= \lvert\Psi(1)\rvert \frac{\sigma_{\steady}^{2} - \Delta^{2}}{\sigma^{2}} = \frac{\Delta^{2} \lvert\Psi(1)\rvert}{2\sigma^{2}} \left(\sqrt{1 + \frac{4\sigma^{2}}{\Delta^{2} \lambda_{s^{(n)}(t)}(t)} } - 1 \right).
\end{equation}
(This can also be derived by setting $c_{i}(t) = \sigma^{2} / \sigma_{\steady}^{2}$ in \eqref{eqn:myopicCost2} followed by some algebra.)  For the uniform policy, $\lambda_{s^{(n)}(t)}(t) = \Lambda / Q$ and the expectation over $\lvert\Psi(1)\rvert$ gives  
\begin{equation}\label{eqn:myopicUniSS}
\lim_{t\to\infty} \E \left\{ M_{t}(\boldsymbol{\lambda}^{\uni}) \right\} = \frac{\Delta^{2} p_{0} Q}{2\sigma^{2}} \left(\sqrt{1 + \frac{4\sigma^{2} Q}{\Delta^{2} \Lambda} } - 1 \right).
\end{equation}
For the omniscient policy, $\lambda_{s^{(n)}(t)}(t) = \Lambda / \lvert\Psi(1)\rvert$ and \eqref{eqn:gSS} takes the form $x (\sqrt{1+x} - 1)$, where $x$ is proportional to $\lvert\Psi(1)\rvert$.  Since $x (\sqrt{1+x} - 1)$ is a convex function, applying Jensen's inequality to \eqref{eqn:gSS} yields 
\begin{align}
\lim_{t\to\infty} \E \left\{ M_{t}(\boldsymbol{\lambda}^{\omni}) \right\} &\geq 
\frac{\Delta^{2} p_{0} Q}{2\sigma^{2}} \left(\sqrt{1 + \frac{4\sigma^{2} p_{0} Q}{\Delta^{2} \Lambda} } - 1 \right).\label{eqn:myopicOmniSS}
\end{align}
%
Taking the ratio of \eqref{eqn:myopicOmniSS} and \eqref{eqn:myopicUniSS} yields the bound in the proposition statement, which holds for all SNR. 

For high SNR (small $\sigma^{2} Q / (\Delta^{2} \Lambda)$), we may apply the Taylor expansion $\sqrt{1 + x} = 1 + x/2 - x^{2}/8 + O(x^{3})$ to \eqref{eqn:myopicUniSS}, yielding 
\begin{equation}\label{eqn:myopicUniSS2}
\lim_{t\to\infty} \E \left\{ M_{t}(\boldsymbol{\lambda}^{\uni}) \right\} = \frac{p_{0} Q^{2}}{\Lambda} \left[ 1 - \frac{\sigma^{2} Q}{\Delta^{2} \Lambda} + O\left( \left( \frac{\sigma^{2} Q}{\Delta^{2} \Lambda} \right)^{2} \right) \right].
\end{equation}
Using the same expansion in \eqref{eqn:gSS} with $\lambda_{s^{(n)}(t)}(t) = \Lambda / \lvert\Psi(1)\rvert$ and dropping the higher-order terms, we obtain an alternative lower bound on the steady-state per-stage cost of the omniscient policy: 
\[
\lim_{t\to\infty} M_{t}(\boldsymbol{\lambda}^{\omni}) \geq \frac{\lvert\Psi(1)\rvert^{2}}{\Lambda} \left( 1 - \frac{\sigma^{2} \lvert\Psi(1)\rvert}{\Delta^{2} \Lambda} \right).
\]
Hence 
\begin{equation}\label{eqn:myopicOmniSS2}
\lim_{t\to\infty} \E\left\{ M_{t}(\boldsymbol{\lambda}^{\omni}) \right\} \geq \frac{p_{0} Q (p_{0} Q + 1-p_{0})}{\Lambda} \left( 1 - \frac{\ptilde \sigma^{2} Q}{\Delta^{2} \Lambda} \right)
\end{equation}
using the moments of the binomial distribution. The high-SNR expression for the gain follows from the ratio of \eqref{eqn:myopicUniSS2} and \eqref{eqn:myopicOmniSS2}. 
\end{IEEEproof}
}

Figure \ref{darap-fig:bounds-comparison}(a) is a heat map representing the bound in Proposition \ref{prop:omni+} as a function of SNR and $p_0$, where all other parameters are given in Table \ref{darap-table:simulation-parameters}.  The upper horizontal axis indicates the equivalent values of $r_{+}$, which is inversely proportional to SNR when $\Delta^{2}$ is fixed as in Table \ref{darap-table:simulation-parameters}.  Besides confirming that the potential gain increases as $p_{0}$ decreases, the heat map shows that there are three regimes with respect to SNR.  In Region (A), the SNR is insufficient to offset the degradation due to $\Delta^2$ and gains scale only as $1/\sqrt{p_0}$.  In Region (C), the SNR is high and knowledge of target locations, which increases the observation effort per target by $1/p_{0}$ on average, also increases the gain by approximately the same factor, $1/[p_0+(1-p_0)/Q]$.  In Region (B), the gain ranges between the two extremes.
\ignore{To further interpret Proposition \ref{prop:omni+}, normalize the effort budget by setting $\Lambda = Q$ and scale $\sigma^{2}$ as needed to compensate.  Since the effective noise variance in \eqref{darap-eq:obs} is given by the ratio $\sigma^{2} / \lambda_{i}(t)$, this normalization has no effect on the model.  The expressions in Proposition \ref{prop:omni+} then become functions of the ratio $\sigma^{2} / \Delta^{2}$.  For small $\sigma^{2} / \Delta^{2}$ 
and large $Q$, 
the proposition shows that the gain approaches $1/p_{0}$.  On the other hand, if $\sigma^{2} / \Delta^{2} \gg 1/(4 p_{0})$, 
the upper bound on the gain simplifies to $1/\sqrt{p_{0}}$.  For intermediate values of $\sigma^{2} / \Delta^{2}$, the bound ranges between the two extremes.  This dependence on $\sigma^{2} / \Delta^{2}$ may be interpreted as follows:  If $\sigma^{2} / \Delta^{2}$ is small, i.e., the noise variance is much smaller than the target variance increment, then observations are important in maintaining the steady-state equilibrium.  In this case, knowledge of the target locations, which increases the observation effort per target by $1/p_{0}$ on average, also increases the gain by the same factor.  If $\sigma^{2} / \Delta^{2}$ is large, then observations are less important and the gain due to resource concentration decreases to $1/\sqrt{p_{0}}$.  
}

\begin{figure}
\centering
\subfloat[{\footnotesize Prop. \ref{prop:omni+} (SNR vs $p_0$)}] {
\includegraphics[height=\policyFigHeight]{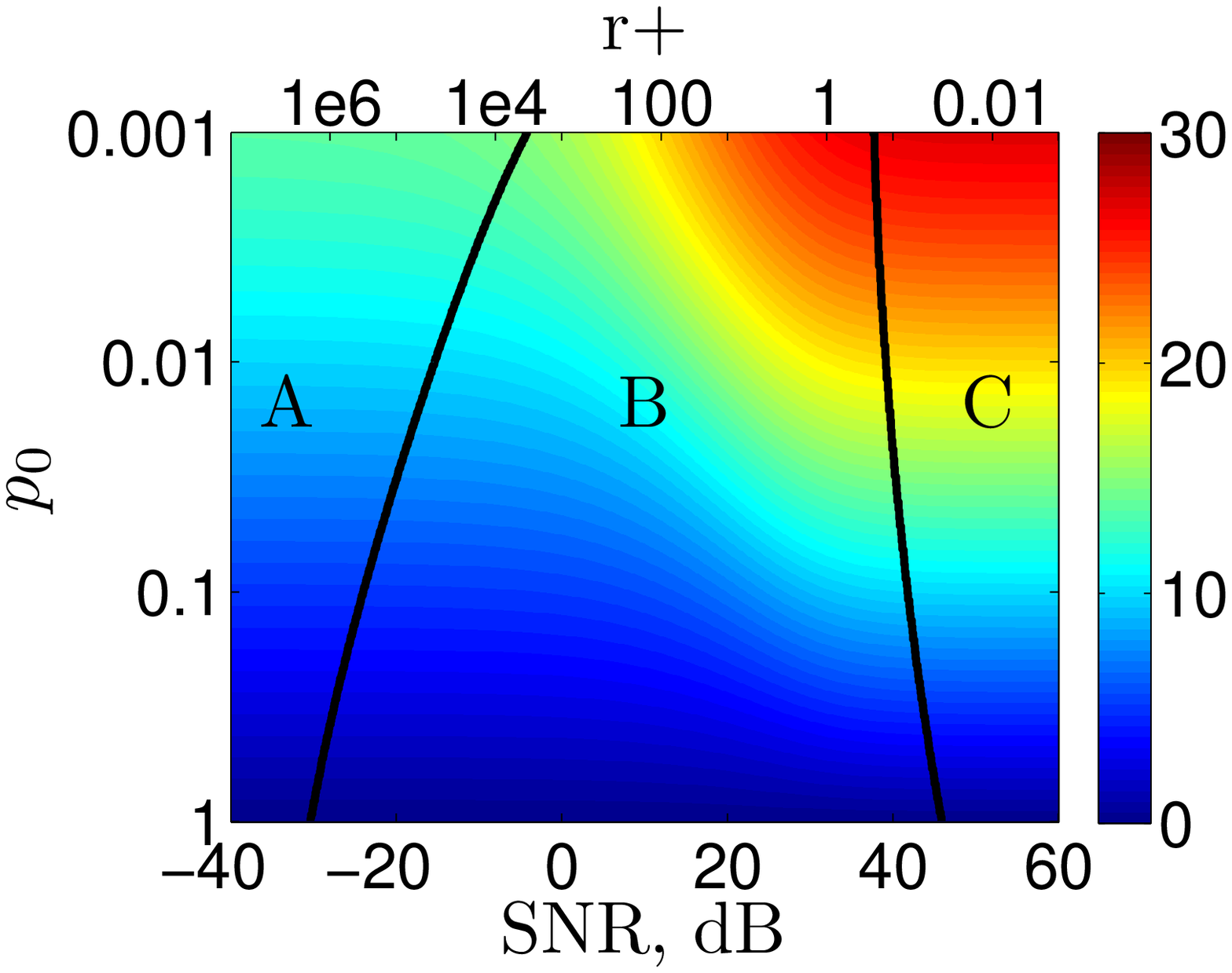}
}
\subfloat[{\footnotesize Prop. \ref{prop:omni+} (SNR vs $\pi_0$, $p_0=0.01$)}]{
\includegraphics[height=\policyFigHeight]{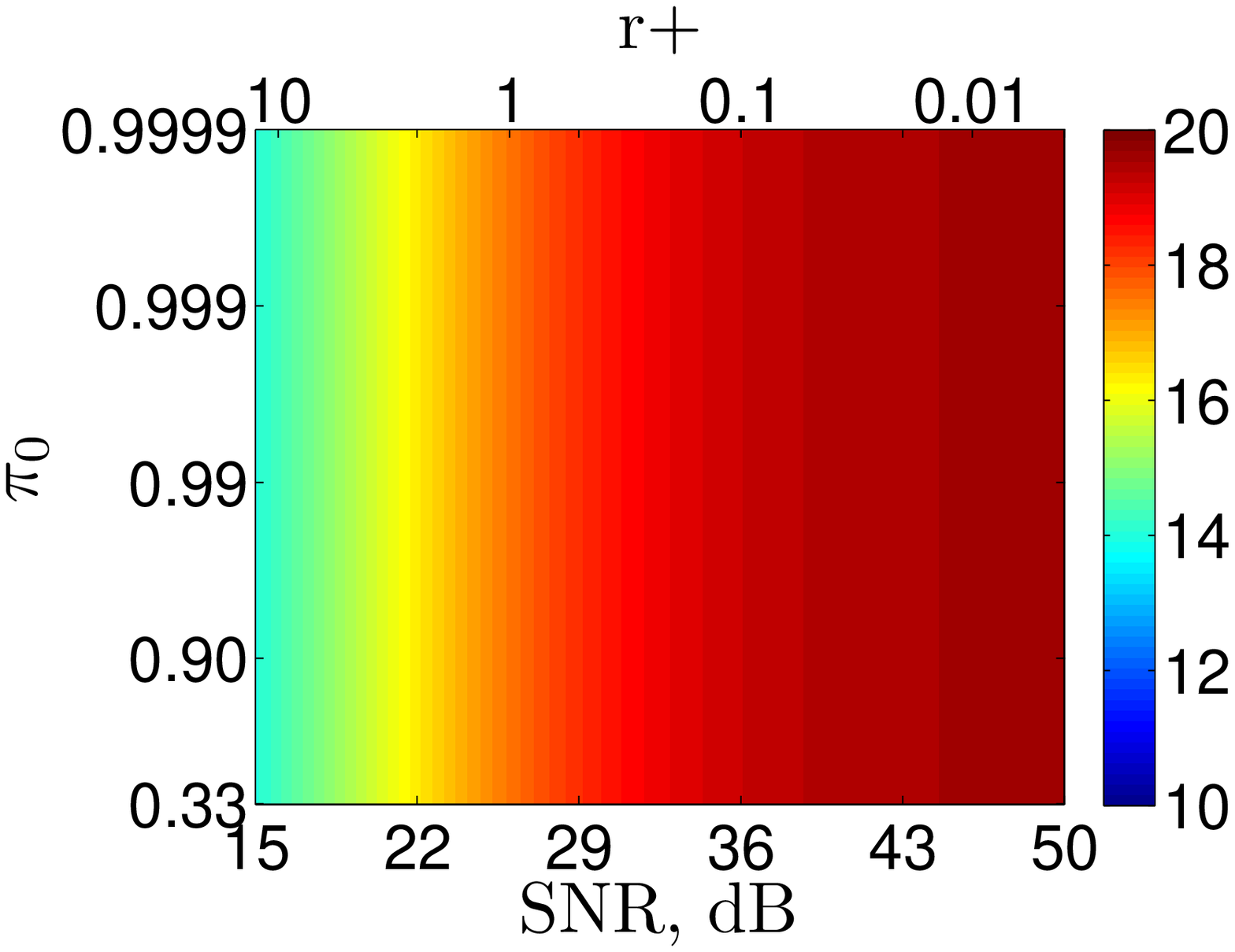}
}
\\\subfloat[{\footnotesize Prop. \ref{prop:semiLarge} (SNR vs $\pi_0$,  $p_0=0.01$)}]{ \includegraphics[height=\policyFigHeight]{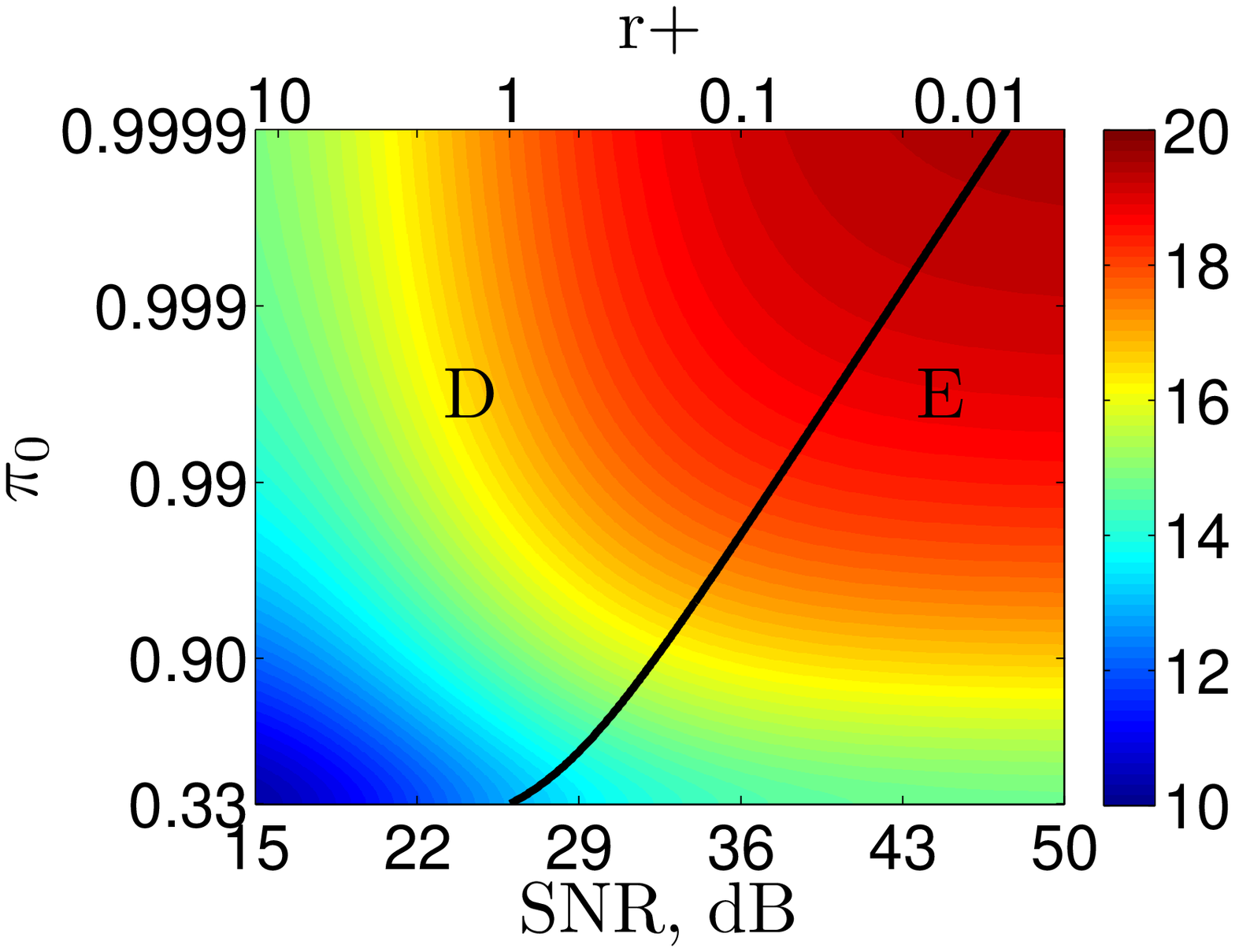}
}
\caption{
Bounds on the gain of the omniscient policy (from Proposition \ref{prop:omni+}, panels (a) and (b)) and the semi-omniscient policy (from Proposition \ref{prop:semiLarge}, panel (c)) with respect to the uniform allocation policy. 
The left plot (a) confirms that potential gains increase as the sparsity parameter $p_{0}$ decreases and indicates three regimes with respect to SNR: in Region (A), gains are relatively low at $1/\sqrt{p_0}$; in Region (C), gains are near their maximum value $1/[p_0 + (1-p_0)/Q]$; in Region (B), they are in between.  The middle and right plots (b, c) compare Props. \ref{prop:omni+} and \ref{prop:semiLarge} over the same values of SNR and $\pi_{0}$; (E) denotes the region where the sufficient condition \eqref{eqn:condSemiLarge} is satisfied and Prop. \ref{prop:semiLarge} gives a provable upper bound.  This bound (c) is tighter than the omniscient bound (b) because it accounts for the effect of having only causal knowledge of target locations, resulting in decreasing gains as $\pi_{0}$ decreases.} 
\label{darap-fig:bounds-comparison}
\end{figure}

\subsection{Semi-omniscient policy, $\Delta^{2} > 0$}
\label{subsec:semi+}

Next we consider the steady-state behavior of the semi-omniscient policy.  
As discussed in Section \ref{subsec:semi0}, for $t > 1$ the posterior precisions $c_{i}(t)$ become non-uniform and random.  However, in the regime 
of small $r_{+}$, where $r_{+}$ is defined in Proposition \ref{prop:omni+}, all $c_{i}(t)$ are guaranteed to be small. 
We can then derive the following bound on the gain of the semi-omniscient policy under the same assumptions as in Proposition \ref{prop:omni+}. 
\begin{prop}\label{prop:semiLarge}
Define $q_{+} = (1 + \lvert G \rvert) r_{+}$. 
Given Assumptions \ref{ass:numTargets}, \ref{ass:gamma}, and \ref{ass:Lambda}, 
assume in addition that 
%
\begin{equation}\label{eqn:condSemiLarge}
r_{+} (e(\pi_0,G)+1) \leq 1.
\end{equation}
Then in the steady-state limit $T \to \infty$, the gain of the semi-omniscient policy relative to uniform allocation is bounded from above as 
\begin{align*}
\lim_{T\to\infty} \Gamma_{T}(\boldsymbol{\lambda}^{\semi}) \leq 
&\left. \Biggl( \frac{\sqrt{1 + 4r_{+}} - 1}{2r_{+} \bigl(\sqrt{\pi_{0}} + \sqrt{\lvert G \rvert (1-\pi_{0})} \bigr)^{2}} \Biggr) \right/\\
&\qquad
\left( \frac{p_{0}}{1+p_{0} q_{+}} + \frac{1-p_{0}}{Q} \frac{1}{(1 + p_{0} q_{+})^{3}} 
- \frac{(1-p_{0}) (1-2p_{0})}{Q^{2}} 
\frac{q_{+}}{(1 + p_{0} q_{+})^{4}} \right)\\
= &\frac{1 - O(r_{+})}
{(p_{0} + (1-p_{0})/Q) \left(\sqrt{\pi_{0}} + \sqrt{\lvert G \rvert (1-\pi_{0})} \right)^{2} },
\end{align*}
with equality in the 
limit 
$r_{+} \to 0$.
\end{prop}
\begin{IEEEproof}
See Appendix \ref{app:semiLarge}. 
\ignore{
As in Proposition \ref{prop:omni+}, under Assumption \ref{ass:gamma} the gain reduces to the ratio of steady-state expected per-stage costs.  To compute the per-stage cost for the semi-omniscient policy, we first show that the precisions $c_{i}(t)$ are small as claimed.  Rewriting the evolution equation \eqref{eqn:sigma+} in terms of $c_{s^{(n)}(t)}(t)$ gives 
\begin{equation}\label{eqn:cSmall}
c_{s^{(n)}(t+1)}(t+1) = \frac{\sigma^{2}}{\Delta^{2}} \frac{c_{s^{(n)}(t)}(t) + \lambda_{s^{(n)}(t)}(t)}{c_{s^{(n)}(t)}(t) + \lambda_{s^{(n)}(t)}(t) + \sigma^{2}/\Delta^{2}} < \frac{\sigma^{2}}{\Delta^{2}} 
\end{equation}
for targets, and the same bound holds for non-targets as well.  This bound together with assumption \eqref{eqn:condSemiLarge} imply that the semi-omniscient policy allocates nonzero effort to all locations in $H(\Psi(t-1))$, i.e., the sequence $g(k)$ \eqref{darap-eq:g(k)} satisfies $g(k) < \Lambda$ for $k = \lvert H(\Psi(t-1)) \rvert - 1$. 
Using \eqref{eqn:pSemi}, the sum of probability ratios $\sqrt{p_{\pi(i)}(t) / p_{\pi(k+1)}(t)}$ in \eqref{darap-eq:g(k)} can be bounded by $Q \sqrt{\lvert G \rvert \pi_{0} / (1-\pi_{0})}$.  Combining this with $c_{i}(t) < \sigma^{2} / \Delta^{2}$ and \eqref{eqn:condSemiLarge} yields 
\[
g\left( \lvert H(\Psi(t-1)) \rvert - 1 \right) < \frac{\sigma^{2} Q}{\Delta^{2}} \sqrt{\frac{\lvert G \rvert \pi_{0}}{1-\pi_{0}}} \leq \Lambda
\]
as desired.

Given that $\lambda_{i}(t) > 0$ for all $i \in H(\Psi(t-1))$, the per-stage cost for the semi-omniscient policy can be computed from \eqref{darap-eq:myopic-cost}, \eqref{darap-eq:myopic-solution} and \eqref{eqn:pSemi} as 
\begin{align*}
M_{t}(\boldsymbol{\lambda}^{\semi}) 
&= \frac{\left( \sum_{i\in H(\Psi(t-1))} \sqrt{p_{i}(t)} \right)^{2}}{\Lambda + \sum_{i\in H(\Psi(t-1))} c_{i}(t)}\\
&\geq \frac{\left( \sum_{i\in H(\Psi(t-1))} \sqrt{p_{i}(t)} \right)^{2}}{\Lambda + Q \sigma^{2} / \Delta^{2}}\\
&= \frac{\lvert\Psi(1)\rvert^{2} \left( \sqrt{\pi_{0}} + \sqrt{\lvert G \rvert (1-\pi_{0})} \right)^{2}}{\Lambda \left(1 + \sigma^{2} Q / (\Delta^{2} \Lambda)\right)}, 
\end{align*}
where the inequality follows from \eqref{eqn:cSmall}.  Taking the expectation with respect to $\lvert\Psi(1)\rvert$ gives 
\begin{equation}\label{eqn:myopicSemiLarge}
\E\left\{ M_{t}(\boldsymbol{\lambda}^{\semi}) \right\} \geq \frac{p_{0}Q (p_{0}Q + 1 - p_{0}) \left( \sqrt{\pi_{0}} + \sqrt{\lvert G \rvert (1-\pi_{0})} \right)^{2}}{\Lambda \left(1 + \sigma^{2} Q / (\Delta^{2} \Lambda)\right)}. 
\end{equation}
%
%
%
The result is obtained from the ratio of \eqref{eqn:myopicSemiLarge} and \eqref{eqn:myopicUniSS2}. 
}
\end{IEEEproof}

\ignore{
Proposition \ref{prop:semiLarge} may be compared to Proposition \ref{prop:omni+} in the regime $r_{+} \ll 1$, the analogous result for the omniscient policy.  
The comparison shows that the semi-omniscient vs. omniscient degradation factor due to not knowing future target locations is 
%
$\left(\sqrt{\pi_{0}} + \sqrt{\lvert G \rvert (1-\pi_{0})} \right)^{-2}$, 
%
which is strictly less than $1$ for $\pi_{0} < 1$.
}

Figures \ref{darap-fig:bounds-comparison}(b) and (c) compare the omniscient and semi-omniscient bounds in Propositions \ref{prop:omni+} and \ref{prop:semiLarge} 
as functions of SNR and $\pi_0$.  All other parameters are fixed as in Table \ref{darap-table:simulation-parameters}.  The sufficient condition \eqref{eqn:condSemiLarge} for Proposition \ref{prop:semiLarge} is satisfied in region (E) in Fig.~\ref{darap-fig:bounds-comparison}(c).  The resulting semi-omniscient bound is tighter than the omniscient bound because it accounts for the lack of knowledge of future target locations, reflected in decreasing gains as $\pi_{0}$ decreases. In the limit $r_{+} \to 0$, a comparison of Propositions \ref{prop:omni+} and \ref{prop:semiLarge} shows that the semi-omniscient vs.\ omniscient degradation factor is 
%
$\left(\sqrt{\pi_{0}} + \sqrt{\lvert G \rvert (1-\pi_{0})} \right)^{-2}$, which is strictly less than $1$ for $\pi_{0} < 1$. 
Comparisons of Propositions \ref{prop:omni+} and \ref{prop:semiLarge} to the proposed D-ARAP policies are presented in Section \ref{darap-sec:performance-analysis}. 
\ignore{In Section \ref{darap-sec:performance-analysis}, we compare our proposed policies to Propositions \ref{prop:omni+} and \ref{prop:semiLarge} in the following way:  We compare the bounds as stated in the regimes where they apply, and extrapolate in other cases by taking the minimum of the two bounds.}

For the general case $\Delta^{2} > 0$, we take a similar approach as in Section \ref{subsec:semi0}, invoking Assumptions \ref{ass:pi0}--\ref{ass:Lambda} to obtain 
an upper bound $\cbar(t)$ on the conditional expected posterior precisions $\E\left[ c_{i}(t) \mid \lvert\Psi(1)\rvert \right]$. 
As before, there are two regimes to consider, $\cbar(t) < \ccrit$ and $\cbar(t) \geq \ccrit$, where $\ccrit$ is given in \eqref{eqn:ccrit}.  For simplicity, we restrict attention to 
the second regime and establish the following result to propagate the upper bound $\cbar(t)$ forward in time. 
\begin{lemma}\label{lem:cbarHigh+}
Under Assumptions \ref{ass:numTargets} and \ref{ass:pi0}--\ref{ass:Lambda}, 
if the expected posterior 
precisions under the semi-omniscient policy satisfy 
\begin{equation}\label{eqn:cbarHigh+}
\E \left[ c_{i}(t) \mid \lvert\Psi(1)\rvert \right] \leq \cbar(t), \quad i \in H(\Psi(t-1)),
\end{equation}
for some $t$ with $\cbar(t) \geq \ccrit$, then \eqref{eqn:cbarHigh+} also holds for stage $t+1$ with 
\begin{equation}\label{eqn:cbarHigh}
\cbar(t+1) = \pi_{0} \frac{(\sigma^{2}/\Delta^{2}) (\cbar(t) + \Lambda/\lvert\Psi(1)\rvert)}{(\sigma^{2}/\Delta^{2}) + \cbar(t) + \Lambda/\lvert\Psi(1)\rvert} + (1-\pi_{0}) \frac{(\sigma^{2}/\Delta^{2}) \cbar(t)}{(\sigma^{2}/\Delta^{2}) + \cbar(t)}.
\end{equation}
\end{lemma}
\begin{IEEEproof}
See Appendix \ref{app:cbarHigh+}. 
\end{IEEEproof}
\begin{remark}
As $\Delta^{2} \to 0$, the above recursion \eqref{eqn:cbarHigh} reduces to \eqref{eqn:cbar} in the case $\cbar(t) \geq \ccrit$.  Hence \eqref{eqn:cbarHigh} can be seen as a generalization of \eqref{eqn:cbar} to the case $\Delta^{2} > 0$.  
\end{remark}

Lemma \ref{lem:cbarHigh+} can be used to derive an upper bound on the expected posterior precisions 
in the steady state limit $t \to \infty$.  Define $c_{\steady} = \lim_{t\to\infty} \E \left[ c_{s^{(n)}(t)}(t) \mid \lvert\Psi(1)\rvert \right]$ to be the steady-state precision for targets.
\begin{lemma}\label{lem:cbarInf+}
Under Assumptions \ref{ass:numTargets} and \ref{ass:pi0}--\ref{ass:Lambda}, the steady-state precision $c_{\steady}$ for targets is bounded from above by a root $\cbar_{\steady}$ of the cubic equation 
\begin{equation}\label{eqn:cbarSSCubic}
\pi_{0} \cbar_{\steady}^{-3} - \frac{\Delta^{2}}{\sigma^{2}} \left(\frac{\Delta^{2}}{\sigma^{2}} + \frac{\lvert\Psi(1)\rvert}{\Lambda} \right) \cbar_{\steady}^{-1} - \left( \frac{\Delta^{2}}{\sigma^{2}} \right)^{2} \frac{\lvert\Psi(1)\rvert}{\Lambda} = 0,
\end{equation}
provided that $\cbar_{\steady} \geq \ccrit$.
\end{lemma}
\begin{IEEEproof}
Initially, let $\cbar_{\steady}$ denote an upper bound on $c_{\steady}$ such that $\cbar_{\steady} \geq \ccrit$. Then Lemma \ref{lem:cbarHigh+} implies that $c_{\steady}$ is also bounded by 
\[
\pi_{0} \frac{(\sigma^{2}/\Delta^{2}) (\cbar_{\steady} + \Lambda/\lvert\Psi(1)\rvert)}{(\sigma^{2}/\Delta^{2}) + \cbar_{\steady} + \Lambda/\lvert\Psi(1)\rvert} + (1-\pi_{0}) \frac{(\sigma^{2}/\Delta^{2}) \cbar_{\steady}}{(\sigma^{2}/\Delta^{2}) + \cbar_{\steady}}. 
\]
To obtain a stationary upper bound on $c_{\steady}$, we equate $\cbar_{\steady}$ with the above expression, resulting in \eqref{eqn:cbarSSCubic} after some algebraic manipulations. 
This stationary bound is valid provided that a root of \eqref{eqn:cbarSSCubic} satisfies the initial assumption $\cbar_{\steady} \geq \ccrit$. 
\end{IEEEproof}

In general, it is difficult to obtain a tractable expression for $\cbar_{\steady}$ from \eqref{eqn:cbarSSCubic}.  We consider two special cases.  In the case $\pi_{0} = 1$, the cubic polynomial can be factored into 
\[
\left( \cbar_{\steady}^{-1} + \frac{\Delta^{2}}{\sigma^{2}} \right) \left( \cbar_{\steady}^{-2} - \frac{\Delta^{2}}{\sigma^{2}} \cbar_{\steady}^{-1} - \frac{\Delta^{2}}{\sigma^{2}} \frac{\lvert\Psi(1)\rvert}{\Lambda} \right).
\]
The first factor yields an infeasible negative root while the second factor can be shown to be proportional to the quadratic polynomial in \eqref{eqn:sigmaSSQuad} with $\lambda_{s^{(n)}(t)}(t) = \Lambda / \lvert\Psi(1)\rvert$, the effort allocation for targets under the omniscient policy.  Thus setting $\pi_{0} = 1$ recovers the omniscient case analyzed in Section \ref{subsec:omni+}.  For $\pi_{0} < 1$, a tractable solution can also be extracted if $\Delta^{2}$ is close to zero, as detailed in the following lemma.  
\begin{lemma}\label{lem:cubic}
In the limit $\sqrt{\Delta^{2} \Lambda / \sigma^{2}} \to 0$, the cubic equation \eqref{eqn:cbarSSCubic} has a single positive root given by 
\begin{equation}\label{eqn:cbarSSSmall}
\cbar_{\steady}^{-1} = \sqrt{\frac{\Delta^{2} \lvert\Psi(1)\rvert}{\pi_{0} \sigma^{2} \Lambda}} \left( 1 + O\left( \sqrt{\frac{\Delta^{2} \Lambda}{\sigma^{2}}} \right) \right). 
\end{equation}
\end{lemma}
\begin{IEEEproof}
See Appendix \ref{app:cubic}. 
\end{IEEEproof}

Combining Lemmas \ref{lem:cbarInf+} and \ref{lem:cubic}, we arrive at the following steady-state characterization of the semi-omniscient policy. 
\begin{prop}\label{prop:semiSmall}
Given Assumptions \ref{ass:numTargets} and \ref{ass:gamma}--\ref{ass:Lambda}, assume in addition that $\sqrt{\Delta^{2} \Lambda / \sigma^{2}} \ll 1$ and $\pi_{0}$ is large enough so that 
\begin{equation}\label{eqn:condSemiSmall}
\sqrt{\frac{\pi_{0} \sigma^{2}}{\Delta^{2} \Lambda}} e(\pi_{0}, G) 
\geq 1.
\end{equation}
Then in the steady-state limit $T \to \infty$, the gain of the semi-omniscient policy relative to uniform allocation is bounded from above as 
\[
\lim_{T\to\infty} \Gamma_{T}(\boldsymbol{\lambda}^{\semi}) \leq \sqrt{\frac{\pi_{0}}{p_{0}}} \left( 1 + O\left( \sqrt{\frac{\Delta^{2} \Lambda}{\sigma^{2}}} \right) \right).
\]
\end{prop}
\begin{IEEEproof}
See Appendix \ref{app:semiSmall}.
\end{IEEEproof}

We again compare the above result to Proposition \ref{prop:omni+} on the omniscient policy, this time in the regime $r_{+} \gg \sigma^{2} / (\Delta^{2} \Lambda) \gg 1$.  In this case, the gain of the omniscient policy is approximately given by $1/\sqrt{p_{0}}$ and the performance loss due to causal knowledge of target motion is therefore $\sqrt{\pi_{0}}$.

\fi

\section{Numerical simulations}
\label{darap-sec:performance-analysis}
\ifjournal
\else
	\subsection{Simulation set-up}
In this section, we analyze the performance of the proposed rollout and myopic+ D-ARAP policies in a variety of situations that include model mismatch and missing measurements.  We further examine the policies over a variety of performance metrics (MSE and probability of detection). With regard to rollout policies, we investigate the effects of using different base policies for offline rollout, and also compare the performance of offline and online rollout.  We continue by investigating the sensitivity of the dynamical model by varying birth/death probabilities and transition probabilities. Simulation parameters are given by Table \ref{darap-table:simulation-parameters} unless stated otherwise.

\subsection{Comparison to semi-omniscient/uniform policies}
\begin{figure}[!t] 
\centering
\subfloat[MSE, 0dB SNR]{
\label{darap-fig:mse-0}
\includegraphics[width=\figwidth]{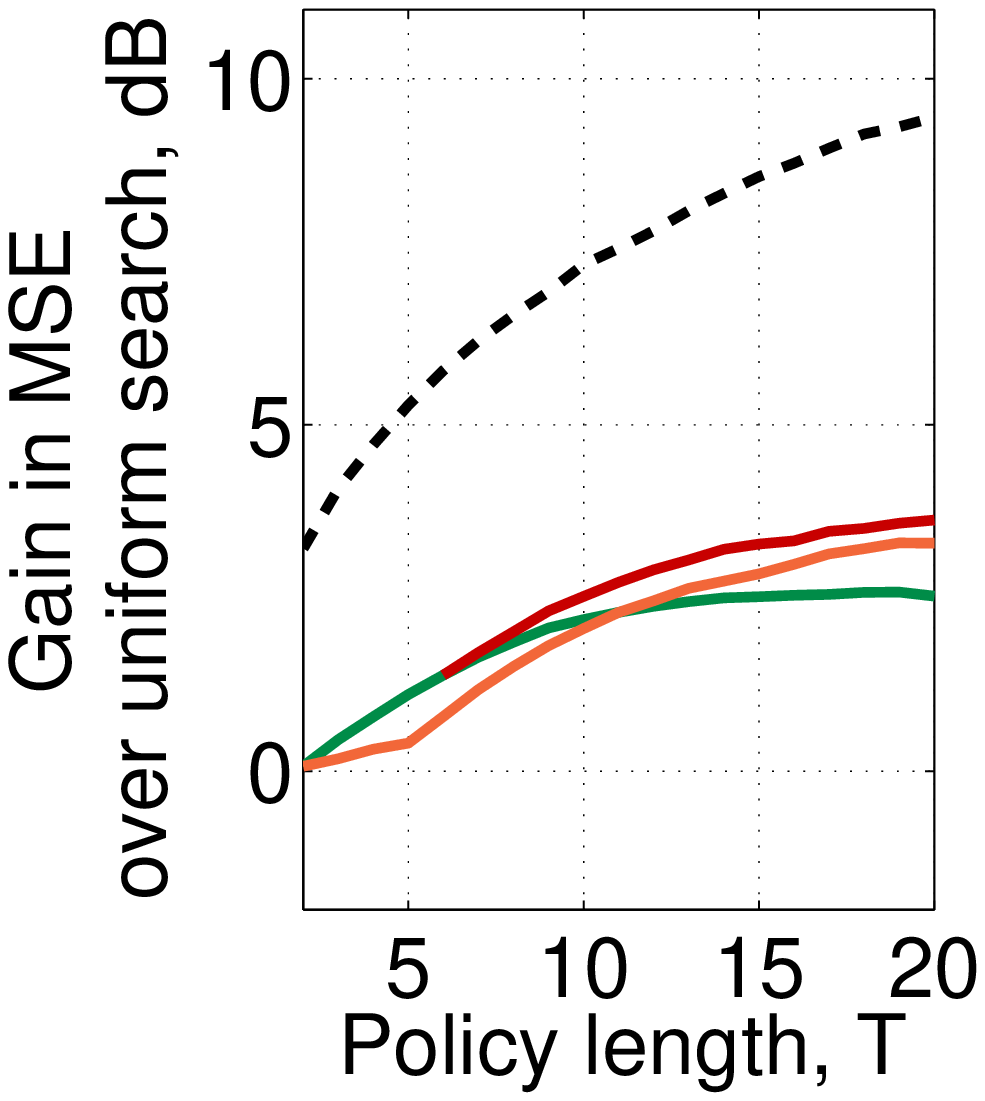}
}
\subfloat[MSE, 10dB SNR]{
\label{darap-fig:mse-10}
\includegraphics[width=\figwidth]{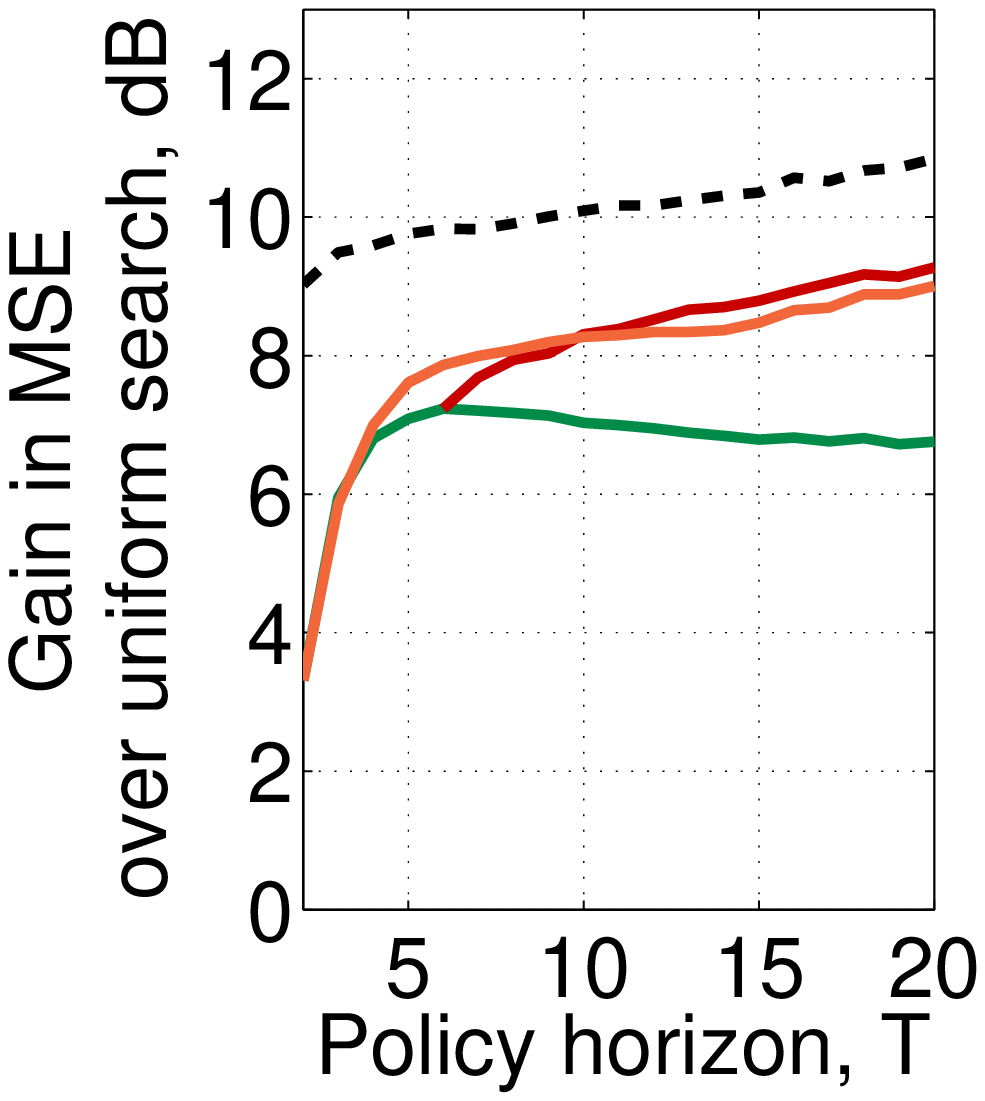}
}
\subfloat[Cost, 10dB SNR]{
\label{darap-fig:cost-10}
\includegraphics[width=\figwidth]{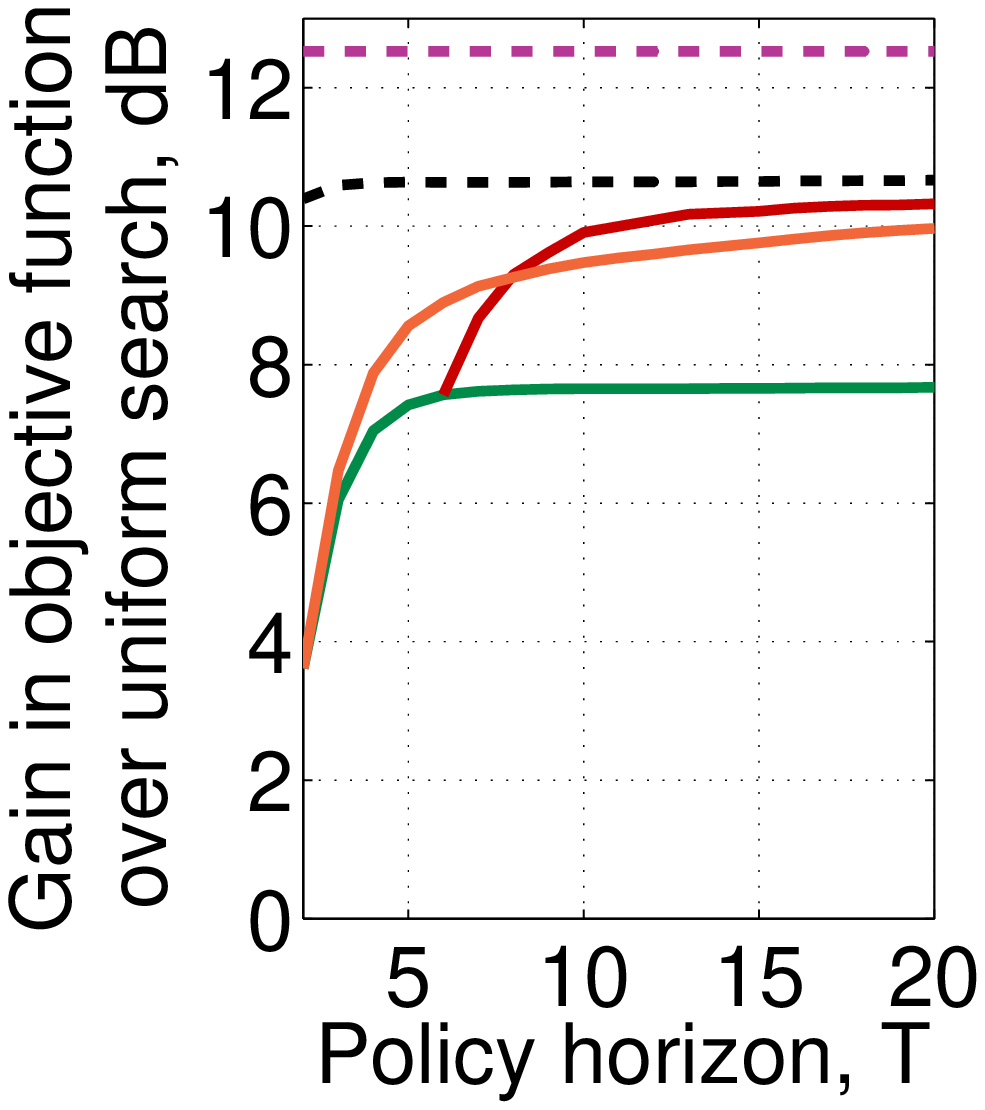}
}\\
\subfloat[Prob. Det., 0dB SNR]{
\label{darap-fig:det-0}
\includegraphics[width=\figwidth]{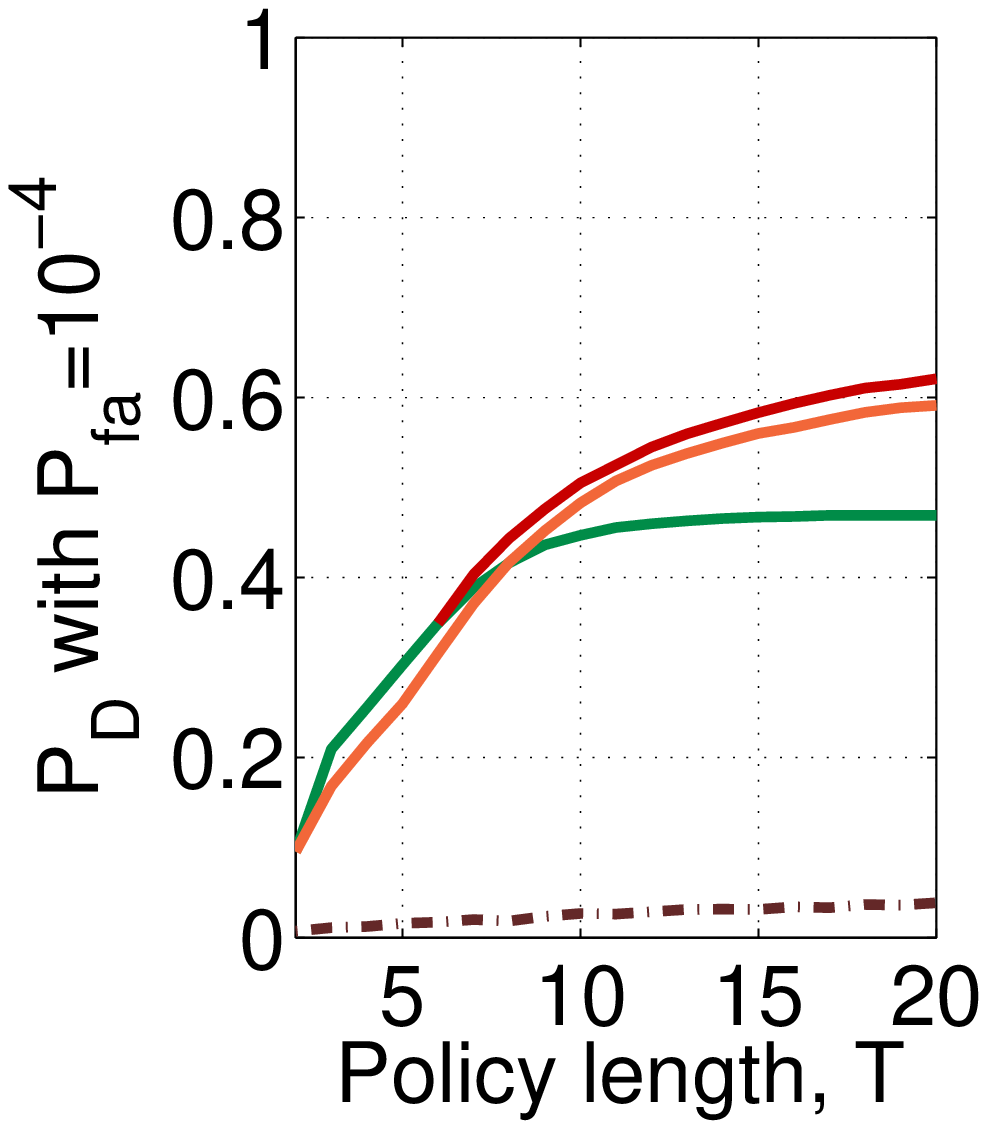}
}
\subfloat[Prob. Det., 10dB SNR]{
\label{darap-fig:det-10}
\includegraphics[width=\figwidth]{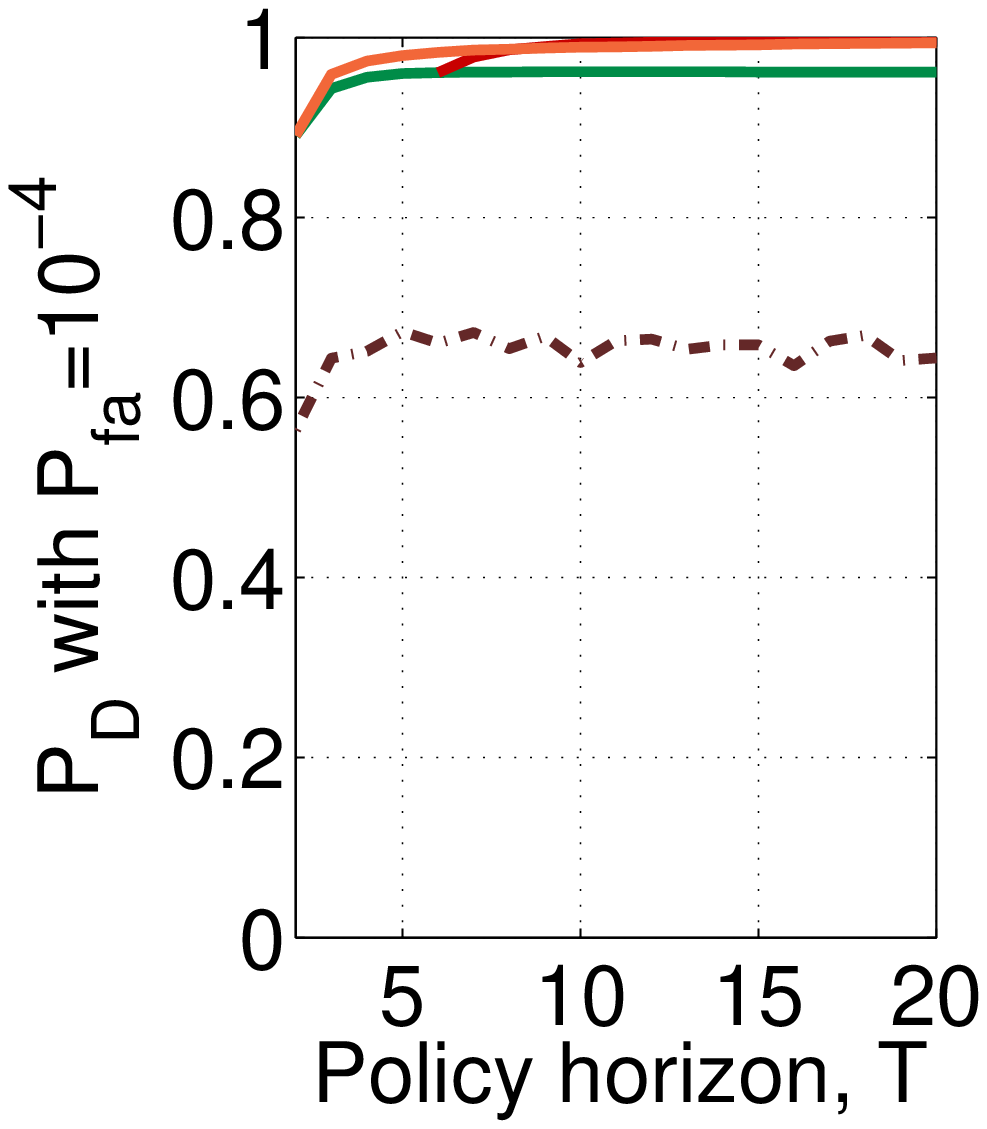}
}
\subfloat{
\begin{minipage}[b]{\figwidth}
\includegraphics[width=\textwidth]{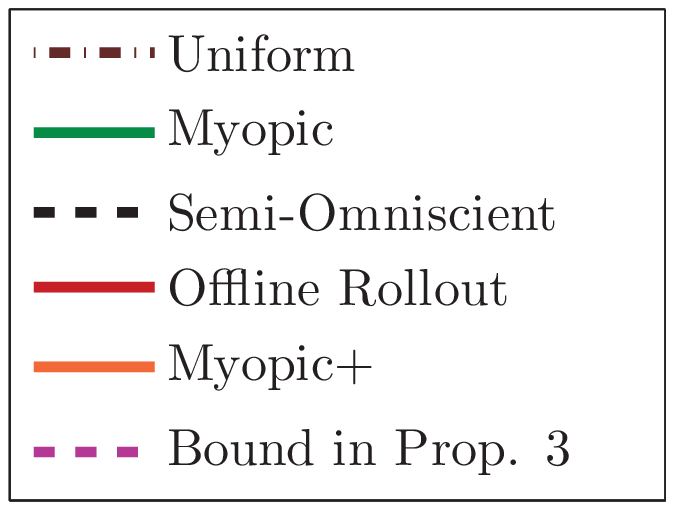}
\vspace{0.1in}
\end{minipage}
}
\caption{These plots compare estimation and detection performance with 0 dB and 10 dB SNR for various values of $T=1,2,\dots,20$.  In (a) and (b), gains in MSE are plotted with respect to a uniform allocation policy (on a dB scale) for 4 alternative policies: the myopic policy (green), the myopic+ policy (orange),\ignore{ the functional policy (purple),} the offline rollout policy with $T_0=5$ (red), the semi-omniscient oracle policy (black dashed), and the upper bound from Proposition \ref{prop:omni+} (purple dashed). In (c), the cost in \eqref{darap-eq:DARAP-cost} is plotted with respect to a uniform allocation policy.  In (d) and (e), the detection probability with fixed false alarm rate of $10^{-4}$ is shown for the same policies as well as the uniform allocation policy (brown dash-dotted). Observe that the proposed myopic+ policy and the offline rollout policy perform best, but the myopic+ policy has significantly lower implementation complexity.}
\label{darap-fig:estimation-detection-performance}
\end{figure}

In this section, we examine the performance of all of the proposed policies (offline rollout, myopic+, and myopic) as well as the semi-omniscient oracle, which provides an upper bound on performance. 

Fig. \ref{darap-fig:estimation-detection-performance} (a) and (b) show the MSE gains (with respect to a uniform policy) for estimating $\set{\theta_i(T)}_{i\in\Psi(T)}$ for different values of $T$ with 0 dB and 10 dB SNR, respectively.  Generally the offline rollout policy has the highest gains in MSE among non-oracle policies, with performance close to the semi-omniscient policy as $T$ gets large.  However, the performance gain of the offline rollout is small with respect to the myopic+ policy.  Fig.  \ref{darap-fig:estimation-detection-performance}(c) provides the gain in the objective function in (\ref{darap-eq:DARAP-cost}) with respect to the uniform policy.  Recall that (\ref{darap-eq:DARAP-cost}) is used as a surrogate optimization objective for amplitude estimation MSE.  Comparing (b) and (c), it is clear that improvements in cost generally lead to improvements in MSE, suggesting that (\ref{darap-eq:DARAP-cost}) is a good surrogate function.  In (c), the bound in Prop. \ref{prop:omni+} is also plotted (note that the condition \eqref{eqn:condSemiLarge} in Prop. \ref{prop:semiLarge} is not satisfied).  In the next section, we empirically analyze the conditions which lead to tight oracle bounds as a function of model parameters and SNR.

Figs. \ref{darap-fig:estimation-detection-performance} (d) and (e) show the probability of detection for a fixed probability of false alarm ($P_{fa}=10^{-4}$) as a function of $T$ for 0 dB and 10 dB SNR, respectively.  The probability of detection for D-ARAP (offline rollout and myopic+ policies) consistently approaches 1 as $T$ gets large and does so significantly faster than for the uniform and myopic policies. Moreover, D-ARAP achieves perfect detection $P_d=1$ within just a few stages.

\begin{figure*}
\centering
\subfloat[Cost vs. $\pi_0$, 20dB]{
\label{darap-fig:cost-vs-pi0-20db}
\includegraphics[height=\theoryfigheight]{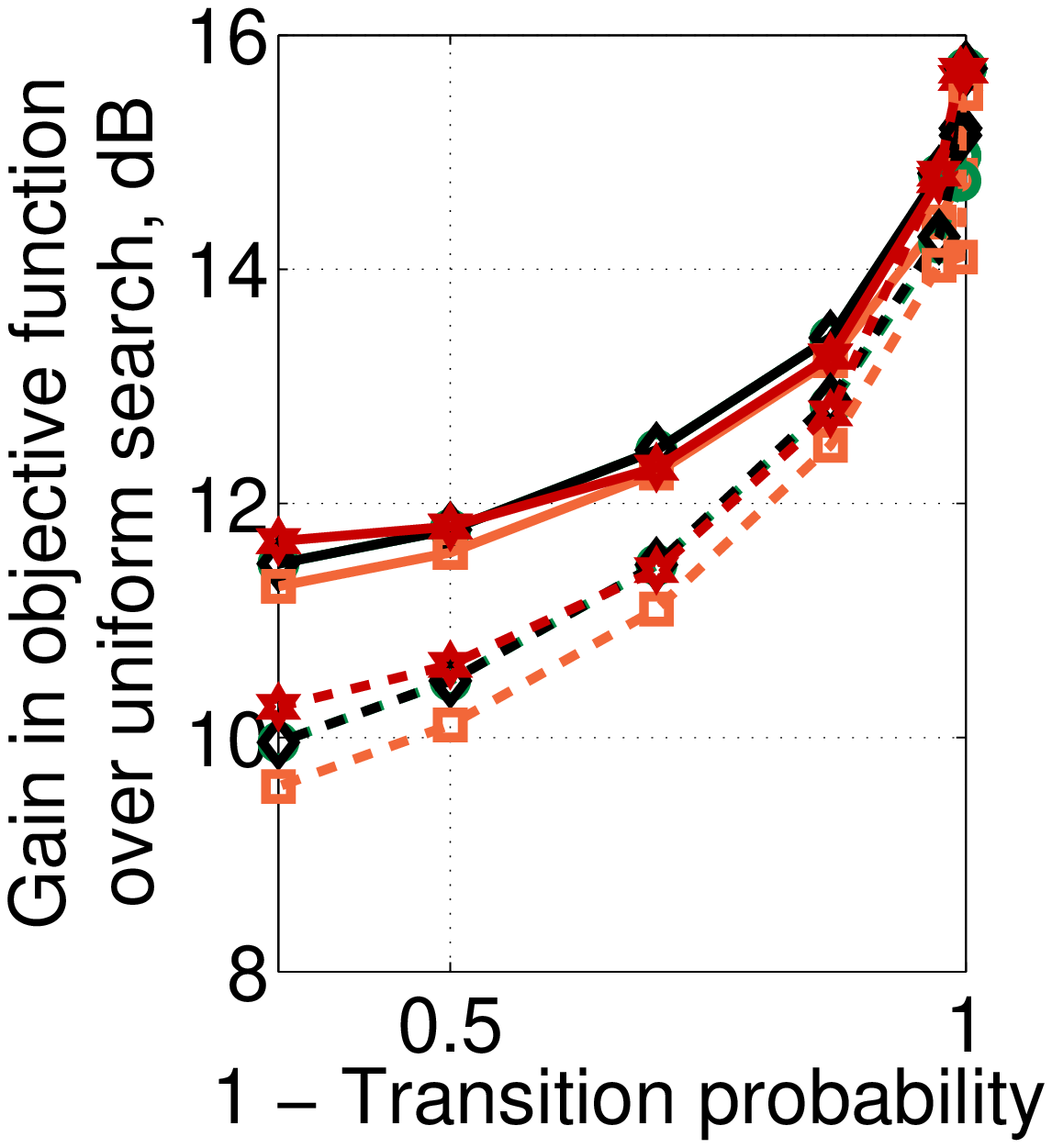}
}
\subfloat[\footnotesize Theory vs. SNR, $\pi_0$=1/3]{
\label{darap-fig:pi0-vs-snr}
\includegraphics[height=\theoryfigheight]{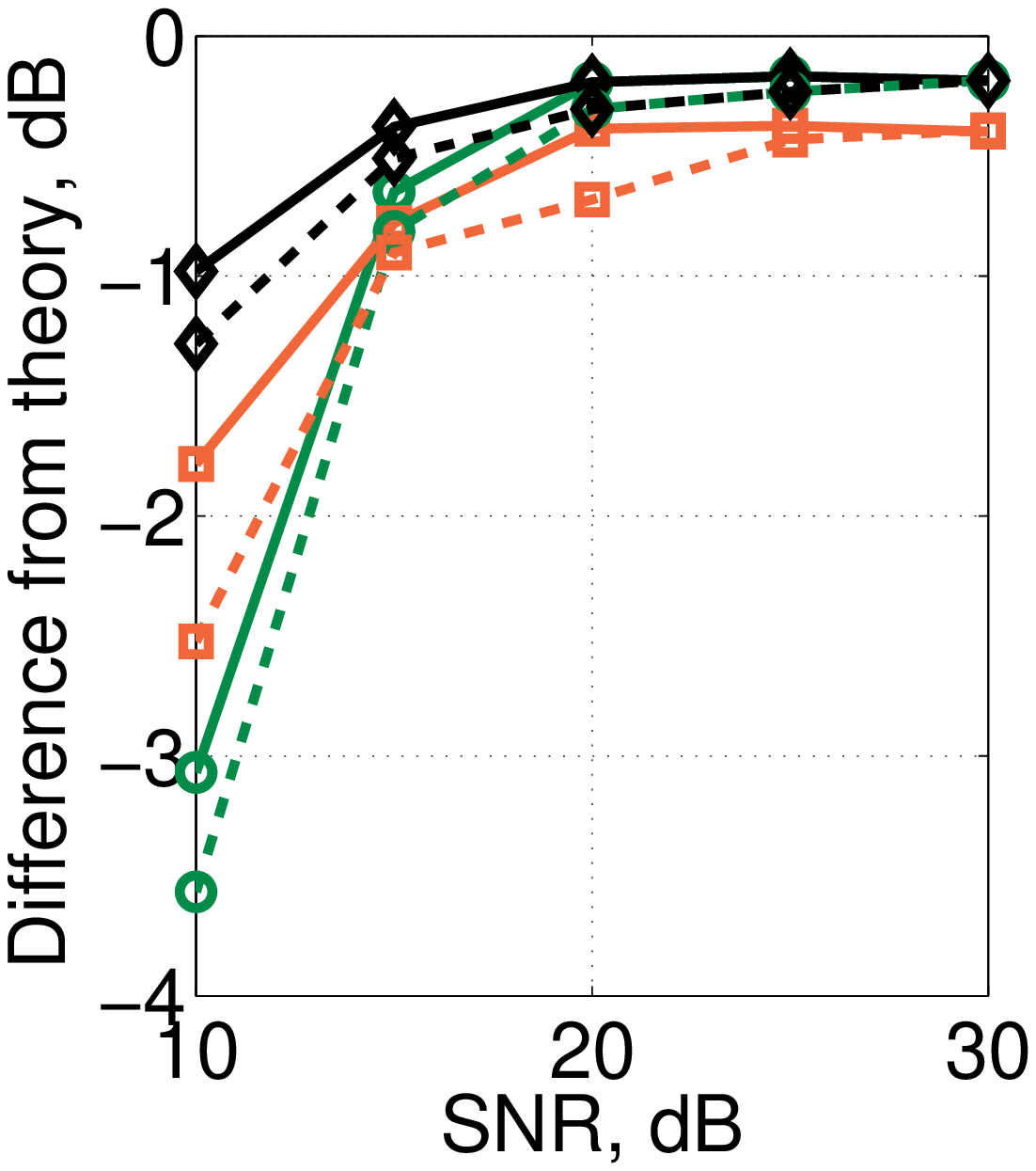}
}
\\
\subfloat[Cost vs. $\beta$, 20dB]{
\label{darap-fig:cost-vs-beta}
\includegraphics[height=\theoryfigheight]{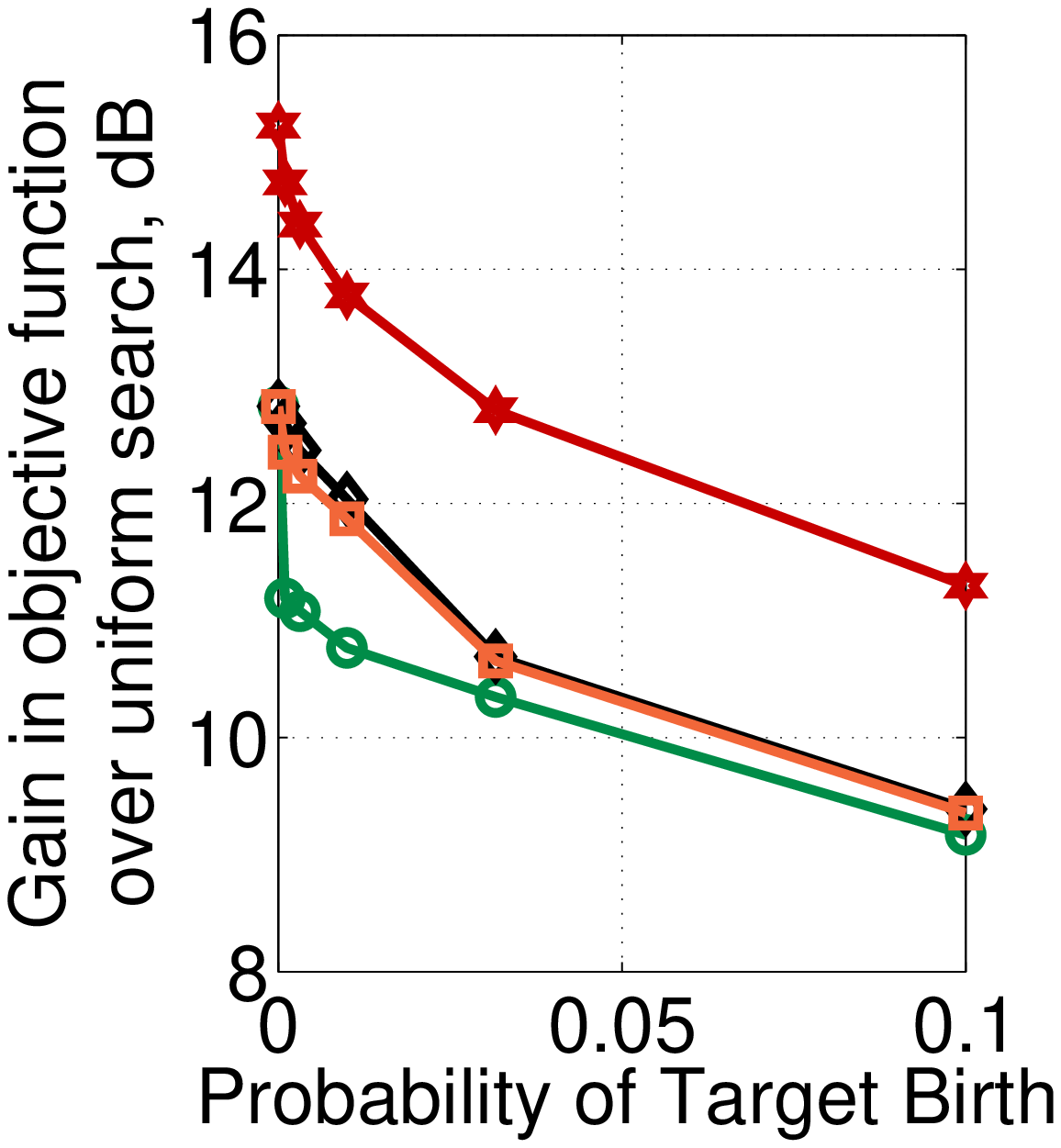}
}
\subfloat[Theory vs. SNR, $\beta$=0.01]{
\label{darap-fig:beta-vs-snr}
\includegraphics[height=\theoryfigheight]{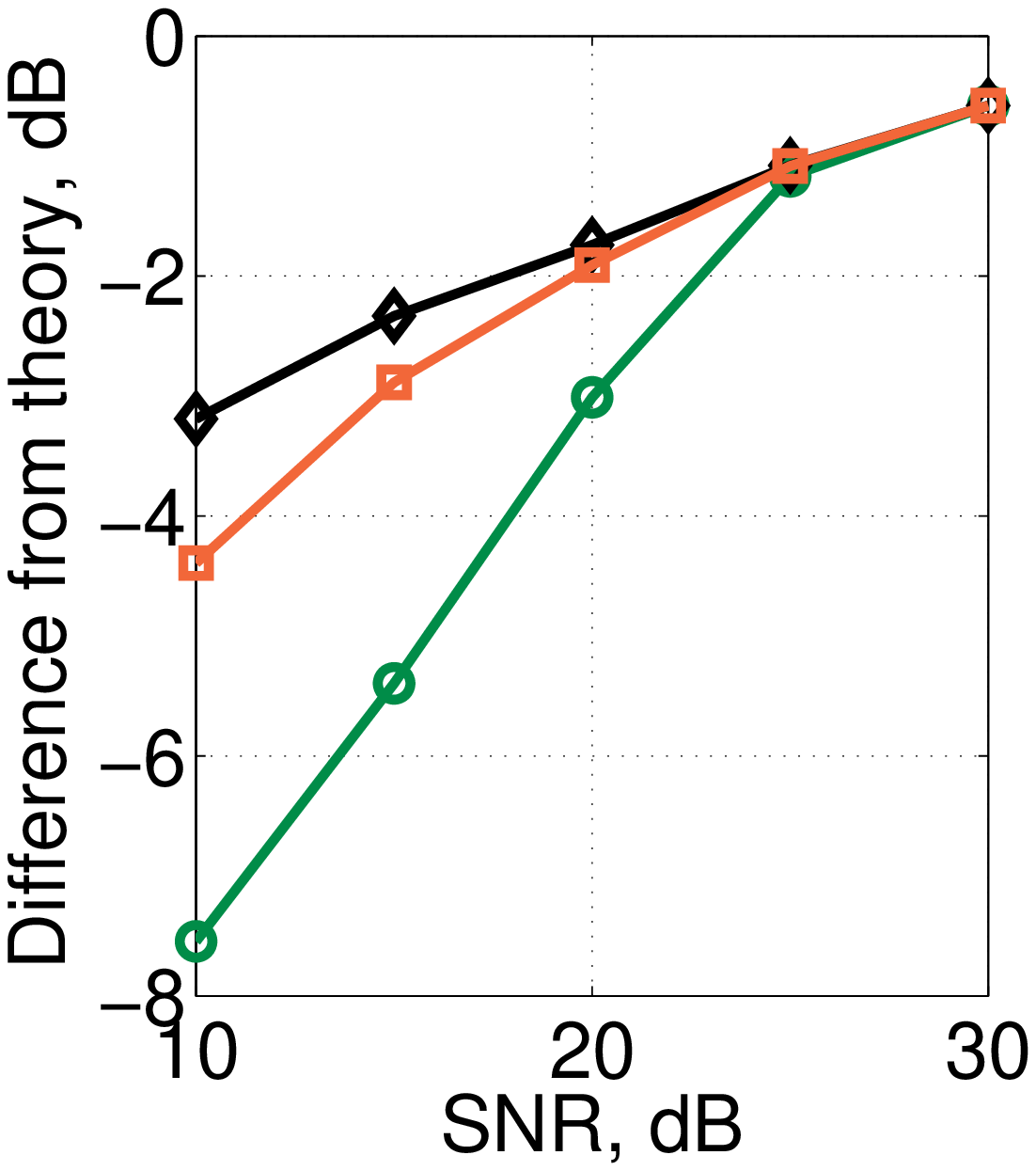}
}\\
\subfloat{
\includegraphics[width=0.7\textwidth]{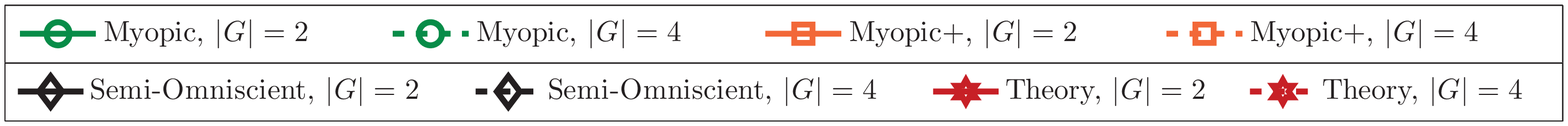}
}
\caption{These figures compare the performance of policies as a function of the dynamic model parameters. (a) and (b) show comparisons with variations in the transition probability ($1-\pi_0$) and the number of neighbors $|G|$.  Solid lines indicate the performance when $|G|=2$, while dashed lines indicate performance when $|G|=4$. (b) compares to the minimum of the theoretical bounds in Propositions 3 and 4 when fixing $\pi_0=1/3$ and varying SNR. (c) and (d) provide similar analysis as a function of the target birth probability, $\beta$. (c) shows results when fixing SNR=20 dB, while (d) compares the policies to the upper bound given in Chapter 3 of \cite{newstadtThesis2013} as a function of SNR while fixing $\beta=0.01$. In (a), the theoretical bounds are tight in comparison to both the numerical semi-omniscient policy and the adaptive policies.  In (b), the theoretical bounds are tight when the sufficient SNR condition \eqref{eqn:condSemiLarge} of Prop. \ref{prop:semiLarge} is nearly satisfied.  Similarly, in (d), the error between theoretical and numerical performance is also non-zero for small SNR, yet decreases near to zero as SNR increases.}
\label{darap-fig:comparison-dynamic-model}
\end{figure*}

\subsection{Comparison across dynamic model parameters}
We continue by comparing the performance of the myopic, myopic+, and semi-omniscient policies as a function of the dynamic model parameters.  Figs. \ref{darap-fig:comparison-dynamic-model}(a) and (b) analyze performance as a function of the transition probability $(1-\pi_0)$ and the number of potential neighbors $|G|$. In (a), we fix SNR=20 dB and vary $\pi_0$.  Solid lines represent the case where $|G|=2$ while dashed lines represent $|G|=4$.  Moreover, we compare to Propositions 3 and 4 in the regimes where they apply, and extrapolate in between by taking the minimum of the two bounds. In (b) we compare across the semi-omniscient, myopic+ and myopic polices when fixing $\pi_0=1/3$ and varying SNR. In (a), the theoretical bounds are tight in comparison to both the numerical semi-omniscient policy and the adaptive policies.  In (b), the theoretical bounds are tight when the sufficient SNR condition \eqref{eqn:condSemiLarge} of Prop. \ref{prop:semiLarge} is nearly satisfied.  
Also, we see that the myopic policy tends to have poorer performance than the myopic+ policy when either SNR is low or $\pi_0$ is low.  

Figs. \ref{darap-fig:comparison-dynamic-model}(c) and (d) provide a comparison as a function of the target birth probability, $\beta$.  In these simulations, we set the target death probability, $\alpha$, as a function of $\beta$ to keep the expected number of targets the same across stages.  If $\beta$ increases, then more resources are allocated uniformly across the scene to search for new targets. This results in a reduction in gains compared to the uniform search.  This pattern is verified in Fig. \ref{darap-fig:comparison-dynamic-model}(c) for fixed SNR=20 dB.  The semi-omniscient policy performs the best, followed by the myopic+ and myopic policies.  We also provide a theoretical upper bound as given in Chapter 3 of \cite{newstadtThesis2013}, which was derived under stronger assumptions than given in Section \ref{darap-sec:theory}. There is an apparent bias between the semi-omniscient policy and this theoretical upper bound.  In (d), we analyze this bias as a function of SNR while fixing $\beta=0.01$.  As SNR gets large, all policies approach the theoretical upper bound on performance.  

\subsection{Model Mismatch}
\def\subfigwidth{0.2\textwidth}
\begin{figure}[!t]
\centering
\includegraphics[width=\bigfigwidth]{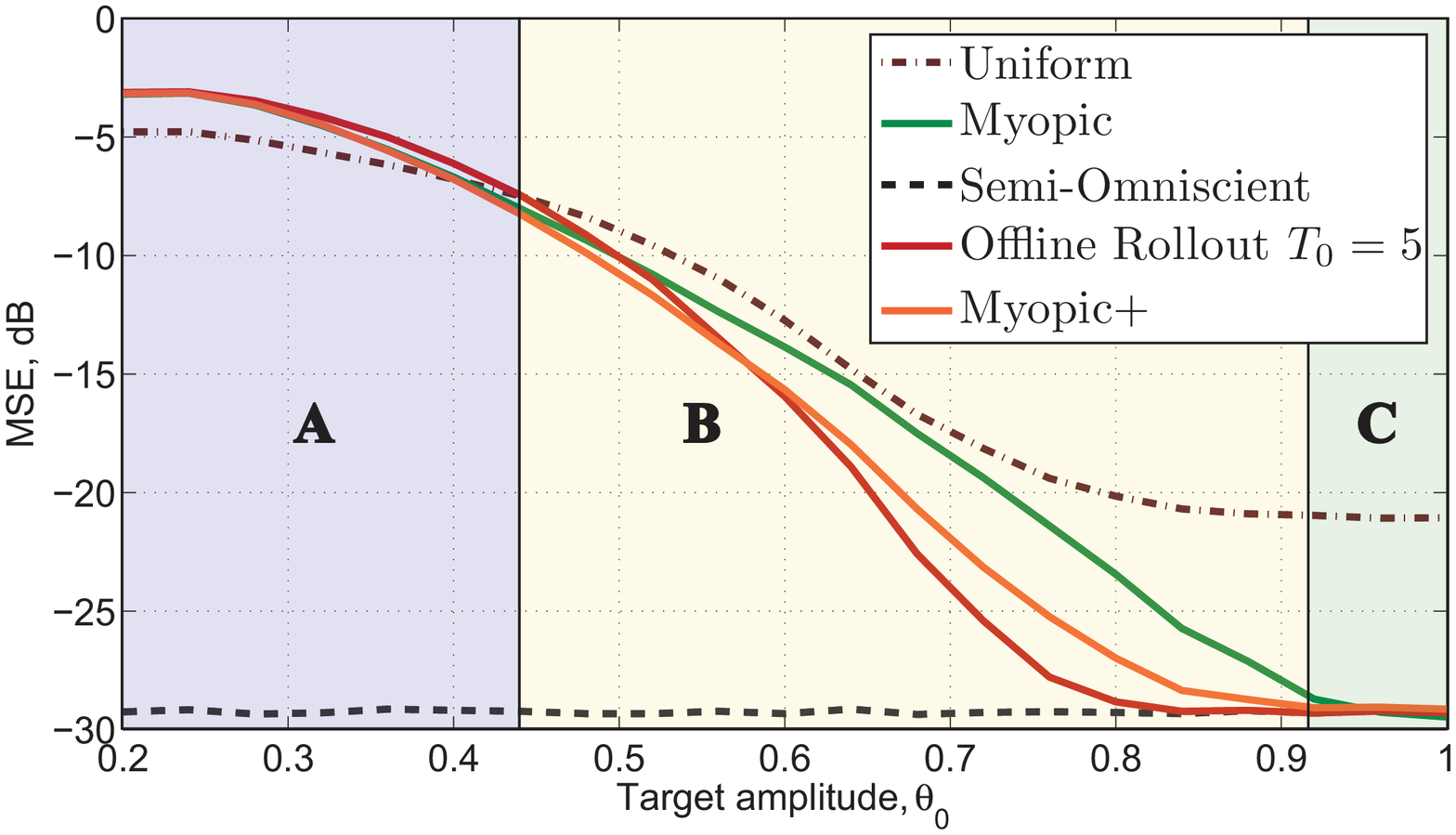}
\caption{This figure compares the proposed non-myopic policies (offline rollout and myopic+) to the myopic and semi-omniscient policies in a model mismatch scenario.  Policies are derived under the assumption that the mean target amplitude $\mu_0=1$.  However, these policies are mismatched to the actual target amplitudes $\theta_i(t)$, which are constant and lower than expected with  $\theta_i(t)=\theta_0<1$.   The figure plots MSE for various $\theta_0$ (x-axis) and policies (curves).  This figure is divided into 3 regions: in Region (A), there is not sufficient signal for adaptive policies to outperform the uniform alternative; in Region (B), adaptive policies perform better, yet there is significant benefit in using non-myopic strategies; in Region (C), all adaptive policies perform similarly.}
\label{darap-fig:minTargetValue}
\end{figure}

In this section, we compare the non-myopic policies to the myopic and semi-omniscient oracle policy in cases where there might be model mismatch. In particular, we consider the case where the policy is derived under the model given in Table \ref{darap-table:simulation-parameters} with prior mean amplitude $\mu_0=1$. This model is mismatched to the actual measurements where the target amplitudes are all identical and lower than expected
\begin{equation}
\theta_i(t) = \theta_0 < \mu_0=1
\end{equation}
for $\theta_0\in[0.2,1]$. For $\theta_0\ll 1$, noisy measurements from cells containing targets can be easily confused with the background noise.  In these situations, the myopic policy will be more adversely affected by small posterior probability $p_i(t)$ in the ROI $i\in \Psi$ as compared to the non-myopic policies.   Fig. \ref{darap-fig:minTargetValue} shows the performance as a function of MSE (lower is better).  When $\theta_0<0.45$ (Region A), all policies perform worse than the uniform search, indicating severe model mismatch.  Conversely all policies show positive performance for $\theta_0>0.45$ (Regions B and C).  Moreover, the performances of the non-myopic policies improve at a faster rate than the myopic policy, and quickly approach the theoretical bound as given by the semi-omniscient oracle policy.  \ignore{The nested policy from related work  \cite{NewstadtWeiHeroCAMSAP13} performs better for lower $\theta_0$, but does not attain the oracle performance when $\theta_0\rightarrow 1$.}

\subsection{Complex dynamic behavior: missing measurements}\label{subsec:missed}
In the next simulation, we test the policies in the scenario where the sensor turns off periodically for several consecutive stages, creating time periods of missing data.  This is representative of a modern radar system that must multi-task between different modes of operation, e.g., tracking, automated target recognition, and synthetic aperture radar \cite{ashInterruptedSAR2013}, which compete for radar resources.  In this simulation, 6 stages of data are collected followed by 3 stages of no measurements.  Fig. \ref{darap-fig:missed-obs} shows the resultant MSE (on a dB scale, lower is better) for the uniform, myopic, offline rollout, myopic+, and semi-omniscient policies and for various SNR (per-stage budgets of 0, 5, 10 and 15 dB).

All adaptive policies perform better than uniform search which has growing errors with $t$. In Fig. \ref{darap-fig:missed-obs}(a) and Fig. \ref{darap-fig:missed-obs}(b) (very low and low SNR), the non-myopic policies outperform the myopic and uniform policies, though their errors still grow with $t$.  With sufficient SNR, in Fig. \ref{darap-fig:missed-obs}(c) (medium SNR) the non-myopic policies approach a stable point over each period of measurements, while the myopic policy progressively increases over those time points. For high SNR (15 dB) in (d), all adaptive policies perform similarly and close to the semi-omniscient oracle.
\begin{figure}[!t]
\centering
\subfloat[SNR = 0 dB]{
\includegraphics[width=\figwidth]{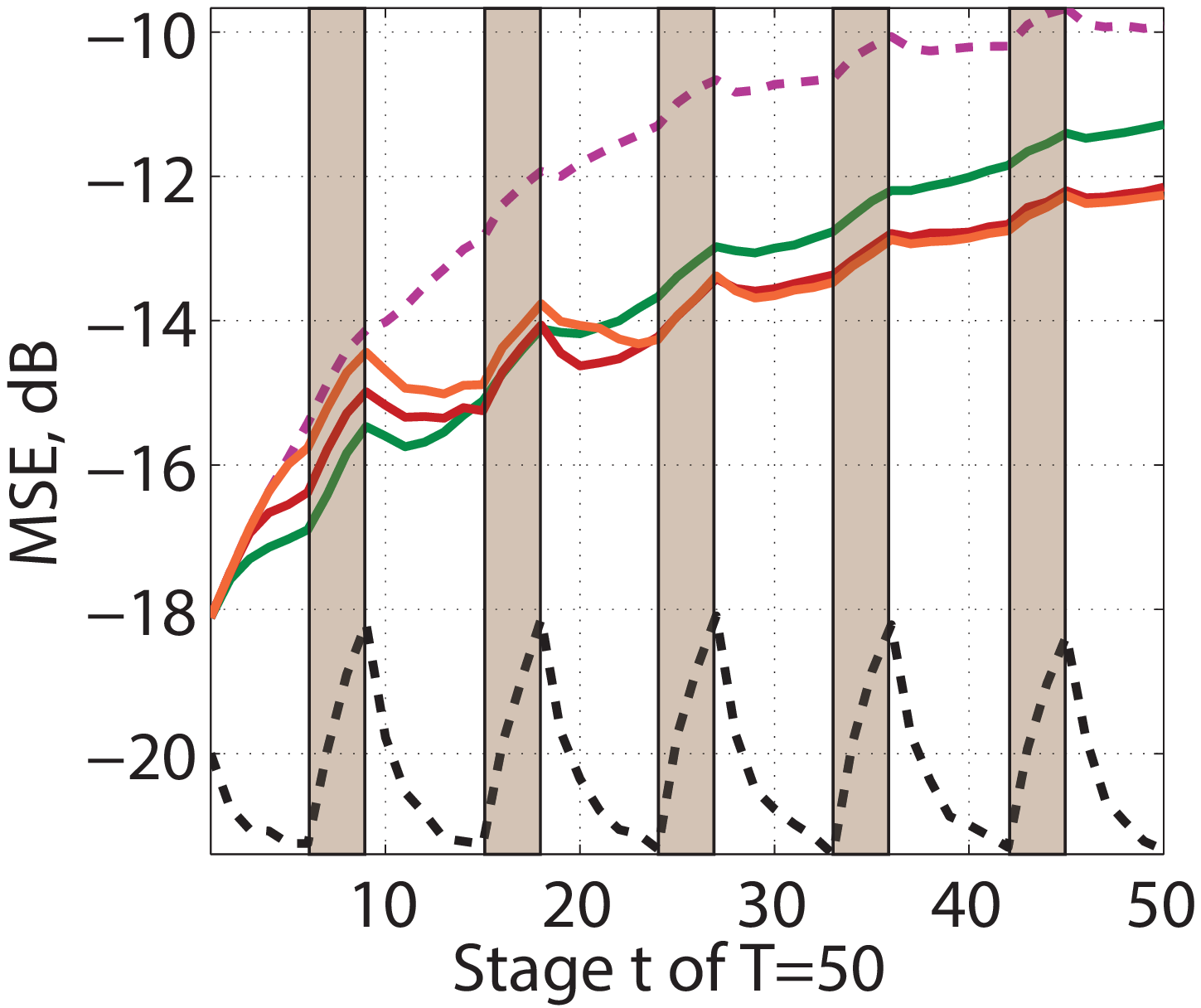}
}
\subfloat[SNR = 5 dB]{
\includegraphics[width=\figwidth]{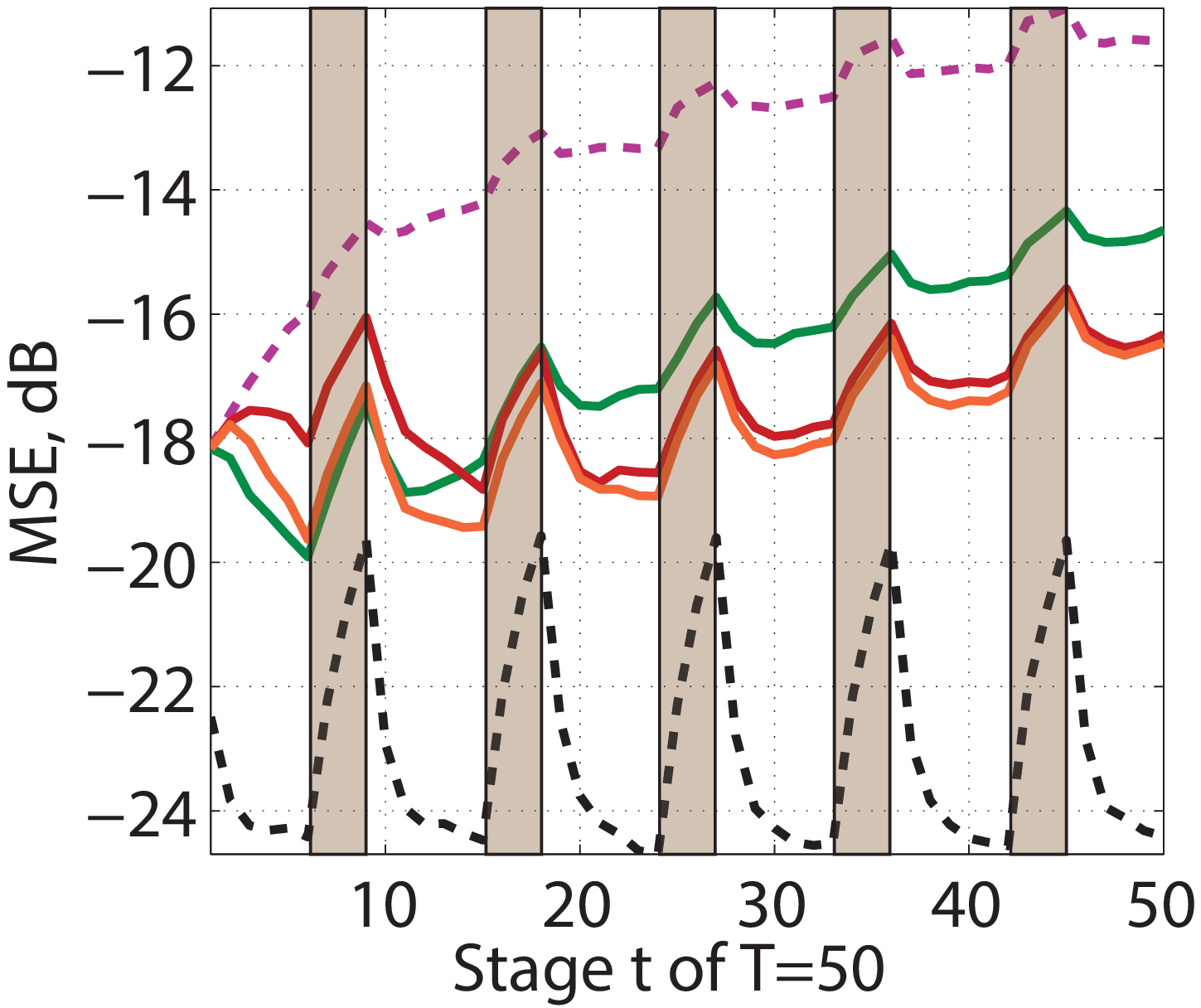}
}
\\
\subfloat[SNR = 10 dB]{
\includegraphics[width=\figwidth]{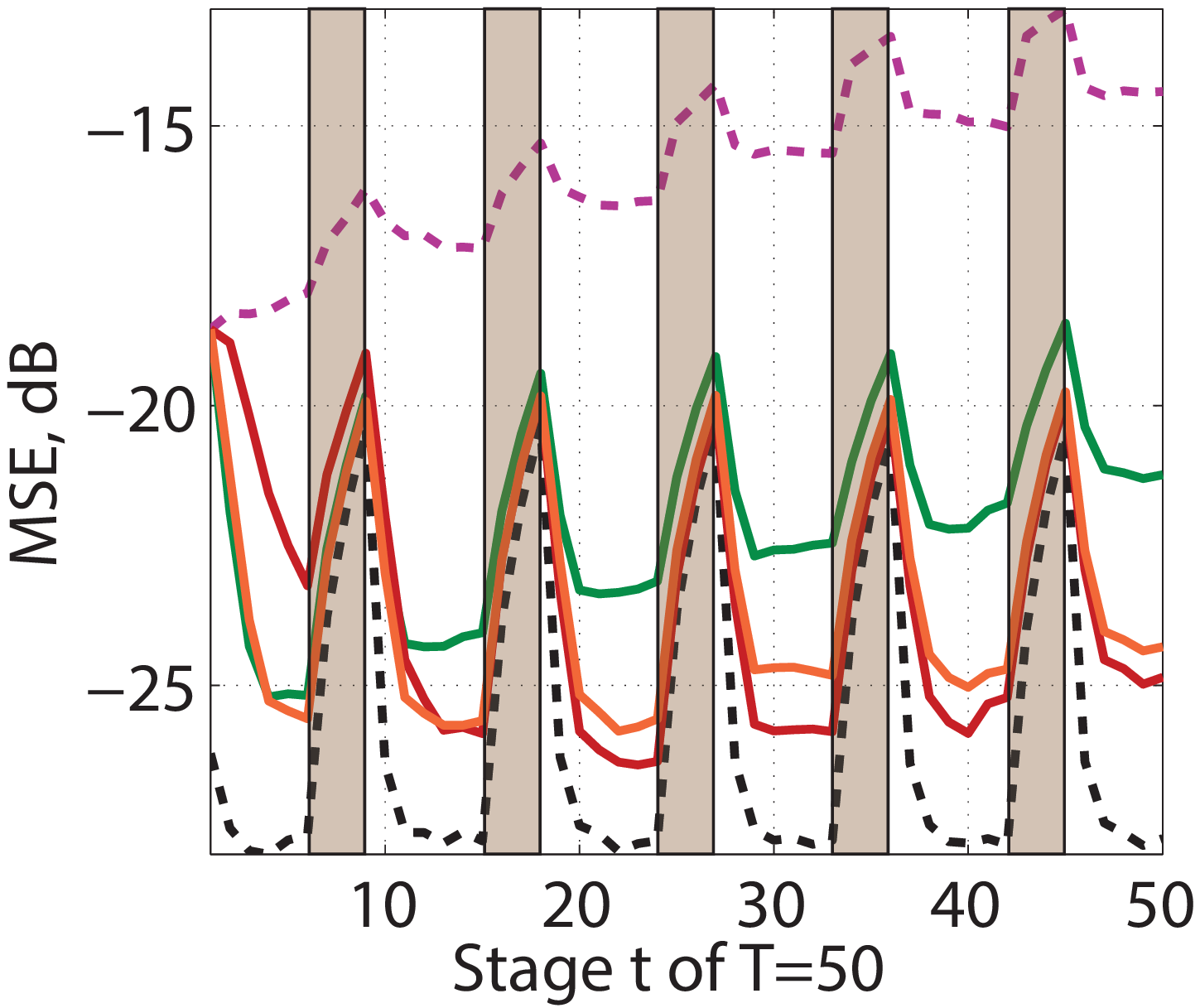}
}
\subfloat[SNR = 15 dB]{
\includegraphics[width=\figwidth]{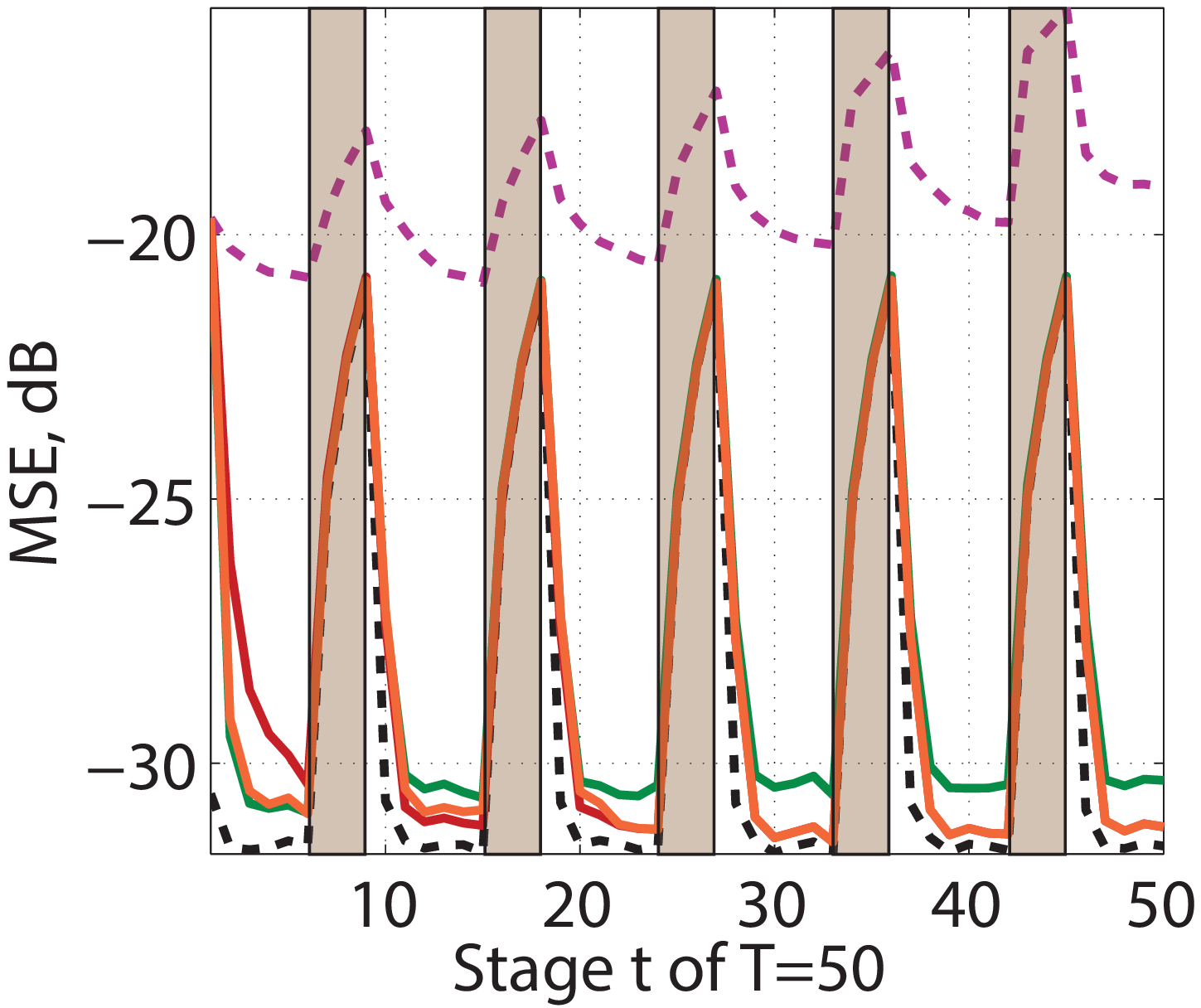}
}
\\
\subfloat{
\includegraphics[width=\bigfigwidth]{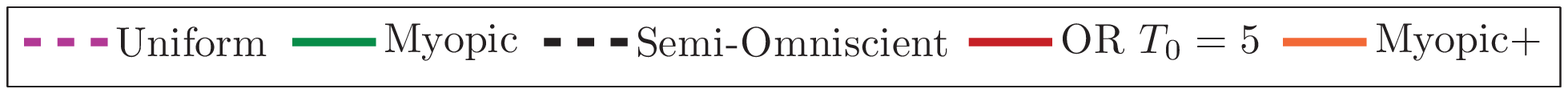}
}
\caption{This figure illustrates the effect of missing data on MSE performance of the allocation policies. In this scenario, 6 stages of observations are followed by no measurements for three consecutive stages (shaded regions). We report MSE on a dB scale (lower is better) for non-myopic, myopic, semi-omniscient, and uniform polices. Performance is compared for four SNR levels in 0, 5, 10, and 15 dB. Note that adaptive policies enable MSE to stabilize in the absence of measurements when there is sufficient SNR as in (c) and (d), whereas the uniform search has growing errors.}
\label{darap-fig:missed-obs}
\end{figure}
\ignore{\renewcommand{\arraystretch}{1.5}
\begin{table*}
\centering
\begin{minipage}{0.85\textwidth}
\caption{Computational cost comparison}
\label{darap-tab:computational-cost}
\begin{center}
\begin{tabular}{|l|c|c|c|c|c|}
\hline
\multirow{2}{*} {Algorithm} & {Offline Simulation} & {Online Simulation} & \multicolumn{3}{c|}{Simulated Parameters$^*$}\\
{} & {Time (Big-$\mathcal{O}$)} &{Time (Big-$\mathcal{O}$)} & {Offline} & {Online} & {Total}\\
\hline
\hline
Myopic, $\kappa(t)=0$ & $\mathcal{O}(1)$ & $\mathcal{O}(N_{sim})$ & $1\times 10^0$ & $1\times 10^2$ & $1.01\times 10^2$\\
\hline
Nested, $\kappa(t)$ & $\mathcal{O}(T^2KN_{mc})$ & $\mathcal{O}(N_{sim})$ & $4\times 10^5$ & $1\times 10^2$ & $4.00\times 10^5$\\
\hline
Heuristic, $\kappa(t)$ & $\mathcal{O}(TKN_{mc})$ & $\mathcal{O}(N_{sim})$ & $2\times 10^4$ & $1\times 10^2$ & $2.01\times 10^4$\\
\hline
Functional, $\kappa(t)$ & $\mathcal{O}(KN_{mc})$ & $\mathcal{O}(N_{sim})$ & $1\times 10^3$ & $1\times 10^2$ & $1.10\times 10^3$\\
\hline
$T_0$-stage Rollout, $\kappa(t)$ & $\mathcal{O}(1)^{**}$ & $\mathcal{O}(TT_0KN_{mc}N_{sim})$ & $1\times 10^0$ & $1\times 10^7$ & $1.00\times 10^7$\\
\ignore{\hline
$T_0$-stage Rollout, $\kappa(t)$ and $z(t)$ & $\mathcal{O}(1)^{**}$ & $\mathcal{O}(TT_0KZN_{mc}N_{sim})$ & $1\times 10^0$ & $1\times 10^8$ & $1.00\times 10^8$\\}
\hline
\end{tabular}
\end{center}
$^*$ For parameters, $N_{sim} = 100$, $T=20$, $K=10$ (number of possibilities for $\kappa(t)$), $N_{mc} = 100$, $T_0=5$.\\
$^{**}$ Using a myopic base policy.  Otherwise, include the offline simulation time for nested/heuristic/functional policies.
\end{minipage}
\end{table*}
\renewcommand{\arraystretch}{1.0}
}

\subsection{Offline and online rollout policies}

Fig. \ref{darap-fig:linear-pols} compares estimation and detection performance for the offline rollout policies as a function of the base policy.  We consider myopic base policies which set the exploration parameter $\kappa(t)=0$ for $T_0=1,2,5$ consecutive stages.  We compare performance for two SNR levels, with SNR=10 dB given by diamonds, and SNR=0 dB given by circles.  All policies perform at least as well as the myopic policy.  Moreover, increasing $T_0$ generally improves performance in both estimation and detection.  It should be noted that for low SNR, using $T_0=1$ did not noticeably improve performance over the myopic policy.

As discussed in the introduction, the full POMDP solution to the adaptive sensing problem is generally intractable due to the size of belief state and action spaces.  As an alternative, we consider an approximate POMDP solution, namely the (online) rollout policy, in order to compare D-ARAP to online solutions.  Note that in online solutions, the optimal action at each stage must be chosen separately for each realization of the model.  Thus, the online rollout policy likely will incur significant computational costs in comparison to the offline policies presented in this paper. 

\ignore{Next we compare the performance of offline to online rollout policies.  Recall that in an online rollout policy, we must search for $\kappa(t)$ for each $t=2,3,\dots,T-1$ for every instantiation of the problem. Moreover, online rollout requires Monte Carlo estimation to evaluate the objective function where each Monte Carlo trial is initialized based on our current belief state (i.e., the current posterior distribution).  This process tends to be significantly more computationally demanding than the offline case. }

We compare offline and online policies using myopic base policies of various stage lengths, $T_0$. In Fig. \ref{darap-fig:online-vs-offline}(a), we compare the performance of the offline and online rollout policies in the SNR=10dB case for parameters given in Table \ref{darap-table:simulation-parameters}. The online and offline rollout policies perform similarly in the standard model (a) without model mismatch or missed observations.  The online policy has significantly noisier results, which is partly caused by computational limits on the number of realizations from which the average performance is computed.  Nevertheless, the online policy clearly performs better than the myopic policy.  In Fig. \ref{darap-fig:online-vs-offline}(b), we compare the performance of the offline and online rollout policies in the SNR=10 dB case where stages of measurements are missing as in Section \ref{subsec:missed}.  It is seen that the online $T_0=2$ rollout policy performs similarly to the offline $T_0=5$ policy.  On the other hand, the online $T_0=5$ policy performs significantly worse (and approximately the same as the myopic policy).  

In our experience, the online rollout policies tend to be significantly noisier than their offline counterparts.  This may be due to (a) necessary tradeoffs in computational (Monte Carlo) effort vs. accuracy or (b) difficulties in sampling from the belief state, particularly in sparse scenarios where the probability of targets existing at given locations is small.  This indicates one advantage of the offline policies, which tend to be more robust to complex environments.\ignore{ such as missed measurements.}

\begin{figure}[t] 
\centering
\subfloat[MSE]{
\label{darap-fig:mse-0-rollout}
\includegraphics[width=\policyFigHeight]{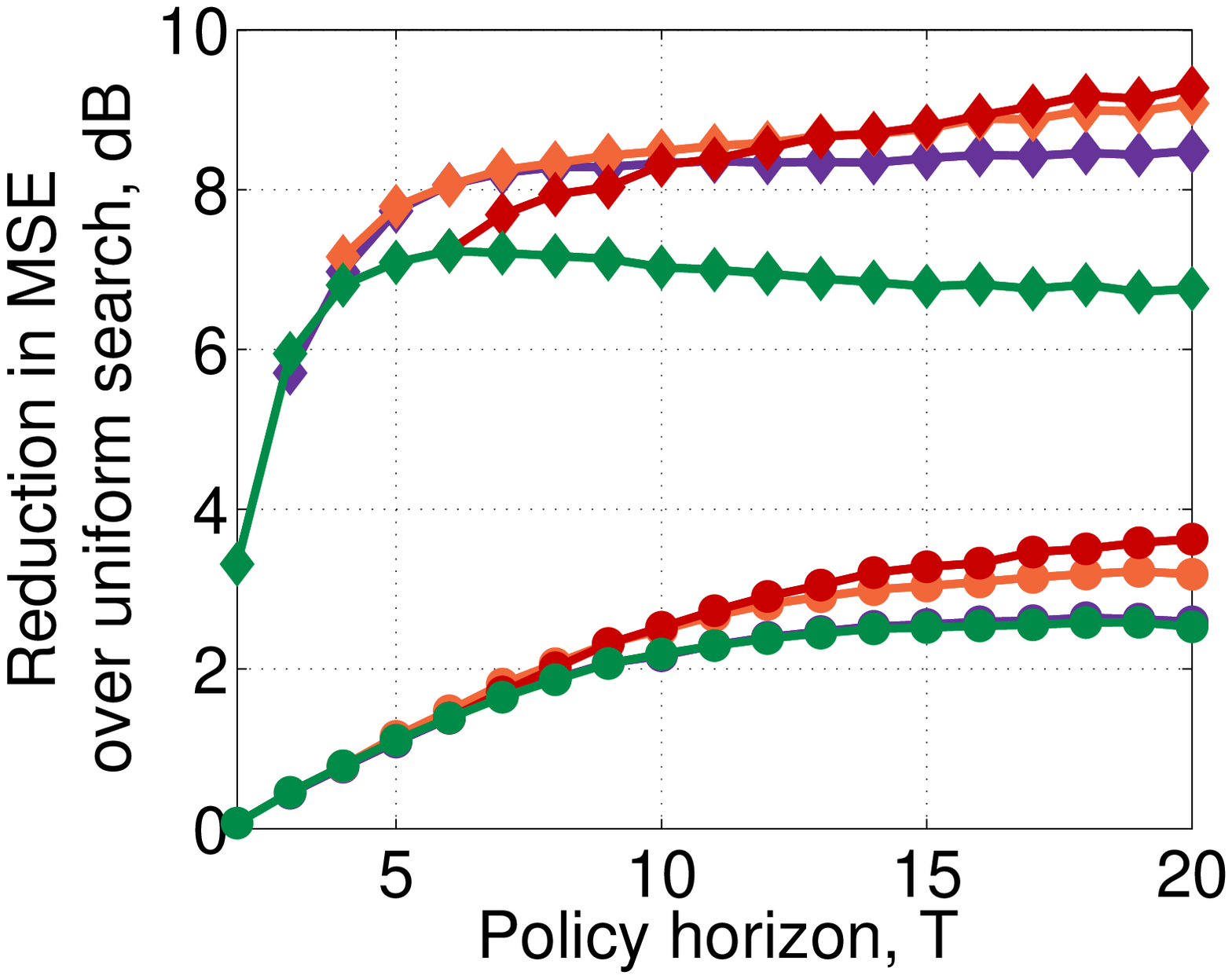}
}
\subfloat[Prob. Detection]{
\label{darap-fig:det-0-rollout}
\includegraphics[width=\policyFigHeight]{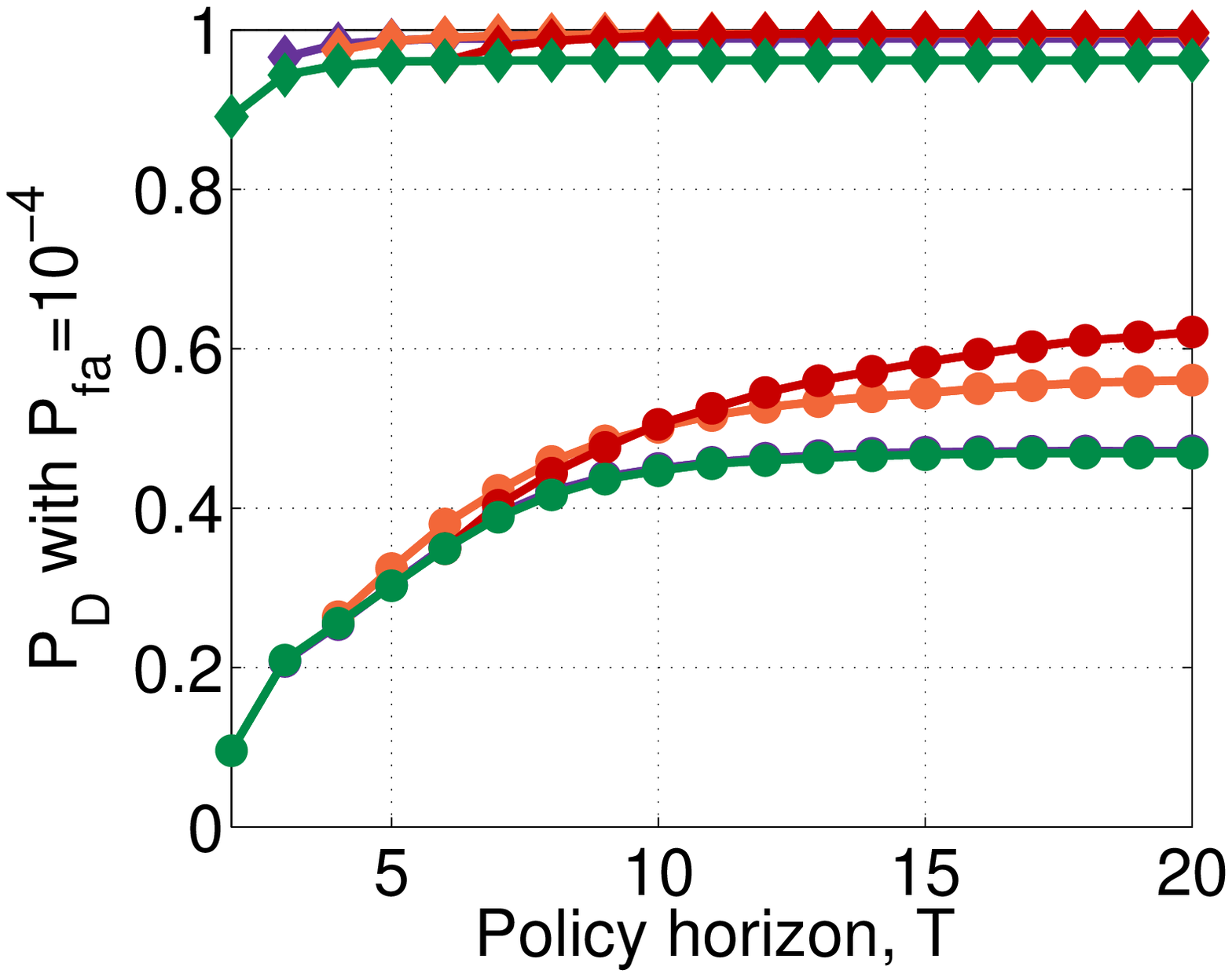}
}
\\
\subfloat{
\includegraphics[width=\bigfigwidth]{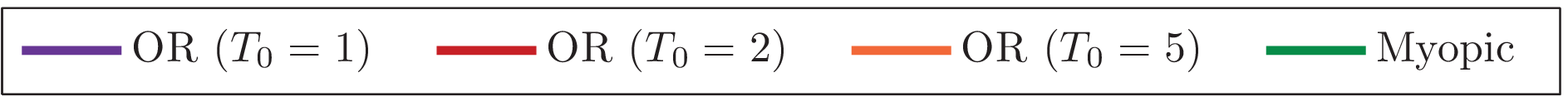}
}
\caption{These plots compare estimation and detection performance for the offline rollout policies as a function of the base policy.  We consider myopic policies which set $\kappa(t)=0$ for $T_0=1,2,5$ consecutive stages.  Performance is shown for two SNR levels, with SNR=10 dB given by diamonds, and SNR=0 dB given by circles. It is seen that higher values of $T_0$ (i.e., less myopic policies) tend to perform better in terms of both estimation and detection error.}
\label{darap-fig:linear-pols}
\end{figure}

\begin{figure}[t]
\centering
\centering
\subfloat[Standard Model]{
\label{darap-fig:online-offline-standard}
\includegraphics[width=\policyFigHeight]{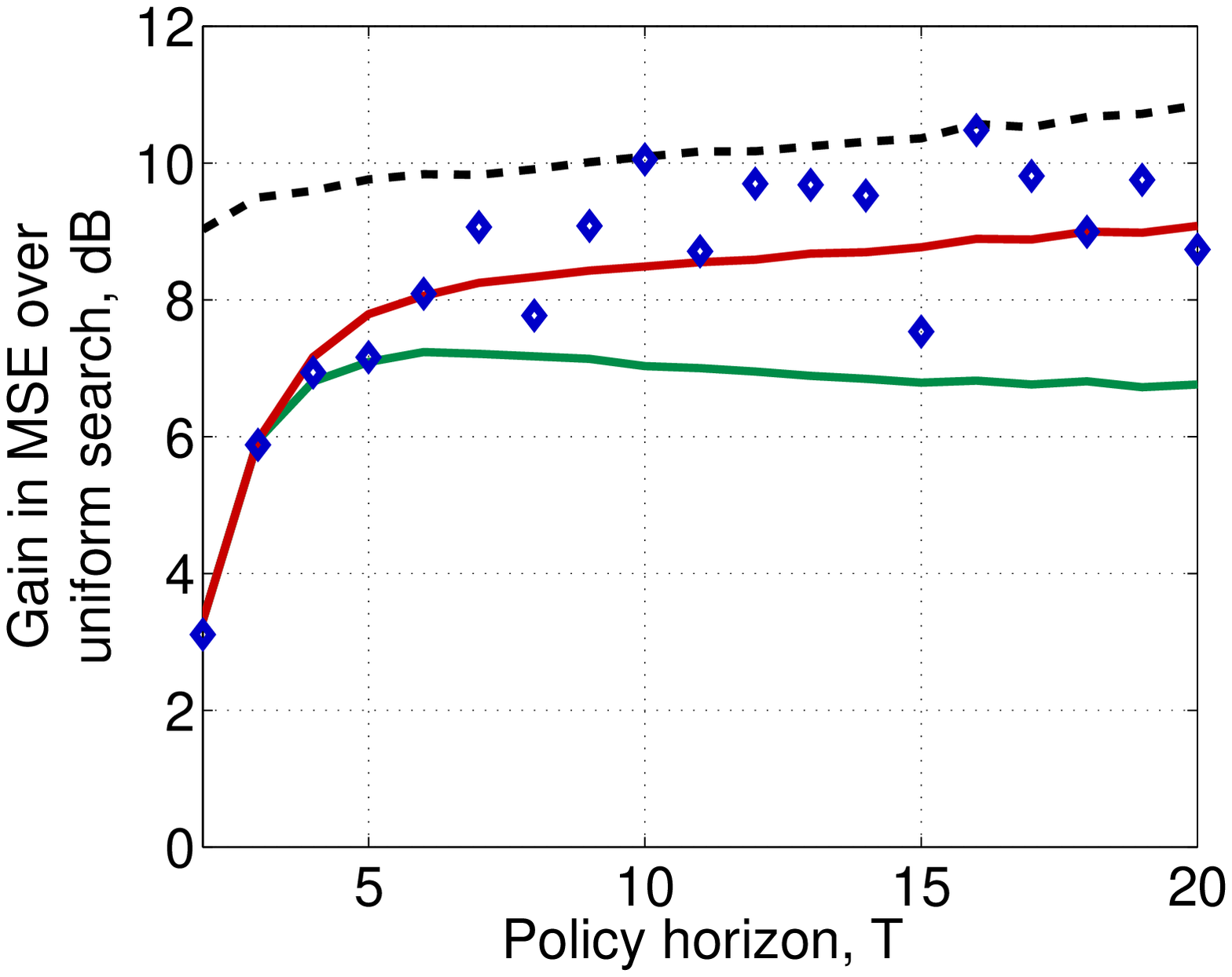}
}
\subfloat[Missed Observations]{
\label{darap-fig:online-offline-missed}
\includegraphics[width=\policyFigHeight]{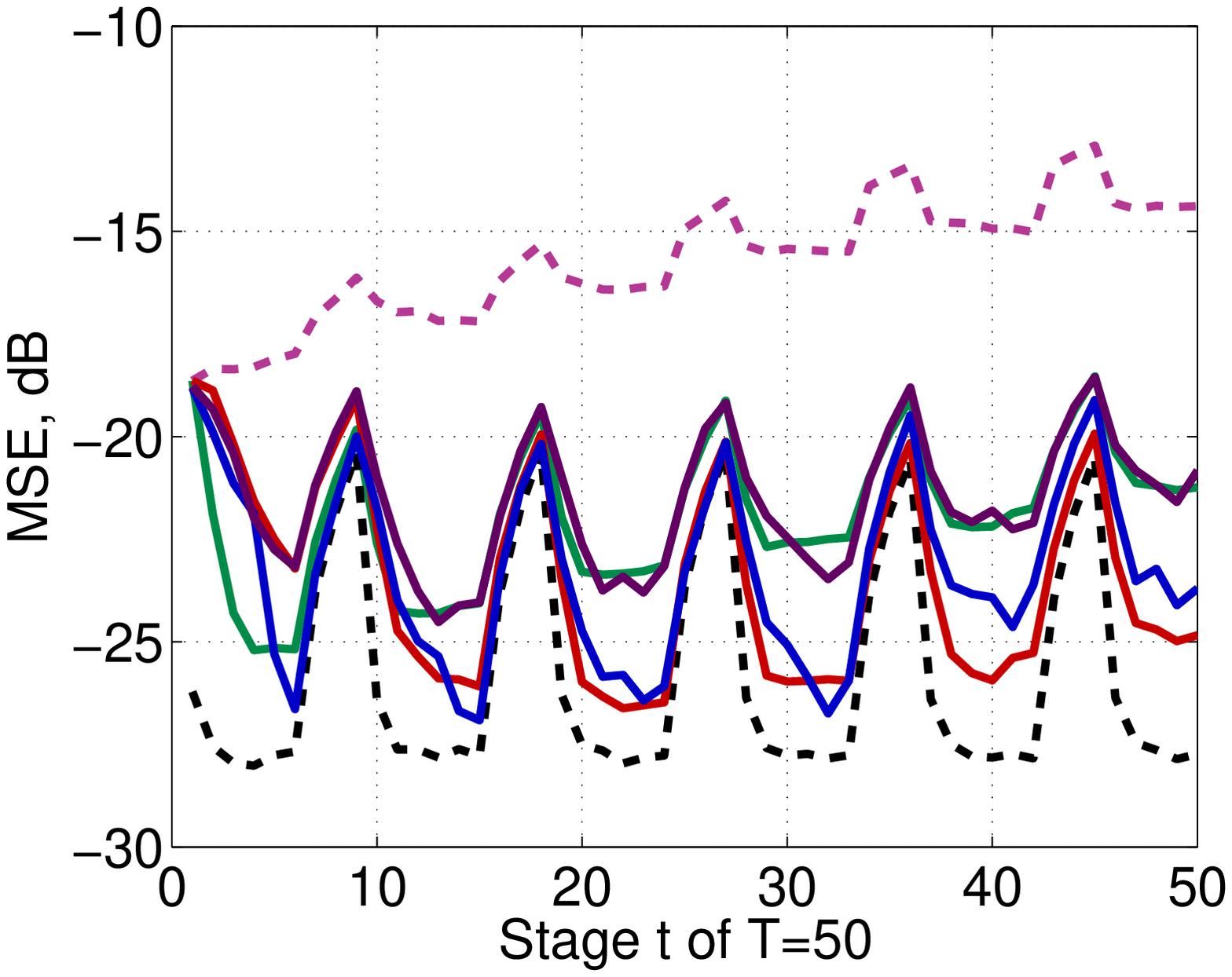}
}\\
\subfloat{
\includegraphics[width=\bigfigwidth]{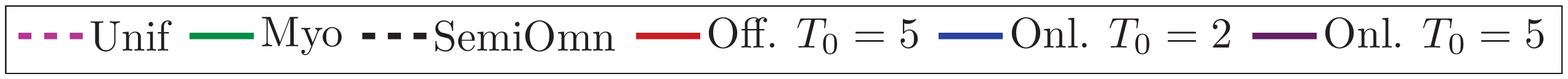}
}
\caption{These plots compare offline and online rollout policies in two scenarios.  In (a), we compare the performance of the offline and online rollout policies in the SNR=10dB case for parameters given in Table \ref{darap-table:simulation-parameters}. In (b), we compare the performance of the offline and online rollout policies in the SNR=10 dB case where stages of measurements are missing as in Section \ref{subsec:missed}.  In the standard model (a), the online policy performs similarly (albeit with more noise) to the offline version.  In (b), however, the online policies performs worse than the offline version, most likely due to the additional computational complexity of the online policy.  Nevertheless, the online policy with $T_0=2$ still performs better than the myopic policy.}
\label{darap-fig:online-vs-offline}
\end{figure}

\fi

\section{Discussion and future work}
\label{darap-sec:conclusion}
This paper presented a framework for adaptive sampling that significantly extends previous work \cite{Bashan-08-opt-2-stage-search} to incorporate dynamic 
targets while providing a computationally tractable solution, namely D-ARAP. 
A cost function related to mean squared error was proposed and upper bounds on the performance of adaptive sensing were derived through analysis of oracle policies.  These bounds shed light on the impact of target motion and amplitude variation in addition to target sparsity. 
In terms of implementable policies, a myopic solution is given that has an analytical form, but suffers from being overly aggressive in the allocation of resources in cases of 
model mismatch or faulty measurements.  We offer a non-myopic extension of the myopic policy that balances 
exploration of the scene and exploitation of prior observations in a tractable manner. 
Numerical evidence suggests that the proposed D-ARAP policies (a) have significant performance gains over the baseline policy that uniformly allocates resources across the scene, (b) perform similarly to the gold-standard POMDP approximate solutions, albeit at a fraction of the computational cost, and (c) improve upon the myopic policy, especially in terms of robustness to model mismatch and faulty measurements.

Future research directions include consideration of constraints on the number of measurements, which may include coarse-scale or compressed sensing measurements.  Moreover, 
further analytical results are of interest, for example convergence rates (in comparison to exhaustive search) and/or minimum detectable amplitudes, and performance bounds that are more refined than the oracle bounds presented herein.  Finally, 
online policies that are computed as measurements are taken may be worthy of continued study 
since they could improve performance dramatically in some cases, including cases where targets will be obscured in the near future.

\section*{Acknowledgments}
The authors would like to thank Dr. Eran Bashan for his valuable insights and assistance in the development of this research.

\appendices
\ifjournal
\else
	\section{Efficient posterior estimation for given dynamic state model}
	\label{darap-appendix:efficient-posterior-estimation}
	In order to use the algorithms provided in this work to adaptively estimate the state $\vxi(t)$ given the measurements, we need to be able to calculate the posterior probabilities for the indicator variables, $\{I_i(t)\}_{i=1}^{Q}$ given the measurements up until time $t$.  To do this efficiently (which is required for planning purposes), we make the following assumptions:

\begin{assumption}\label{ass:interactingTargets}
There is at most 1 target in the vicinity of any target: 
\begin{equation*}
(s^{(n)}=i)\rightarrow |G(i)|=0,\quad \forall n 
\end{equation*}
where $G(i)$ is the set of neighbors of location $i$.
\end{assumption}

Define the measurement vectors
\begin{equation}
\vy(t) = \{y_1(t),y_2(t),\dots,y_Q(t)\}
\end{equation}
and
\begin{equation}
\vY(t) = \{\vy(1),\vy(2),\dots,\vy(t)\}
\end{equation}
Let
\begin{equation}
\Pr(I_i(t)=1|\vY(t-1)) \qquad i=1,2,\dots,Q, \quad t=1,2,\dots,T
\end{equation}
be the posterior probabilities that need to be calculated.  For $t=1$, we have 
\begin{equation}
\Pr(I_i(1)=1) = p,
\end{equation}
under assumption \ref{ass:interactingTargets}.   For $t>1$, we have
\begin{equation}
\begin{split}
\Pr(I_i(t)=1|\vY(t-1))&=\int \Pr(I_i(t)=1,\vS(t)|\vY(t-1))d\vS(t)\\
&= \int \Pr(I_i(t)=1|\vS(t))f(\vS(t)|\vY(t-1))d\vS(t)\\
&= \sum_{n=1}^{N(t)} \Pr(\sn(t)=i|\vY(t-1)),
\end{split}
\label{darap-eq:indic-prob-given-state-measurements}
\end{equation}
where the last equation can be derived noting that 
\begin{equation}
\Pr(I_i(t)=1|\vS(t)) = 
\begin{cases}
1,&\exists n: \sn(t)=i\\
0,&\mathrm{else}
\end{cases}
\end{equation}
Thus, in order to compute equation (\ref{darap-eq:indic-prob-given-state-measurements}), we need to be able to estimate the state $\vxi(t)$ given $\vY(t-1)$.
\subsection{Recursive equations for updating $\vxi(t)$}
In general, we can compute the posteriors using the equations:
\begin{align}
f(\vxi(t)|\vY(t-1)) &= \int f(\vxi(t)|\vxi(t-1))f(\vxi(t-1)|\vY(t-1))d\vxi(t-1)\\
f(\vxi(t)|\vY(t)) &= \frac{f(\vy(t)|\vxi(t)) f(\vxi(t)|\vY(t-1))} {\int f(\vy(t)|\tilde{\vxi}(t)) f(\tilde{\vxi}(t)|\vY(t-1))d\tilde{\vxi}(t)}
\end{align}
Note that each target has an associated real-valued amplitude $\xn(t)$ and a location on a large discrete grid $\sn(t)\in\{1,2,\dots,Q\}$ for large $Q$.  Thus, the joint densities $f(\vxi(t)|\vY(t-1))$ and $f(\vxi(t)|\vY(t-1))$ are in general very high-dimensional functions that may be intractable to estimate exactly.  Under certain assumptions, however, it may be possible to derive exact equations for these updates.
\subsection{Static case}
In the static case when $\alpha=\beta=0$ and $\pi_0=1$, we have the simple situation where
\begin{equation}
\vS(t)=\vS(t-1).
\end{equation}
Since targets are fixed in position and cannot occupy the same cell by Assumption \ref{ass:interactingTargets}, we can easily show that the joint density factors into:
\begin{equation}
f(\vxi(t)|\vY(t')) = f(\Psi(t),\Theta(t)|\vY(t'))=f(\Theta(t)|\vY(t'),\Psi(t)) f(\Psi(t)|\vY(t'))
\end{equation}
for $t' = t,t-1$, $\Psi(t) = \{I_i(t)\}_{i=1}^Q$, and $\Theta(t)=\{\theta_i(t)\}_{i=1}^Q$.  Moreover, $\Psi(t) = \Psi$ and since the targets are independent across cells, we have:
\begin{align}
f(\Theta(t)|\vY(t'),\Psi(t)) &= \prod_{i=1}^Q f\left(\theta_i(t)|\vyi(t'), I_i\right)\\
f(\Psi(t)|\vY(t')) &= \prod_{i=1}^Q f\left(I_i|\vyi(t')\right)
\end{align}
where $\vyi(t) = \{y_i(t_0)\}_{t_0=1}^{t}$.  Note that $\theta_i(t)$ is only defined if $I_i=1$.  Conditioned on this event, we furthermore note that $\theta_i(t)$ and $y_i(t) | \theta_i(t)$ are normally distributed given the allocations $\lambda_i(t)$.  Thus, the posteriors $ f\left(\theta_i(t)|\vyi(t'), I_i\right)$ for $t'=t,t-1$ can be updated exactly through the Kalman filter equations:
\begin{align}
\label{darap-eq:KF-first}
\delta_i(t) &= y_i(t) - \sqrt{\lambda_i(t)}\hat{\theta}_i (t|t-1)\\
s_i(t) &= \lambda_i(t)\hat{\sigma}_i^2(t|t-1) + \sigma^2\\
\Gamma_i(t) &= \frac{\hat{\sigma}_i^2(t|t-1)\sqrt{\lambda_i(t)}}{s_i(t)}\\
\hat{\theta}_i (t|t) &= \hat{\theta}_i (t|t-1) + \Gamma_i(t)\delta_l(t),\\
\label{darap-eq:KF-static-variance-t-t}
\hat{\sigma}_i^2(t|t) &= [1-\Gamma_i(t)\sqrt{\lambda_i(t)}]\hat{\sigma}_i^2(t|t-1),
\end{align}
where $\delta_i(t)$ is the residual measurement error, $s_i(t)$ is the update measurement error, $\Gamma_i(t)$ is the Kalman gain, and $(\hat{\theta}_i (t|t),\hat{\sigma}_i^2(t|t))$ are the updated state estimates.  The predict equations are given by:
\begin{align}
\hat{\theta}_i(t|t-1) &= \hat{\theta}_i(t-1|t-1),\\
\label{darap-eq:KF-static-variance-t-t1}
\hat{\sigma}_i^2(t|t-1) &= \hat{\sigma}_i^2(t-1|t-1) + \Delta_\theta^2.
\end{align}
Moreover, the posteriors on the indicator functions can be easily computed recursively as
\begin{equation}
f(I_i=1|\vyi(t)) = \frac{f(y_i(t)|I_i=1,\vyi(t-1)) f(I_i=1|\vyi(t-1))}{\sum\limits_{j=0,1}f(y_i(t)|I_i=j,\vyi(t-1)) f(I_i=j|\vyi(t-1))},
\end{equation}
where we note that when $I_i=0$
\begin{equation}
y_i(t)|I_i=0,\vyi(t-1) \sim \mathcal{N}(0,\sigma^2)
\end{equation}
and when $I_i=1$
\begin{equation}
\begin{split}
f(y_i(t)&|I_i=1,\vyi(t-1))\\
&= \int f(y_i(t)|\theta_i(t),I_i=1)f(\theta_i(t)|I_i=1,\vyi(t-1))d\theta_i(t)\\
&= \int \phi(y_i(t); \sqrt{\lambda_i(t)}\theta_i(t),\sigma^2) \phi(\theta_i(t); \hat{\theta}_i(t|t-1), \hat{\sigma}_i^2(t|t-1)) d\theta_i(t)\\
&= \phi(y_i(t); \sqrt{\lambda_i(t)}\hat{\theta}_i(t|t-1), \lambda_i(t)\hat{\sigma}_i^2(t|t-1)+\sigma^2)
\end{split}
\end{equation}
where $\phi(x;\mu,\sigma^2)$ is the Gaussian pdf with mean $\mu$ and variance $\sigma^2$ evaluated at $x$.  From this equation we see that 
\begin{eqnarray}
y_i(t)|I_i=1,\vyi(t-1) \sim \mathcal{N}(\sqrt{\lambda_i(t)}\hat{\theta}_i(t|t-1), \lambda_i(t)\hat{\sigma}_i^2(t|t-1)+\sigma^2)
\end{eqnarray}
In the static case, we see that updating the posteriors for $I_i$ for all $t=1,2,\dots,T$ involves (a) updating the conditional mean and variances for $\theta_i(t)$ given the measurements, and (b) updating the posterior probability for $I_i=1$.  This gives insight into an approximate method that we will use in the general case when the targets are allowed to move, enter, or leave the scene.

\subsection{Approximations in the general case}
Similar to the static case, we assume that there are no interacting targets so that we can factor our posterior density into a form that makes it tractable to estimate directly.  In order to do this, we use Assumption \ref{ass:neighborTargets}:
\begin{assumption}\label{ass:neighborTargets} There is at most one target in the vicinity of a location
\begin{equation}
\label{darap-eq:no-interacting-targets-by-neighborhood}
|\{n:\sn(t)\in H(i)\}|\leq 1
\end{equation}
for all $i=1,2,\dots,Q$. 
\end{assumption} 
This is clearly more restrictive than Assumption \ref{ass:interactingTargets}.  Under this assumption, we have for $t'=t,t-1$
\begin{equation}
\begin{split}
f(\vxi(t)|\vY(t')) &= f(\Psi(t),\Theta(t)|\vY(t'))\\
&= f(\Theta(t)|\Psi(t),\vY(t'))f(\Psi(t)|\vY(t'))\\
&= \prod\limits_{i=1}^Q f(\theta_i(t)|\Psi(t),\vY(t'))f(I_i(t)|\vY(t'))
\end{split}
\end{equation}
Beginning with the target amplitudes, we note that
\begin{equation}
\begin{split}
f(\theta_i(t)|\Psi(t),\vY(t-1)) &= \int f(\theta_i(t)|I_i(t)=1,\Psi_{H(i)}(t-1),\vY(t-1)) \\
&\cdot f(\Psi_{H(i)}(t-1)|I_i(t)=1,\vY(t-1)) d\Psi_{H(i)(t-1)},
\end{split}
\end{equation}
where $\Psi_{H(i)}(t) = \{I_j(t)\}_{j\in H(i)}$.  Define $E_{i,j}(t)$ to be the event that assigns $\Psi_{H(i)}(t)$ as
\begin{equation}
E_{i,j}(t) \triangleq
\begin{cases}
I_k(t) = 1,& j=k,\\
I_k(t) = 0,& j\neq k,
\end{cases}
\hspace{.5 in} \forall k\in H(i)
\end{equation}
The event $E_{i,0}(t)$ refers to the case where a target is added to the scene at location $i$ at time $t$.  Then, under the assumption that at most one target exists in the vicinity of a cell, we have
\begin{equation}
\begin{split}
f(\theta_i(t)&|\Psi(t),\vY(t-1))\\
&= \sum\limits_{j\in H(i)\cup\{0\}} f(\theta_i(t)|I_i(t)=1,E_{i,j}(t-1),\vY(t-1))\\
& \qquad\cdot f(E_{i,j}(t-1)|I_i(t)=1,\vY(t-1))\\
&= \sum\limits_{j\in H(i)\cup\{0\}} \int f(\theta_i(t)|\theta_j(t-1))\\
& \qquad \cdot f(E_{i,j}(t-1)|I_i(t)=1,\vY(t-1)) \\
& \qquad \cdot f(\theta_j(t-1)|I_j(t-1)=1,\vY(t-1))d\theta_j(t-1),
\end{split}
\label{darap-eq:integral-dynamic-target-amplitude}
\end{equation}
where it is understood that in the case where a new target is added to the scene
\begin{equation}
f(\theta_0(t-1)|I_0(t-1)=1,\vY(t-1))=f(\theta_0(t-1)) \sim \mathcal{N}(\mu_\theta,\sigma_\theta^2)
\end{equation}
and
\begin{equation}
\theta_i(t)=\theta_0(t-1)
\end{equation}
In the static case, both $f(\theta_i(t)|\theta_j(t-1))$ and $f(\theta_j(t-1)|I_j(t-1)=1,\vY(t-1))$ are Gaussian which makes it possible to analytically integrate equation (\ref{darap-eq:integral-dynamic-target-amplitude}).  Indeed, at time $t=1$, it can be easily seen that $\theta_j(1)\sim \mathcal{N}(\mu_\theta,\sigma_\theta^2)$.  However, for $t>1$, equation (\ref{darap-eq:integral-dynamic-target-amplitude}) shows that we get a Gaussian mixture model with mixing coefficients given by
\begin{equation}
f(E_{i,j}(t-1)|I_i(t)=1,\vY(t-1))
\end{equation}
In order to make the estimation of the posterior distributions very simple, we make the assumption that
\begin{equation}
\label{darap-eq:single-gaussian-mode-target-equations}
f(E_{i,j}(t-1)|I_i(t)=1,\vY(t-1)) = \indic{j=j^*}
\end{equation}
for a single $j^*\in H(i)\cup\{0\}$.  In other words, conditioned on the event that a target exists at cell $i$, it is known with probability 1 that the target transitioned from either a single neighboring cell or entered the scene at time $t$.  This assumption is restrictive except at high SNR.  However, it allows us to simplify equation (\ref{darap-eq:integral-dynamic-target-amplitude}) as
\begin{equation}
\begin{split}
f(\theta_i(t)|\Psi(t),\vY(t-1))= &\int f(\theta_i(t)|\theta_{j^*}(t-1))\\
& \cdot f(\theta_{j^*}(t-1)|I_{j^*}(t-1)=1,\vY(t-1)) d\theta_{j^*}(t-1),
\end{split}
\label{darap-eq:predict-eq-amplitudes-simplification}
\end{equation}
which can easily seen to be Gaussian distributed as long as $f(\theta_{j}(t)|I_{j}(t)=1,\vY(t))$ is Gaussian.  Indeed, we see the recursion
\begin{equation}
\label{darap-eq:update-eq-amplitudes-simplification}
f(\theta_i(t)|I_i(t)=1,\vY(t))\propto f(y_i(t)|I_i(t),\theta_i(t))f(\theta_i(t)|I_i(t)=1,\vY(t-1))
\end{equation}
Using equations (\ref{darap-eq:predict-eq-amplitudes-simplification}) and (\ref{darap-eq:update-eq-amplitudes-simplification}), it is simple to show that a simply modified Kalman filter will give the exact recursion required to update the posterior densities.  In fact, it is the same recursion given in the static case, except that we have
\begin{align}
\hat{\theta}_i(t|t-1) &= \hat{\theta}_{j_i^*(t-1)}(t-1|t-1),\\
\hat{\sigma}_i^2(t|t-1) &= \hat{\sigma}_{j_i^*(t-1)}^2(t-1|t-1) + \Delta_\theta^2\\
\label{darap-eq:jstari}
j_i^*(t-1) &= \arg\max\limits_{j\in H(i)\cup\{0\}} f(E_{i,j}(t-1)|I_i(t)=1,\vY(t-1)).
\end{align}

\begin{prop}\label{prop:MSE-equivalence}
When (\ref{darap-eq:single-gaussian-mode-target-equations}) holds for all $i\in\Psi(t)$ and for all $t=1,2,\dots,T$, then the cost function (\ref{darap-eq:DARAP-cost}) is proportional to the minimum weighted mean squared error
\begin{equation*}
J_T(\vlamu{T}) = \mathbb{E}\left[\sum\limits_{t=1}^T \gamma(t)\sum\limits_{i=1}^Q I_i(t)\left(\theta_i(t)-{\hat{\theta}}_i(t)\right)^2\right],
\end{equation*}
where ${\hat{\theta}}_i(t) = \E\set{\theta_i(t)\Big|I_i(t)=1,\vY(t)}$.
\end{prop}
\begin{IEEEproof}
We first note that
\begin{equation}
\begin{split}
\mathbb{E}\left[\sum\limits_{i=1}^Q I_i(t)\left(\theta_i(t)-{\hat{\theta}}_i(t)\right)^2\right]
&= \sum\limits_{i=1}^QI_i(t)\text{var}\set{\theta_i(t)\Big|I_i(t)=1,\vY(t)},\\
&= \sum\limits_{i=1}^QI_i(t) \hat{\sigma}_{j_i^*(t-1)}^2(t|t) 
\end{split}
\end{equation}
since ${\hat{\theta}}_i(t)$ is the conditional mean estimator by definition and using the definitions in (\ref{darap-eq:single-gaussian-mode-target-equations}) and (\ref{darap-eq:jstari}).  Moreover, by the update equations (\ref{darap-eq:KF-first})-(\ref{darap-eq:KF-static-variance-t-t}), it can easily be shown that
\begin{equation}
\hat{\sigma}_i^2(t|t) = \sigma^2\left[\frac{\sigma^2}{\hat{\sigma}_i^2(t|t-1)} + \lambda_i(t)\right]^{-1}.
\end{equation}
Thus we have
\begin{equation}
\begin{split}
\mathbb{E}\left[\sum\limits_{t=1}^T \gamma(t)\sum\limits_{i=1}^Q I_i(t)\left(\theta_i(t)-{\hat{\theta}}_i(t)\right)^2\right] 
&= \mathbb{E}\left[\sum\limits_{t=1}^T \gamma(t)\sum\limits_{i=1}^Q I_i(t) \hat{\sigma}_{j_i^*(t-1)}^2(t|t)\right]\\
&\propto \mathbb{E}\left[\sum\limits_{t=1}^T \gamma(t)\sum\limits_{i=1}^Q I_i(t) \left[\frac{\sigma^2}{\hat{\sigma}_i^2(t|t-1)} + \lambda_i(t)\right]^{-1}\right]\\
&= \mathbb{E}\left[\sum\limits_{t=1}^T \gamma(t)\sum\limits_{i=1}^Q \frac{p_i(t)}{{\sigma^2}/{\hat{\sigma}_i^2(t|t-1)} + \lambda_i(t)}\right]=J_T(\vlamu{T}),\\
\end{split}
\end{equation}
where the last equality occurs because $\E\set{I_i(t)|\vY(t)}=p_i(t)$.
\end{IEEEproof}

Looking at the update equations for the target indicators, we get
\begin{equation}
f(I_i(t)=1|\vY(t-1)) = \sum\limits_{j\in H(i)\cup \{0\}} f(I_i(t)=1|E_{i,j}(t-1)) f(E_{i,j}(t-1) | \vY(t-1))
\end{equation}
and
\begin{equation}
f(I_i(t)=1|\vY(t)) \propto f(y_i(t)|I_i(t)=1,\vY(t-1)) f(I_i(t)=1|\vY(t-1)),
\end{equation}
where
\begin{equation}
\begin{split}
f(y_i(t)&|I_i(t)=1,\vY(t-1))\\
& = \int f(y_i(t)|I_i(t)=1,\theta_i(t)\vY(t-1))\\
& \qquad\cdot f(\theta_i(t)|I_i(t)=1,\vY(t-1))d\theta_i(t)
\end{split}
\end{equation}
Similar to the derivation in the static case, it can easily be seen that 
\begin{equation}
y_i(t)|I_i(t)=1,\vY(t-1) \sim \mathcal{N}(\sqrt{\lambda_i(t)}\mu_i(t|t-1), \lambda_i(t)\sigma_i^2(t|t-1)+\sigma^2)
\end{equation}
and
\begin{equation}
y_i(t)|I_i(t)=0,\vY(t-1) \sim \mathcal{N}(0,\sigma^2)
\end{equation}

\subsection{Discussion of generalizations of state model and posterior estimation methods}
\label{darap-sec:generalized-dynamic-models}
As mentioned earlier, it is a difficult, if not intractable, problem to exactly estimate the posterior distribution of $\vxi(t)$ given $\vY(t-1)$ that is required for our adaptive algorithms.  We have provided a simple algorithm that approximates the posterior distribution under some restrictive assumptions.  A simple way to alleviate these restrictions is to use a particle filter implementation for $\vxi(t)$ or other approximate method (e.g., the extended and unscented Kalman filters).  

Moreover, we have provided a particular state model that builds on our previous work with the inclusion of transition, birth, and death probabilities.  However, there are many other models for dynamic state models, including linear and nonlinear motion models, targets that may occupy multiple adjacent cells, and various noise models.  In any of these cases, one would have to use a different posterior estimation algorithm to provide estimates of $\Pr(I_i(t)=1|\vY(t-1))$.

Future work plans to compare other posterior estimation algorithms such as the JMDP particle filter \cite{Kreucher-05-multi-tar-track-jpdf} to the one presented in this work, as well as generalizations to more interesting dynamic state models.

\subsection{Unobservable targets}
One particular generalization of the measurement model that is used in this
work is the inclusion of indicator variables for observable/unobservable targets.  In many applications, certain locations may be obscured for short durations, such as locations in the null of a radar beam.  Define
\begin{equation}
O_i(t)=\begin{cases}1,& {\rm Location}\ i \rm{\ is\ observable}\\0,&{\rm
Otherwise}\end{cases} \hspace{0.1in} \mathrm{for}\ i\in\{1,2,\dots,Q\}.
\end{equation}
to be an indicator variable for the observability of the $i$-th location.  Then the measurement model becomes
\begin{equation}
y_i(t) = \sqrt{\lambda_i(t)}I_i(t)O_i(t)\theta_i(t)+\varepsilon_i(t),
\end{equation}
It is assumed that $\vO = \set{O_i(t)}_{i,t}$ is known to the user a priori. Thus, we are required to estimate the densities:
\begin{equation}
f(\vxi(t)|\vY(t'),\vO)
\end{equation}
for $t' = t, t-1$.  
We make the simplifying assumption that if $I_i(t-1)=1$, then
\begin{equation}
O_i(t) = 1 \Leftrightarrow O_j(t) = 1, \forall j\in H(i)
\end{equation}
It can easily be seen that when $O_i(t)=1$, we have the identical update equations to the fully observable case.  However, when $O_i(t)=0$, the predict equations remain the same as before, but the update equations are changed in the following manner:
\begin{equation}
f(I_i(t)=1|\vY(t),O_i(t)=0) = f(I_i(t)=1|\vY(t-1))
\end{equation}
and the target amplitudes when $O_i(t)=0$:
\begin{align}
\hat{\theta}_i(t|t) &= \hat{\theta}_i(t|t-1)\\
\hat{\sigma}_i^2(t|t) &= \hat{\sigma}_i^2(t|t-1)
\end{align}

\fi

\ifjournal
\else
	\section{Proof of Proposition \ref{prop:omni0}}
\label{app:omni0}

Conditioned on the number of targets $\lvert\Psi(1)\rvert$, the per-stage cost of the omniscient policy can be determined by setting $p_{i}(t) = I_{i}(t)$ and substituting \eqref{darap-eq:omn-policy-defn} and \eqref{eqn:cOmni0} into \eqref{darap-eq:myopic-cost}, yielding 
\[
M_{t}(\boldsymbol{\lambda}^{\omni}) 
= \frac{\lvert\Psi(1)\rvert^{2}}{\lvert\Psi(1)\rvert (\sigma^{2} / \sigma_{\theta}^{2}) + \Lambdab(t)}.
\]
\ignore{
Since this is a convex function of $\lvert\Psi(1)\rvert$, by Jensen's inequality we have 
\[
\E\left\{ M_{t}(\boldsymbol{\lambda}^{\omni}) \right\} \geq \frac{(p_{0} Q)^{2}}{p_{0} Q (\sigma^{2} / \sigma_{\theta}^{2}) + \Lambdab(t)},
\]
noting that $\lvert\Psi(1)\rvert \sim \mathrm{Binomial}(Q,p_{0})$ and $\E\left\{ \lvert\Psi(1)\rvert \right\} = p_{0} Q$.
}
As a function of $\lvert\Psi(1)\rvert$, $M_{t}(\boldsymbol{\lambda}^{\omni})$ is proportional to $h_{0}(z) = z^{2} / (z + \Lambdab(t))$, where $z = (\sigma^{2}/\sigma_{\theta}^{2}) \lvert\Psi(1)\rvert$.  The derivatives of the function $h_{0}$ are given by 
\begin{subequations}\label{eqn:h0Deriv}
\begin{align}
h_{0}'(z) &= \frac{z(z + 2\Lambdab(t))}{(z + \Lambdab(t))^{2}}\\
h_{0}''(z) &= \frac{2\Lambdab(t)^{2}}{(z + \Lambdab(t))^{3}}\\
h_{0}'''(z) &= -\frac{6\Lambdab(t)^{2}}{(z + \Lambdab(t))^{4}}\\
h_{0}^{(4)}(z) &= \frac{24\Lambdab(t)^{2}}{(z + \Lambdab(t))^{5}},
\end{align}
\end{subequations}
%
showing that $h_{0}$ has a positive fourth derivative for $z \geq 0$.  Hence by Taylor's theorem, $h_{0}$ is lower bounded by its third-order expansion,
\[
h_{0}(z) \geq h_{0}(z_{0}) + h_{0}'(z_{0}) (z - z_{0}) + \frac{h_{0}''\left( z_{0} \right)}{2} (z - z_{0})^{2} + \frac{h_{0}'''\left( z_{0} \right)}{3!} \left( z - z_{0} \right)^{3}.
\]
Letting the center of expansion $z_{0} = \E[z]$ and taking expectations with respect to $z$, we obtain
\begin{equation}\label{eqn:genJensen}
\E[h_{0}(z)] \geq h_{0}\left( \E[z] \right) + \frac{h_{0}''\left( \E[z] \right)}{2} \var(z) + \frac{h_{0}'''\left( \E[z] \right)}{3!} \E\left[ \left( z - \E[z] \right)^{3} \right].
\end{equation}
Noting that $\lvert\Psi(1)\rvert \sim \mathrm{Binomial}(Q,p_{0})$, the moments of $z$ are 
\begin{subequations}\label{eqn:zMom}
\begin{align}
\E[z] &= (\sigma^{2}/\sigma_{\theta}^{2}) p_{0} Q\\
\var(z) &= (\sigma^{2}/\sigma_{\theta}^{2})^{2} p_{0} Q (1 - p_{0})\\
\E\left[ \left( z - \E[z] \right)^{3} \right] &= (\sigma^{2}/\sigma_{\theta}^{2})^{3} p_{0} Q (1 - p_{0}) (1 - 2p_{0}).
\end{align}
\end{subequations}
Substituting \eqref{eqn:h0Deriv} and \eqref{eqn:zMom} into \eqref{eqn:genJensen} and using the definition $r_{0}(t) = \sigma^{2} Q / (\sigma_{\theta}^{2} \Lambdab(t))$ to simplify, we obtain 
\[
\E\left\{ M_{t}(\boldsymbol{\lambda}^{\omni}) \right\} \geq \frac{(p_{0}Q)^{2}}{\Lambdab(t) (1 + p_{0} r_{0}(t))} + \frac{p_{0}Q (1-p_{0})}{\Lambdab(t) (1 + p_{0} r_{0}(t))^{3}} - \frac{(1-p_{0}) (1 - 2p_{0}) p_{0} r_{0}(t)}{\Lambdab(t) (1 + p_{0} r_{0}(t))^{4}}.
\]
This bound holds for each $t$ and therefore the total cost is bounded as 
\begin{equation}\label{eqn:JOmni}
J_{T}(\boldsymbol{\lambda}^{\omni}) 
\geq \sum_{t=1}^{T} 
\frac{\gamma(t)}{\Lambdab(t)} \left[ \frac{(p_{0}Q)^{2}}{1 + p_{0} r_{0}(t)} + \frac{p_{0}Q (1-p_{0})}{(1 + p_{0} r_{0}(t))^{3}} - \frac{(1-p_{0}) (1 - 2p_{0}) p_{0} r_{0}(t)}{(1 + p_{0} r_{0}(t))^{4}} \right].
\end{equation}

For the uniform allocation policy defined by $\lambda_i^{\uni}(t) = \Lambda(t)/Q$ for all $i$,
%
%
a formula similar to \eqref{eqn:cOmni0} holds for $c_{i}(t)$ and the per-stage cost \eqref{darap-eq:myopic-cost} is 
\begin{equation}\label{eqn:myopicUni0}
M_{t}(\boldsymbol{\lambda}^{\uni}) = \frac{Q \lvert\Psi(1)\rvert}{Q (\sigma^{2} / \sigma_{\theta}^{2}) + \Lambdab(t)} = \frac{Q \lvert\Psi(1)\rvert}{\Lambdab(t) (1 + r_{0}(t))},
\end{equation}
again conditioned on $\lvert\Psi(1)\rvert$.  Hence the total cost is 
\begin{equation}\label{eqn:JUni}
J_{T}(\boldsymbol{\lambda}^{\uni}) = \sum_{t=1}^{T} 
\frac{\gamma(t)}{\Lambdab(t)} \frac{p_{0} Q^{2}}{1 + r_{0}(t)}.
\end{equation}
The result follows from the ratio of \eqref{eqn:JOmni} and \eqref{eqn:JUni}.

\section{Proof of Lemma \ref{lem:cbar}}
\label{app:cbar}

By definition of $\cbar(1)$, the lemma is true for $t = 1$.  We proceed by induction, considering first the case $\cbar(t) < \ccrit$.  
First it is shown that under the assumptions of the lemma and $\cbar(t) < \ccrit$, all locations in $H(\Psi(t-1))$ receive nonzero allocations in stage $t$.  Given \eqref{eqn:pSemi} and Assumptions \ref{ass:pi0} and \ref{ass:c}, 
the permutation $\chi$ in \eqref{darap-eq:pi} ranks all of the indices $i \in \Psi(t-1)$ in the ROI equally, followed by $i \in G(\Psi(t-1))$, again all equally.  It is then straightforward to see that the sequence $g(k)$ \eqref{darap-eq:g(k)} that determines the number of nonzero allocations is as follows: 
\begin{equation}\label{eqn:gSemi}
g(k) = 
\begin{cases}
0, & k = 1, \ldots, \lvert\Psi(1)\rvert - 1,\\
\cbar(t) \lvert\Psi(1)\rvert e(\pi_0,G), & k = \lvert\Psi(1)\rvert, \ldots, \lvert H(\Psi(1)) \rvert - 1,\\
\infty & k \geq \lvert H(\Psi(1)) \rvert. 
\end{cases}
\end{equation}
%
%
%
In particular, given Assumption \ref{ass:Lambda} and $\cbar(t) < \ccrit$, we have 
\[
g\left( \lvert H(\Psi(1))\rvert - 1 \right) < \ccrit \lvert\Psi(1)\rvert e(\pi_0,G) = \Lambda < g\left( \lvert H(\Psi(1))\rvert \right),
\] 
implying that the number of nonzero allocations $k^{*} = \lvert H(\Psi(1))\rvert$. 

We now use \eqref{eqn:c0} and the fact that $\lambda_{i}(t) > 0$ for $i \in H(\Psi(t-1))$ 
to propagate the posterior precisions forward in time. 
Combining \eqref{eqn:c0} and \eqref{darap-eq:myopic-solution}, we have 
\begin{align}
c_{i}(t+1)
&= \left( \Lambda + \sum_{j\in H(\Psi(t-1))} c_{j}(t) \right) \frac{\sqrt{p_{s^{(n)}(t)}(t)}}{ \sum_{j\in H(\Psi(t-1))} \sqrt{p_{j}(t)} }\nonumber\\
&= \left( \Lambda + \sum_{j\in H(\Psi(t-1))} c_{j}(t) \right) \frac{\sqrt{p_{s^{(n)}(t)}(t)}}{\lvert\Psi(1)\rvert \left( \sqrt{\pi_{0}} + \sqrt{\lvert G\rvert (1-\pi_{0})} \right)}, \quad i \in H\left(s^{(n)}(t)\right),\label{eqn:cLow}
\end{align}
using \eqref{eqn:pSemi} in the second equality.  
%
%
Conditioned on $\lvert\Psi(1)\rvert$, there are two random quantities in \eqref{eqn:cLow}: the probability $p_{s^{(n)}(t)}(t)$, which depends on target motion 
between stages $t-1$ and $t$, and the precisions $c_{j}(t)$, which depend on target motion up to stage $t-1$.  These quantities are independent according to the target model.  Thus taking the conditional expectation of both sides of \eqref{eqn:cLow} yields 
\begin{subequations}
\begin{align}
\E \left[ c_{i}(t+1) \mid \lvert\Psi(1)\rvert \right] &= \left( \Lambda + \sum_{j\in H(\Psi(t-1))} \E \left[ c_{j}(t) \mid \lvert\Psi(1)\rvert \right] \right) \frac{\E \left[ \sqrt{p_{s^{(n)}(t)}(t)} \mid \lvert\Psi(1)\rvert \right]}{\lvert\Psi(1)\rvert \left( \sqrt{\pi_{0}} + \sqrt{\lvert G\rvert (1-\pi_{0})} \right)}\nonumber\\
&= \left( \Lambda + \sum_{j\in H(\Psi(t-1))} \E \left[ c_{j}(t) \mid \lvert\Psi(1)\rvert \right] \right) \frac{\pi_{0}^{3/2} + \frac{1}{\sqrt{\lvert G \rvert}} (1-\pi_{0})^{3/2}}{\lvert\Psi(1)\rvert \left( \sqrt{\pi_{0}} + \sqrt{\lvert G\rvert (1-\pi_{0})} \right) }\label{eqn:cbarLow1}\\
&\leq \left( \frac{\Lambda}{\lvert\Psi(1)\rvert} + \left(1 + \lvert G \rvert\right) \cbar(t) \right) \frac{\pi_{0}^{3/2} + \frac{1}{\sqrt{\lvert G \rvert}} (1-\pi_{0})^{3/2}}{\sqrt{\pi_{0}} + \sqrt{\lvert G\rvert (1-\pi_{0})} }\label{eqn:cbarLow2}\\
&\equiv \cbar(t+1), \qquad i \in H\left(s^{(n)}(t)\right).\label{eqn:cbarLow3}
\end{align}
\end{subequations}
The second line \eqref{eqn:cbarLow1} follows from \eqref{eqn:pSemi} and because $s^{(n)}(t) = s^{(n)}(t-1)$ with probability $\pi_{0}$ and $s^{(n)}(t) \in G(s^{(n)}(t-1))$ with probability $1 - \pi_{0}$.  In the third line \eqref{eqn:cbarLow2}, we have used 
the inductive assumption $\cbar(t) \geq 
\E \left[ c_{j}(t) \mid \lvert\Psi(1)\rvert \right]$, $j \in H(\Psi(t-1))$, and the equality $\lvert H(\Psi(t-1)) \rvert = (1 + \lvert G \rvert) \lvert\Psi(1)\rvert$. 
The last line \eqref{eqn:cbarLow3} follows from the recursion \eqref{eqn:cbar}, thus completing the induction.  

Next we consider the case $\cbar(t) \geq \ccrit$, using induction as before.  
The base case, i.e., the first $t$ such that $\cbar(t) \geq \ccrit$, is either true for $t=1$ or follows eventually from the previous induction for $\cbar(t) < \ccrit$. 
Similar to above, it can be shown that under Assumptions \ref{ass:pi0}--\ref{ass:Lambda} and $\cbar(t) \geq \ccrit$, 
\[
0 = g\left( \lvert\Psi(1)\rvert - 1 \right) < \Lambda \leq g\left( \lvert\Psi(1)\rvert \right),
\]
implying that only the previous target locations $i \in \Psi(t-1)$ are allocated nonzero effort in stage $t$. 
In the case $s^{(n)}(t) = s^{(n)}(t-1)$, which occurs with probability $\pi_{0}$, we combine \eqref{eqn:c0} with \eqref{darap-eq:myopic-solution} 
to obtain 
\[
c_{i}(t+1) = \frac{1}{\lvert\Psi(1)\rvert} \left( \Lambda + \sum_{j\in \Psi(t-1)} c_{j}(t) \right), \quad i \in H\left( s^{n}(t) \right).
\]
In the other case $s^{(n)}(t) \in G(s^{(n)}(t-1))$ with probability $1-\pi_{0}$, $\lambda_{s^{(n)}(t)}(t) = 0$ and $c_{i}(t+1) = c_{s^{(n)}(t)}(t)$.  Therefore the expected precision conditioned on $\lvert\Psi(1)\rvert$ 
is given by 
\begin{align*}
\E[ c_{i}(t+1) \mid \lvert\Psi(1)\rvert ]
&= \frac{\pi_{0}}{\lvert\Psi(1)\rvert} \left( \Lambda + \sum_{j\in \Psi(t-1)} \E \left[ c_{j}(t) \mid \lvert\Psi(1)\rvert \right] \right)
+ (1-\pi_{0}) \E \left[ c_{s^{(n)}(t)}(t) \mid \lvert\Psi(1)\rvert \right]\\
&\leq \cbar(t) + \frac{\pi_{0} \Lambda}{\lvert\Psi(1)\rvert}\\
&= \cbar(t+1), \qquad i \in H\left( s^{(n)}(t) \right), 
\end{align*}
using the inductive assumption on $\cbar(t)$ and \eqref{eqn:cbar} 
to complete the proof.

\section{Proof of Proposition \ref{prop:semi0}}
\label{app:semi0}


We rewrite the expected per-stage cost by combining \eqref{eqn:myopicCost2} and \eqref{eqn:c0} 
and iterating expectations to yield 
\[
\E\left[ M_{t}(\boldsymbol{\lambda}) \right] = \E\left\{ \sum_{n=1}^{\lvert\Psi(1)\rvert} \E\left[ \frac{1}{c_{s^{(n)}(t+1)}(t+1)} \mid \lvert\Psi(1)\rvert \right] \right\}.
\]
Using the convexity of the function $1/x$ and Jensen's inequality, this may be bounded from below as 
\[
\E\left[ M_{t}(\boldsymbol{\lambda}) \right] \geq \E\left\{ \sum_{n=1}^{\lvert\Psi(1)\rvert} \frac{1}{\E \left[ c_{s^{(n)}(t+1)}(t+1) \mid \lvert\Psi(1)\rvert \right]} \right\}.
\]
For $t$ large enough such that $\cbar(t) \geq \ccrit$, Lemma \ref{lem:cbar} applies to provide a further lower bound,
\begin{equation}\label{eqn:myopicSemi0}
\E\left[ M_{t}(\boldsymbol{\lambda}) \right] \geq \E \left\{ \frac{\lvert\Psi(1)\rvert}{\cbar(t+1)} \right\}.
\end{equation}
 The recursion \eqref{eqn:cbar} for $\cbar(t)$ implies that 
\begin{equation}\label{eqn:cbarInf}
\cbar(t) = \frac{\pi_{0} \Lambda}{\lvert\Psi(1)\rvert} t + O(1). 
\end{equation}
%
Substituting \eqref{eqn:cbarInf} into \eqref{eqn:myopicSemi0} and recalling that $\lvert\Psi(1)\rvert$ is binomially distributed with parameters $Q$ and $p_{0}$, we obtain  
\begin{align*}
\E\left[ M_{t}(\boldsymbol{\lambda}^{\semi}) \right] 
&\geq \E\left\{ \frac{\lvert\Psi(1)\rvert^{2}}{\pi_{0} \Lambda t} + O\left(\frac{1}{t^{2}}\right) \right\}\\
&= \frac{p_{0} Q (p_{0} Q + 1-p_{0})}{\pi_{0} \Lambda t} + O\left(\frac{1}{t^{2}}\right).
\end{align*}

\section{Proof of Proposition \ref{prop:omni+}}
\label{app:omni+}

First we derive an expression for the steady-state (large $t$) posterior variance. 
For the uniform and omniscient policies under Assumption \ref{ass:Lambda}, the effort allocation $\lambda_{s^{(n)}(t)}(t)$ for targets is independent of both $n$ and $t$.  Therefore all targets have the same steady-state posterior variance $\sigma_{\steady}^{2}$, which may be determined by setting $i = s^{(n)}(t+1)$ and $\sigma_{s^{(n)}(t+1)}^{2}(t+1) = \sigma_{s^{(n)}(t)}^{2}(t) = \sigma_{\steady}^{2}$ in \eqref{eqn:sigma+} to yield the following quadratic equation:
\begin{equation}\label{eqn:sigmaSSQuad}
\lambda_{s^{(n)}(t)}(t) \left(\sigma_{\steady}^{4} - \Delta^{2} \sigma_{\steady}^{2}\right) - \sigma^{2} \Delta^{2} = 0.
\end{equation}
Taking the positive root results in   
\begin{equation}\label{eqn:sigmaSS}
\sigma_{\steady}^{2} = \frac{\Delta^{2}}{2} \left(1 + \sqrt{1 + \frac{4\sigma^{2}}{\Delta^{2} \lambda_{s^{(n)}(t)}(t)} }\right).
\end{equation}

%
%

To relate the steady-state variance \eqref{eqn:sigmaSS} to the gain \eqref{darap-eq:gain-cost}, 
we take $T \to \infty$ and use Assumption \ref{ass:gamma}, which reduces the gain to a ratio of steady-state expected per-stage costs.  To compute the per-stage cost, we first note that the denominator $c_{i}(t) + \lambda_{i}(t)$ in \eqref{eqn:myopicCost2} corresponds to the posterior variance after measurement but before the increment $\Delta^{2}$, while $\sigma_{\steady}^{2}$ in \eqref{eqn:sigmaSS} is the steady-state variance after the increment.  Hence the steady-state per-stage cost conditioned on $\Psi(t)$ is 
%
\begin{align}
\lim_{t\to\infty} \sum_{i\in\Psi(t)} \frac{1}{c_{i}(t) + \lambda_{i}(t)} 
&= \lvert\Psi(1)\rvert \frac{\sigma_{\steady}^{2} - \Delta^{2}}{\sigma^{2}}\nonumber\\
&= \frac{\Delta^{2} \lvert\Psi(1)\rvert}{2\sigma^{2}} \left(\sqrt{1 + \frac{4\sigma^{2}}{\Delta^{2} \lambda_{s^{(n)}(t)}(t)} } - 1 \right).\label{eqn:gSS}
\end{align}
%
For the uniform policy, $\lambda_{s^{(n)}(t)}(t) = \Lambda / Q$ and the expectation over $\lvert\Psi(1)\rvert$ gives  
\begin{equation}\label{eqn:myopicUniSS}
\lim_{t\to\infty} \E \left\{ M_{t}(\boldsymbol{\lambda}^{\uni}) \right\} = \frac{\Delta^{2} p_{0} Q}{2\sigma^{2}} \left(\sqrt{1 + 
4 r_{+} } - 1 \right).
\end{equation}
For the omniscient policy, $\lambda_{s^{(n)}(t)}(t) = \Lambda / \lvert\Psi(1)\rvert$ and \eqref{eqn:gSS} 
is proportional to $h_{+}(z) = z (\sqrt{1+z} - 1)$, where $z = 4\sigma^{2} \lvert\Psi(1)\rvert / (\Delta^{2} \Lambda)$.  The function $h_{+}$ has derivatives 
\begin{subequations}\label{eqn:h+Deriv}
\begin{align}
h_{+}'(z) &= \sqrt{1+z} - 1 + \frac{z}{2\sqrt{1+z}}\\
h_{+}''(z) &= \frac{1+3z/4}{(1+z)^{3/2}}\\
h_{+}'''(z) &= -\frac{3(2+z)}{8(1+z)^{5/2}}\\
h_{+}^{(4)}(z) &= \frac{3(8+3z)}{16(1+z)^{7/2}}.
\end{align}
\end{subequations}
%
Hence similar to $h_{0}$ in the proof of Proposition \ref{prop:omni0}, $h_{+}$ has a positive fourth derivative and the same lower bound \eqref{eqn:genJensen} applies to $\E[h_{+}(z)]$.  Since $z$ is again proportional to $\lvert\Psi(1)\rvert$, the moments of $z$ are given by similar expressions as in \eqref{eqn:zMom}.  Combining these moments with \eqref{eqn:genJensen}, \eqref{eqn:h+Deriv}, and the definition of $r_{+}$ yields after some simplification 
%
%
%
\begin{multline}
\lim_{t\to\infty} \E \left\{ M_{t}(\boldsymbol{\lambda}^{\omni}) \right\} \geq 
\frac{\Delta^{2} p_{0} Q}{2\sigma^{2}} \left(\sqrt{1 + 
4 p_{0} r_{+} } - 1 \right)
+ \frac{p_{0} Q (1-p_{0})}{\Lambda} \frac{1 + 3p_{0} r_{+}}{(1 + 4p_{0} r_{+})^{3/2}}\\
- \frac{p_{0} (1-p_{0}) (1-2p_{0})}{\Lambda} \frac{r_{+} (1 + 2p_{0} r_{+})}{(1 + 4p_{0} r_{+})^{5/2}}.\label{eqn:myopicOmniSS}
\end{multline}
%
Taking the ratio of \eqref{eqn:myopicOmniSS} and \eqref{eqn:myopicUniSS} yields the result. 

\section{Proof of Proposition \ref{prop:semiLarge}}
\label{app:semiLarge}

As in Proposition \ref{prop:omni+}, under Assumption \ref{ass:gamma} the gain reduces to the ratio of steady-state expected per-stage costs.  To compute the per-stage cost for the semi-omniscient policy, we first show that the precisions $c_{i}(t)$ are small as claimed.  Rewriting the evolution equation \eqref{eqn:sigma+} in terms of $c_{i}(t)$ gives 
\begin{equation}\label{eqn:cSmall}
c_{i}(t+1) = \frac{\sigma^{2}}{\Delta^{2}} \frac{c_{s^{(n)}(t)}(t) + \lambda_{s^{(n)}(t)}(t)}{c_{s^{(n)}(t)}(t) + \lambda_{s^{(n)}(t)}(t) + \sigma^{2}/\Delta^{2}} < \frac{\sigma^{2}}{\Delta^{2}},\quad i \in H\left( s^{(n)}(t) \right).  
\end{equation}
%
This bound together with assumption \eqref{eqn:condSemiLarge} imply that the semi-omniscient policy allocates nonzero effort to all locations in $H(\Psi(t-1))$, i.e., the sequence $g(k)$ \eqref{darap-eq:g(k)} satisfies $g(k) < \Lambda$ for $k = \lvert H(\Psi(t-1)) \rvert - 1$. 
Using \eqref{eqn:pSemi}, the sum of probability ratios $\sqrt{p_{\chi(i)}(t) / p_{\chi(k+1)}(t)}$ in \eqref{darap-eq:g(k)} can be bounded by $Q (e(\pi_0,G)+1)$.  Combining this with 
\eqref{eqn:cSmall}, \eqref{eqn:condSemiLarge}, and the definition of $r_{+}$ yields 
\[
g\left( \lvert H(\Psi(t-1)) \rvert - 1 \right) < \frac{\sigma^{2} Q}{\Delta^{2}} (e(\pi_0,G)+1) \leq \Lambda
\]
as desired.

Given that $\lambda_{i}(t) > 0$ for all $i \in H(\Psi(t-1))$, the per-stage cost for the semi-omniscient policy can be computed from \eqref{darap-eq:myopic-cost}, \eqref{darap-eq:myopic-solution} and \eqref{eqn:pSemi} as 
\begin{align}
M_{t}(\boldsymbol{\lambda}^{\semi}) 
&= \frac{\left( \sum_{i\in H(\Psi(t-1))} \sqrt{p_{i}(t)} \right)^{2}}{\Lambda + \sum_{i\in H(\Psi(t-1))} c_{i}(t)}\nonumber\\
&\geq \frac{\left( \sum_{i\in H(\Psi(t-1))} \sqrt{p_{i}(t)} \right)^{2}}{\Lambda + \bigl(1+\lvert G\rvert\bigr) \lvert\Psi(1)\rvert \sigma^{2} / \Delta^{2}}\nonumber\\
&= \frac{\lvert\Psi(1)\rvert^{2} \left( \sqrt{\pi_{0}} + \sqrt{\lvert G \rvert (1-\pi_{0})} \right)^{2}}{\Lambda + \bigl(1+\lvert G\rvert\bigr) \lvert\Psi(1)\rvert \sigma^{2} / \Delta^{2}}\label{eqn:myopicSemiLarge0}
\end{align}
where the inequality follows from \eqref{eqn:cSmall} and $\lvert H(\Psi(t-1)) \rvert = \bigl(1 + \lvert G \rvert\bigr) \lvert\Psi(1)\rvert$.  
As a function of $\lvert\Psi(1)\rvert$, the right-hand side of \eqref{eqn:myopicSemiLarge0} has the same form as the function $h_{0}(z)$ in the proof of Proposition \ref{prop:omni0} in Appendix \ref{app:omni0}.  Applying the same technique as before and simplifying the resulting expressions, the expectation with respect to $\lvert\Psi(1)\rvert$ can be bounded as 
%
\begin{multline}\label{eqn:myopicSemiLarge}
\E\left\{ M_{t}(\boldsymbol{\lambda}^{\semi}) \right\} \geq 
\frac{p_{0} Q^{2}}{\Lambda} \left( \sqrt{\pi_{0}} + \sqrt{\lvert G \rvert (1-\pi_{0})} \right)^{2}\\
\times \left( \frac{p_{0}}{1+p_{0} q_{+}} + \frac{1-p_{0}}{Q} \frac{1}{(1 + p_{0} q_{+})^{3}} 
- \frac{(1-p_{0}) (1-2p_{0})}{Q^{2}} 
\frac{q_{+}}{(1 + p_{0} q_{+})^{4}} \right).
\end{multline}
%
%
%
The result is obtained from the ratio of \eqref{eqn:myopicSemiLarge} and \eqref{eqn:myopicUniSS}. 

\section{Proof of Lemma \ref{lem:cbarHigh+}}
\label{app:cbarHigh+}

As in the proof of Proposition \ref{prop:semiLarge}, the evolution of the posterior 
precisions is given by \eqref{eqn:cSmall}.  Under Assumptions 
\ref{ass:pi0}--\ref{ass:Lambda} and $\cbar(t) \geq \ccrit$, similar to the proof of Proposition \ref{prop:semi0} we have 
\[
c_{s^{(n)}(t)}(t) + \lambda_{s^{(n)}(t)}(t) = 
\begin{cases}
\frac{1}{\lvert\Psi(1)\rvert} \left( \Lambda + \sum_{j\in \Psi(t-1)} c_{j}(t) \right), & s^{(n)}(t) = s^{(n)}(t-1) \text{\ w.p.\ } \pi_{0},\\
c_{s^{(n)}(t)}(t), & s^{(n)}(t) \in G\left(s^{(n)}(t-1)\right) \text{\ w.p.\ } 1 - \pi_{0}.
\end{cases}
\]
Substituting this into \eqref{eqn:cSmall} results in  
\begin{align*}
\E \left[ c_{i}(t+1) \mid \lvert\Psi(1)\rvert \right] &= \pi_{0} \E\left\{ \frac{(\sigma^{2}/\Delta^{2}) \frac{1}{\lvert\Psi(1)\rvert} \left( \Lambda + \sum_{j\in \Psi(t-1)} c_{j}(t) \right)}{(\sigma^{2}/\Delta^{2}) + \frac{1}{\lvert\Psi(1)\rvert} \left( \Lambda + \sum_{j\in \Psi(t-1)} c_{j}(t) \right)} \mid \lvert\Psi(1)\rvert \right\} \\ 
&\qquad\qquad+ (1-\pi_{0}) \E\left\{ \frac{(\sigma^{2}/\Delta^{2}) c_{s^{(n)}(t)}(t)}{(\sigma^{2}/\Delta^{2}) + c_{s^{(n)}(t)}(t)} \mid \lvert\Psi(1)\rvert \right\}, \quad i \in H\left( s^{(n)}(t) \right).
\end{align*}
The terms on the right-hand side are of the form $ax/(a+x)$ with $a > 0$, which is a concave and increasing function of $x$.  Applying Jensen's inequality and the assumption on $\cbar(t)$, we obtain 
\begin{align*}
\E \left[ c_{i}(t+1) \mid \lvert\Psi(1)\rvert \right] &\leq \pi_{0} \frac{(\sigma^{2}/\Delta^{2}) \frac{1}{\lvert\Psi(1)\rvert} \left( \Lambda + \sum_{j\in \Psi(t-1)} \E\left[ c_{j}(t) \mid \lvert\Psi(1)\rvert \right] \right)}{(\sigma^{2}/\Delta^{2}) + \frac{1}{\lvert\Psi(1)\rvert} \left( \Lambda + \sum_{j\in \Psi(t-1)} \E\left[ c_{j}(t) \mid \lvert\Psi(1)\rvert \right] \right)} \\ 
&\qquad\qquad+ (1-\pi_{0}) \frac{(\sigma^{2}/\Delta^{2}) \E\left[ c_{s^{(n)}(t)}(t) \mid \lvert\Psi(1)\rvert \right]}{(\sigma^{2}/\Delta^{2}) + \E\left[ c_{s^{(n)}(t)}(t) \mid \lvert\Psi(1)\rvert \right]}\\
&\leq \pi_{0} \frac{(\sigma^{2}/\Delta^{2}) (\cbar(t) + \Lambda/\lvert\Psi(1)\rvert)}{(\sigma^{2}/\Delta^{2}) + \cbar(t) + \Lambda/\lvert\Psi(1)\rvert} + (1-\pi_{0}) \frac{(\sigma^{2}/\Delta^{2}) \cbar(t)}{(\sigma^{2}/\Delta^{2}) + \cbar(t)}\\
&= \cbar(t+1), \qquad i \in H\left( s^{(n)}(t) \right), 
\end{align*}
as desired.

\section{Proof of Lemma \ref{lem:cubic}}
\label{app:cubic}

First we show that \eqref{eqn:cbarSSCubic} has three distinct real roots.  This is equivalent to the discriminant of \eqref{eqn:cbarSSCubic} being positive. 
Let $a_{1} = \Delta^{2} / \sigma^{2}$ and $a_{2} = \lvert\Psi(1)\rvert / \Lambda$.  Noting that \eqref{eqn:cbarSSCubic} lacks a quadratic term, the discriminant can be simplified to 
\begin{equation}\label{eqn:discriminant}
D = \pi_{0} a_{1}^{3} \left( 4 a_{1}^{3} + 12 a_{1}^{2} a_{2} + (12 - 27\pi_{0}) a_{1} a_{2}^{2} + 4 a_{2}^{3} \right).
\end{equation}
In the omniscient case $\pi_{0} = 1$, it is known that \eqref{eqn:cbarSSCubic} has three real roots and hence both $D$ and the quantity in the outer parentheses in \eqref{eqn:discriminant} are positive.  As $\pi_{0}$ decreases from $1$, the parenthesized quantity only increases and therefore $D > 0$ for $0 < \pi_{0} < 1$ as well.  Now given that \eqref{eqn:cbarSSCubic} has three real roots, it can be seen that one of the roots is positive and the other two are negative.  This is because the coefficients of \eqref{eqn:cbarSSCubic} constrain the product of the roots to be positive and their sum to be zero.  
The positive root can then be expressed in terms of trigonometric functions as 
\begin{equation}\label{eqn:cbarSS}
\cbar_{\steady}^{-1} = 2 \sqrt{ \frac{a_{1}(a_{1} + a_{2})}{3\pi_{0}} } \cos \left( \frac{1}{3} \arccos \left( \frac{3\sqrt{3\pi_{0}}}{2} \frac{a_{2}}{a_{1} + a_{2}} \sqrt{\frac{a_{1}}{a_{1} + a_{2}}} \right) \right).
\end{equation}

We now use the assumption that $\sqrt{\Delta^{2} \Lambda / \sigma^{2}} \ll 1$, implying that $\sqrt{a_{1} / a_{2}} \ll 1$, to simplify the expression in \eqref{eqn:cbarSS}.  First we expand the argument of the $\arccos$ function to lowest order in $a_{1} / a_{2}$:
\[
\frac{a_{2}}{a_{1} + a_{2}} \sqrt{\frac{a_{1}}{a_{1} + a_{2}}} = \sqrt{\frac{a_{1}}{a_{2}}} \left(1 + O\left( \frac{a_{1}}{a_{2}} \right)\right).
\]
It then follows from further expansions that 
\[
\arccos \left( \frac{3\sqrt{3\pi_{0}}}{2} \frac{a_{2}}{a_{1} + a_{2}} \sqrt{\frac{a_{1}}{a_{1} + a_{2}}} \right) = \frac{\pi}{2} - \frac{3}{2} \sqrt{\frac{3\pi_{0} a_{1}}{a_{2}}} + O\left( \frac{a_{1}}{a_{2}} \right) 
\]
and
\[
\cos \left( \frac{1}{3} \arccos \left( \frac{3\sqrt{3\pi_{0}}}{2} \frac{a_{2}}{a_{1} + a_{2}} \sqrt{\frac{a_{1}}{a_{1} + a_{2}}} \right) \right) = \frac{\sqrt{3}}{2} + \frac{1}{4} \sqrt{\frac{3\pi_{0} a_{1}}{a_{2}}} + O\left( \frac{a_{1}}{a_{2}} \right). 
\]
Combining this with 
\[
2 \sqrt{ \frac{a_{1}(a_{1} + a_{2})}{3\pi_{0}} } = 2 \sqrt{ \frac{a_{1}a_{2}}{3\pi_{0}} } \left(1 + O\left( \frac{a_{1}}{a_{2}} \right)\right),
\]
we obtain 
%
\[
\cbar_{\steady}^{-1} = \sqrt{\frac{a_{1} a_{2}}{\pi_{0}}} \left( 1 + O\left( \sqrt{\frac{a_{1}}{a_{2}}} \right) \right)
= \sqrt{\frac{\Delta^{2} \lvert\Psi(1)\rvert}{\pi_{0} \sigma^{2} \Lambda}} \left( 1 + O\left( \sqrt{\frac{\Delta^{2} \Lambda}{\sigma^{2}}} \right) \right).
\]

\section{Proof of Proposition \ref{prop:semiSmall}}
\label{app:semiSmall}

In the regime $\sqrt{\Delta^{2}\Lambda / \sigma^{2}} \ll 1$, the positive root of the cubic equation \eqref{eqn:cbarSSCubic} is given by \eqref{eqn:cbarSSSmall} in Lemma \ref{lem:cubic}. 
We verify 
that $\cbar_{\steady}$ in \eqref{eqn:cbarSSSmall} satisfies $\cbar_{\steady} \geq \ccrit$ for $\lvert\Psi(1)\rvert \geq 1$. 
Combined with Lemma \ref{lem:cbarInf+}, this will imply that $\cbar_{\steady}$ is a stationary upper bound on the steady-state precision $c_{\steady}$. 
Substituting \eqref{eqn:cbarSSSmall} and \eqref{eqn:ccrit} for $\cbar_{\steady}$ and $\ccrit$ and neglecting higher-order terms, the condition $\cbar_{\steady} \geq \ccrit$ 
is equivalent to 
\[
\sqrt{\frac{\pi_{0} \sigma^{2} \lvert\Psi(1)\rvert}{\Delta^{2} \Lambda}} e(\pi_{0}, G) \geq 1.
\]
The above inequality is most stringent for $\lvert\Psi(1)\rvert = 1$, 
in which case it is ensured by assumption \eqref{eqn:condSemiSmall}.  Hence we conclude that $\cbar_{\steady} \geq c_{\steady}$.

The remainder of the proof uses arguments from the proofs of Propositions \ref{prop:semi0} and \ref{prop:omni+}.  Under Assumption \ref{ass:gamma}, similar to Proposition \ref{prop:omni+} it suffices to compute the steady-state expected per-stage costs.  As in Proposition \ref{prop:semi0}, the expected per-stage cost can be written as 
\[
\E\left[ M_{t}(\boldsymbol{\lambda}) \right] = \E\left\{ \sum_{n=1}^{\lvert\Psi(1)\rvert} \left( \E\left[ \frac{1}{c_{s^{(n)}(t+1)}(t+1)} \mid \lvert\Psi(1)\rvert \right] - \frac{\Delta^{2}}{\sigma^{2}} \right) \right\},
\]
where the additional term $\Delta^{2} / \sigma^{2}$ reflects the fact that the per-stage cost corresponds to the posterior variance before the increment $\Delta^{2}$ whereas $1 / c_{s^{(n)}(t+1)}(t+1)$ includes the increment.  Applying Jensen's inequality as before, taking the steady-state limit $t \to \infty$ and using the bound $c_{\steady} \leq \cbar_{\steady}$, for the semi-omniscient policy we have 
\[
\lim_{t\to\infty} \E\left[ M_{t}(\boldsymbol{\lambda}^{\semi}) \right] \geq \E\left\{ \lvert\Psi(1)\rvert \left( \frac{1}{\cbar_{\steady}} - \frac{\Delta^{2}}{\sigma^{2}} \right) \right\}. 
\]
%
Substituting \eqref{eqn:cbarSSSmall} for $1/\cbar_{\steady}$ gives 
\[
\lim_{t\to\infty} \E\left[ M_{t}(\boldsymbol{\lambda}^{\semi}) \right] \geq \E\left\{ \lvert\Psi(1)\rvert \sqrt{\frac{\Delta^{2} \lvert\Psi(1)\rvert}{\pi_{0} \sigma^{2} \Lambda}} \left( 1 + O\left( \sqrt{\frac{\Delta^{2}\Lambda}{\sigma^{2}}} \right) \right) \right\},
\]
using the approximation $\sqrt{\Delta^{2} \Lambda / \sigma^{2}} \ll 1$.  Since the right-hand side is proportional to $\lvert\Psi(1)\rvert^{3/2}$, a convex function of $\lvert\Psi(1)\rvert \sim \mathrm{Bin}(Q, p_{0})$, another application of Jensen's inequality yields 
%
\begin{equation}\label{eqn:myopicSemiSmall}
\lim_{t\to\infty} \E\left[ M_{t}(\boldsymbol{\lambda}^{\semi}) \right] \geq 
p_{0} Q \sqrt{\frac{\Delta^{2} p_{0} Q}{\pi_{0} \sigma^{2} \Lambda} } \left( 1 + O\left( \sqrt{\frac{\Delta^{2} \Lambda}{\sigma^{2}}} \right) \right).
\end{equation}
%
For the uniform policy, the steady-state expected per-stage cost may be approximated from \eqref{eqn:myopicUniSS} in Proposition \ref{prop:omni+} as 
\begin{equation}\label{eqn:myopicUniSmall}
\lim_{t\to\infty} \E\left[ M_{t}(\boldsymbol{\lambda}^{\uni}) \right] = p_{0} Q \sqrt{\frac{\Delta^{2} Q}{\sigma^{2} \Lambda} } \left( 1 - O\left( \sqrt{\frac{\Delta^{2} \Lambda}{\sigma^{2}}} \right) \right).
\end{equation}
The result is given by the ratio of \eqref{eqn:myopicSemiSmall} and \eqref{eqn:myopicUniSmall}.

\fi

\bibliographystyle{IEEEtran}
\bibliography{References}

\begin{thebibliography}{10}
\providecommand{\url}[1]{#1}
\csname url@samestyle\endcsname
\providecommand{\newblock}{\relax}
\providecommand{\bibinfo}[2]{#2}
\providecommand{\BIBentrySTDinterwordspacing}{\spaceskip=0pt\relax}
\providecommand{\BIBentryALTinterwordstretchfactor}{4}
\providecommand{\BIBentryALTinterwordspacing}{\spaceskip=\fontdimen2\font plus
\BIBentryALTinterwordstretchfactor\fontdimen3\font minus
  \fontdimen4\font\relax}
\providecommand{\BIBforeignlanguage}[2]{{%
\expandafter\ifx\csname l@#1\endcsname\relax
\typeout{** WARNING: IEEEtran.bst: No hyphenation pattern has been}%
\typeout{** loaded for the language `#1'. Using the pattern for}%
\typeout{** the default language instead.}%
\else
\language=\csname l@#1\endcsname
\fi
#2}}
\providecommand{\BIBdecl}{\relax}
\BIBdecl

\bibitem{abidi2008survey}
B.~R. Abidi, N.~R. Aragam, Y.~Yao, and M.~A. Abidi, ``Survey and analysis of
  multimodal sensor planning and integration for wide area surveillance,''
  \emph{ACM Computing Surveys (CSUR)}, vol.~41, no.~1, p.~7, 2008.

\bibitem{Castro-04-coarse-to-fine}
R.~Castro, R.~Willet, and R.~Nowak, ``Coarse-to-fine manifold learning,'' in
  \emph{Proceedings 2004 International Conference on Acoustics, Speech and
  Signal Processing}, vol.~3, May 2004, pp. iii--992--5.

\bibitem{Castro-05-faster-rate-reg-via-act-learn}
R.~Castro, R.~Willett, and R.~Nowak, ``Faster rates in regression via active
  learning,'' in \emph{Proceedings of the Neural Information Processing Systems
  Conference (NIPS) 2005}, Vancouver, Canada, December 2005.

\bibitem{Bashan-08-opt-2-stage-search}
E.~Bashan, R.~Raich, and A.~O. Hero, III, ``{Optimal Two-Stage Search for
  Sparse Targets Using Convex Criteria},'' \emph{IEEE Transactions on Signal
  Processing}, vol.~56, pp. 5389--5402, 2008.

\bibitem{bashan2011marapTSP}
E.~Bashan, G.~Newstadt, and A.~O. Hero, III, ``{Two-Stage Multi-Scale Search
  for Sparse Targets},'' \emph{IEEE Transactions on Signal Processing},
  vol.~59, no.~5, pp. 2331--2341, 2011.

\bibitem{Hitchings_Castanon_AdaptiveSensing2010}
D.~Hitchings and D.~A. Castanon, ``Adaptive sensing for search with continuous
  actions and observations,'' in \emph{Decision and Control (CDC), 2010 49th
  IEEE Conference on}, 2010, pp. 7443--7448.

\bibitem{haupt2011distilled}
J.~Haupt, R.~M. Castro, and R.~Nowak, ``Distilled sensing: Adaptive sampling
  for sparse detection and estimation,'' \emph{Information Theory, IEEE
  Transactions on}, vol.~57, no.~9, pp. 6222--6235, 2011.

\bibitem{haupt2012sequentially}
J.~Haupt, R.~Baraniuk, R.~Castro, and R.~Nowak, ``Sequentially designed
  compressed sensing,'' in \emph{Statistical Signal Processing Workshop (SSP),
  2012 IEEE}.\hskip 1em plus 0.5em minus 0.4em\relax IEEE, 2012, pp. 401--404.

\bibitem{Wei-Hero-multistage-adaptive-estimation-of-sparse-signals}
D.~Wei and A.~O. Hero, III, ``Multistage adaptive estimation of sparse
  signals,'' \emph{{IEEE} Journal of Selected Topics in Signal Processing},
  vol.~7, no.~5, pp. 783--796, Oct. 2013.

\bibitem{malloy2012sequential}
M.~L. Malloy and R.~Nowak, ``Sequential testing for sparse recovery,''
  \emph{arXiv preprint arXiv:1212.1801}, 2012.

\bibitem{MalloyAdaptiveCompressed13}
M.~Malloy and R.~D. Nowak, ``Near-optimal adaptive compressed sensing,''
  \emph{CoRR}, vol. abs/1306.6239, 2013.

\bibitem{Krishnamurthy-01-hmm-mab-beam-sched}
V.~Krishnamurthy and R.~J. Evans, ``Hidden {Markov} model multiarm bandits: a
  methodology for beam scheduling in multitarget tracking,'' \emph{IEEE
  Transactions on Signal Processing}, vol.~49, no.~12, pp. 2893--2908, December
  2001.

\bibitem{chong2008monte}
E.~Chong, C.~Kreucher, and A.~O. Hero, III, ``Monte-{Carlo}-based partially
  observable {Markov} decision process approximations for adaptive sensing,''
  in \emph{Discrete Event Systems, 2008. WODES 2008. 9th International Workshop
  on}.\hskip 1em plus 0.5em minus 0.4em\relax IEEE, 2008, pp. 173--180.

\bibitem{Wei-Hero-guarantees-adaptive-estimation-sparse-signals}
D.~Wei and A.~O. Hero, III, ``Performance guarantees for adaptive estimation of
  sparse signals,'' Nov. 2013, arXiv:1311.6360.

\bibitem{newstadtThesis2013}
G.~E. Newstadt, ``Adaptive sensing techniques for dynamic target tracking and
  detection with applications to synthetic aperture radars,'' Ph.D.
  dissertation, University of Michigan, May 2013.

\bibitem{Kreucher-05-multi-tar-track-jpdf}
C.~Kreucher, K.~Kastella, and A.~O.~H. {III}, ``Multitarget tracking using the
  joint multitarget probability density,'' \emph{IEEE Transactions on Aerospace
  and Electronic Systems}, vol.~41, no.~4, pp. 1396--1414, October 2005.

\bibitem{bertsekas1999rollout}
D.~P. Bertsekas and D.~A. Castanon, ``Rollout algorithms for stochastic
  scheduling problems,'' \emph{Journal of Heuristics}, vol.~5, no.~1, pp.
  89--108, 1999.

\bibitem{NewstadtWeiHeroCAMSAP13}
G.~E. Newstadt, D.~L. Wei, and A.~O. Hero, III, ``Adaptive search for sparse
  dynamic targets,'' in \emph{2013 5th IEEE International Workshop on
  Computational Advances in Multi-Sensor Adaptive Processing (CAMSAP)}, 2013.

\bibitem{ashInterruptedSAR2013}
\BIBentryALTinterwordspacing
J.~N. Ash, ``Joint imaging and change detection for robust exploitation in
  interrupted {SAR} environments,'' vol. 8746, 2013, pp. 87\,460J--87\,460J--9.
  [Online]. Available: \url{http://dx.doi.org/10.1117/12.2019019}
\BIBentrySTDinterwordspacing

\end{thebibliography}
\ignore{
\begin{IEEEbiography}[]{Gregory Newstadt}
received the B.S. degrees (summa cum laude) from Miami University, Oxford, OH, in 2007 in Electrical Engineering and in Engineering Physics.  He also received the M.S.E in Electrical Engineering: Systems (2009), M.A. in Statistics (2012) and Ph.D. in Electrical Engineering: Systems (2013) degrees from the University of Michigan, Ann Arbor, MI.

He is currently a postdoctoral researcher and lecturer at the University of Michigan, Ann Arbor, MI, in Electrical Engineering (Systems).  His research interests include detection, estimation theory, target tracking, sensor fusion, and statistical signal processing.
\end{IEEEbiography}

\begin{IEEEbiography}[]{Dennis Wei}
has a biography.
\end{IEEEbiography}

\begin{IEEEbiography}[]{Alfred O. Hero, III}
received the B.S. (summa cum laude) from Boston University (1980) and the Ph.D from Princeton University (1984), both in Electrical Engineering. Since 1984 he has been with the University of Michigan, Ann Arbor, where he is the R. Jamison and Betty Professor of Engineering. His primary appointment is in the Department of Electrical Engineering and Computer Science and he also has appointments, by courtesy, in the Department of Biomedical Engineering and the Department of Statistics. In 2008 he was awarded the the Digiteo Chaire d'Excellence, sponsored by Digiteo Research Park in Paris, located at the Ecole Superieure d'Electricite, Gif-sur-Yvette, France. He has held other visiting positions at LIDS Massachussets Institute of Technology (2006), Boston University (2006), I3S University of Nice, Sophia-Antipolis, France (2001), Ecole Normale Sup\'erieure de Lyon (1999), Ecole Nationale Sup\'erieure des T\'el\'ecommunications, Paris (1999), Lucent Bell Laboratories (1999), Scientific Research Labs of the Ford Motor Company, Dearborn, Michigan (1993), Ecole Nationale Superieure des Techniques Avancees (ENSTA), Ecole Superieure d'Electricite, Paris (1990), and M.I.T. Lincoln Laboratory (1987 - 1989).

Alfred Hero is a Fellow of the Institute of Electrical and Electronics Engineers (IEEE). He has been plenary and keynote speaker at major workshops and conferences. He has received several best paper awards including: a IEEE Signal Processing Society Best Paper Award (1998), the Best Original Paper Award from the Journal of Flow Cytometry (2008), and the Best Magazine Paper Award from the IEEE Signal Processing Society (2010). He received a IEEE Signal Processing Society Meritorious Service Award (1998), a IEEE Third Millenium Medal (2000) and a IEEE Signal Processing Society Distinguished Lecturership (2002). He was President of the IEEE Signal Processing Society (2006-2007). He sits on the Board of Directors of IEEE (2009-2011) where he is Director Division IX (Signals and Applications).

Alfred Hero's recent research interests have been in detection, classification, pattern analysis, and adaptive sampling for spatio-temporal data. Of particular interest are applications to network security, multi-modal sensing and tracking, biomedical imaging, and genomic signal processing.
\end{IEEEbiography}
}
\end{document}